\documentclass[11pt, a4paper]{scrartcl}
\usepackage{amsmath, amssymb, amsthm, color, graphicx,
            subfigure, paralist} %
\usepackage[bookmarks,colorlinks,breaklinks]{hyperref}
\definecolor{dullmagenta}{rgb}{0.4,0,0.4}   
\definecolor{darkblue}{rgb}{0,0,0.4}
\definecolor{darkgreen}{rgb}{0,0.6,0}
\definecolor{darkred}{rgb}{0.6,0,0}
\hypersetup{linkcolor=darkblue,citecolor=blue,filecolor=dullmagenta,urlcolor=darkblue} 
\newcommand{\eqnumtag}{%
   \refstepcounter{equation}%
   \tag{\theequation}%
}

\newtheorem {theorem}{Theorem}
\newtheorem {lemma}[theorem]{Lemma}
\newtheorem {proposition}[theorem]{Proposition}
\newtheorem {corollary}[theorem]{Corollary}

\newtheorem {defi}[theorem]{Definition}

\newtheorem {remark}[theorem]{Remark}
\newtheorem {remarks}[theorem]{Remarks}

\newcommand{\proofof}[1] {\noindent \textsc{Proof of {#1}.} }

\numberwithin{equation}{section}
\numberwithin{theorem}{section}
\numberwithin{figure}{section}

\usepackage{mathtools}

\newcommand{\ds} {\displaystyle}

\newcommand{\beq}{\begin{equation}}
\newcommand{\eeq}{\end{equation}}
\newcommand{\Leq}[1]{\label{#1}\end{equation}}
\newcommand{\beqn}{\begin{eqnarray}}
\newcommand{\eeqn}{\end{eqnarray}}
\newcommand{\beqno}{\begin{eqnarray*}}
\newcommand{\eeqno}{\end{eqnarray*}}

\renewcommand {\l}{\left}

\newcommand {\ri}{\right}
\newcommand {\vv}{\varphi}

\newcommand {\pa}{\partial}

\newcommand {\bR}{{\mathbb R}}
\newcommand {\bN}{{\mathbb N}}

\newcommand {\bC}{{\mathbb C}}

\newcommand{\rstr}{{\upharpoonright}}
\newcommand{\idty}{{\rm 1\mskip-4mu l}} 

\newcommand {\Id}{\idty}

\newcommand{\cB}{{\mathcal B}} 
\newcommand{\cC}{{\mathcal C}} %
\newcommand{\cE}{{\mathcal E}} %
\newcommand{\cF}{{\mathcal F}} %
\newcommand{\cG}{{\mathcal G}} %
\newcommand{\cH}{{\mathcal H}}

\newcommand{\cK}{{\mathcal K}}
\newcommand{\cL}{{\mathcal L}}

\newcommand{\cO}{{\mathcal O}} 

\newcommand{\cR}{{\mathcal R}}
\newcommand{\cV}{{\mathcal V}}

\newcommand{\cZ}{{\mathcal Z}}

\newcommand{\z}{\setminus\{0\}}

\newcommand{\ov}{\overline}

\newcommand{\bem}{\l(\! \begin{array}}
\newcommand{\eem}{\end{array}\!\ri)}
\newcommand{\bsm}{\left(\begin{smallmatrix}} 
\newcommand{\esm}{\end{smallmatrix}\right)}  

\newcommand{\qmbox}[1]{\quad\mbox{#1}\quad}

\newcommand{\dist}{{\rm dist}}
\newcommand{\te}{\widetilde{\eta}}

\newcommand{\s}{\sin}
\newcommand{\sh}{\sinh}
\newcommand{\co}{\cos}
\newcommand{\ch}{\cosh}

\newcommand{\tv}{\tilde{v}}

\newcommand{\rst}[1]{\ensuremath{{\mathbin\upharpoonright}%
\raise-.5ex\hbox{$\scriptstyle #1$}}}

\newcommand{\Mid}{\;\middle\vert\;}

\DeclareMathOperator{\arcsinh}{arcsinh}
\DeclareMathOperator{\ce}{ce}
\DeclareMathOperator{\se}{se}
\DeclareMathOperator{\sech}{sech}
\DeclareMathOperator{\Log}{Log}

\newcommand {\eh}{{\textstyle \frac{1}{2}}}
\usepackage[normalem]{ulem} 

\begin{document}
\title{Resonances in the\\ Two-Centers Coulomb Systems}
\author{Marcello Seri\thanks{Department of Mathematics and Statistics
University of Reading, Whiteknights, PO Box 220, Reading RG6 6AX (UK),
\texttt{m.seri@reading.ac.uk}}, 
Andreas Knauf\thanks{
Department of Mathematics,
Friedrich-Alexander-University Erlangen-Nuremberg,
Cauerstr.\ 11, D-91058 Erlangen,
Germany, \texttt{knauf@math.fau.de}}, 
Mirko Degli Esposti\thanks{
Dipartimento di Matematica, Universit\`a di Bologna,
Piazza di Porta S. Donato, 5, I-40127 Bologna, Italy,
\texttt{mirko.degliesposti@unibo.it}} 
\ and Thierry Jecko\thanks{
D\'epartement de Math\'ematiques, Universit\'e de Cergy-Pontoise, 
Site de Saint Martin, 2, avenue Adolphe Chauvin, 
F-95000 Cergy-Pontoise, France,
\texttt{thierry.jecko@u-cergy.fr}}}
\date{}

\maketitle

\begin{abstract}
We investigate the existence of resonances for two-centers Coulomb
systems with arbitrary charges in two dimensions, defining
them in terms of generalised complex eigenvalues of a non-selfadjoint
deformation of the two-centers Schr\"odinger operator. We construct
the resolvent kernels of the operators and prove that they can be
extended analytically to the second Riemann sheet. The resonances
are then analysed by means of perturbation theory and numerical methods.

\end{abstract}
\tableofcontents
%

%
\section{Introduction}\label{I}
%

Our work concerns the study of the quantum mechanical two-fixed-centers
 Coulomb systems in two dimensions. The two-dimensional restriction of
the two-centers problem arises naturally in the analysis of the 
three-dimensional problem and, as described in \cite{thesis}, it is essential
to be able to analyse that case.

Since three centuries the two-centers Coulombic systems have been studied,
from a classical and later also from a quantum mechanical point of view,
starting from pioneering works of Euler, Jacobi \cite{Jacobi} and
Pauli \cite{Pauli} and going on until the recent years. For an historical
overview we refer the reader to \cite{thesis}.

The interest for the quantum mechanical version of the problem comes
mainly from molecular physics. Indeed it defines the simplest model for
one-electron diatomic molecules (e.g. the ions ${\rm H}_{2}^{+}$ and
${\rm He}\, {\rm H}^{++}$) and
a first approximation of diatomic molecules in the Born-Oppenheimer
representation.

In fact many of the results in the literature are related to the hard
problem of finding algorithms to obtain good numerical approximations of
the discrete spectrum and of the scattering waves \cite{GG, RecentContTwo, 
Le86, liu, RecentSpecHydro}, and to the asymptotic analysis of spectral properties in the very small or very large center distance \cite{bbs66,lots86,H1RExpEVRel,klaus83}. In contrast, really little is known
on the regularity of the solutions with respect to the parameters of
the system \cite{Slavyanov} and even less on the problem of resonances.

Quantum resonances are a key notion of quantum physics: roughly
speaking these are scattering states (i.e. states of the essential
spectrum) that for long time behave like bound states
(i.e. eigenfunctions). They are usually defined as poles of a
meromorphic function, but note that there is no consensus on their definition and their study
\cite{zwo}. On the other hand, it is known that many of their 
definitions coincide in some settings \cite{hemartinez} and that
their existence is related to the presence of some classical orbits
``trapped'' by the potential.

If a quantum systems has a potential presenting a positive local
minimum above its upper limit at infinity, for example, it is usually
possible to find quantum resonances, called shape resonances. These are
related to the classical bounded trajectories around the local minimum
\cite{H-S}. These are not the only possible ones: it has been
proven in \cite{BCD2, BCD3, GeSj, Sj} that there can be resonances
generated by closed hyperbolic trajectories or by a non-degenerate
maximum of the potential. The main difference is that the shape
resonances appear to be localised much closer to the real axis
with respect to these last ones.

Even the presence or absence of these resonances is strictly related to
the classical dynamics. In fact it is possible to use some classical
estimates, called non-trapping conditions, to prove the existence of
resonance free regions (see for example \cite{BCD1,mart1,mart1b}).

A major shortcoming of the actual theory of resonances is that the existence
and localisation results require the potentials to be smooth or analytic
everywhere, with the exception of few results concerning non-existence
\cite{mart1, mart1b} or restricting to centrally symmetric cases \cite{AgmonKlein}.

In this sense, the two-centers problem represents a very good test
field. In fact, it is not centrally symmetric but presents still enough
symmetries to  be separated (see Theorem~\ref{thm:sepThm}).
This allows us to shift most of the analysis from the theory of
PDEs with singular potentials to the theory of ODEs,
simpler and more explicitly accessible.

Moreover, the two-centers models present all the previously cited
classical features related to the existence of resonances: the
non-trapping condition fails to hold \cite{CJK}, there are closed hyperbolic
trajectories with positive energies \cite{KK, S14} and there is a
family of bounded trajectories with positive energies \cite{S14}.
At the same time, the energy ranges corresponding to the closed hyperbolic
trajectories and to the bounded ones are explicitly known \cite{S14}.

In general the relation between different definitions of resonances is not fully understood, even for smooth symbols.
In this work we define a notion of resonances for the two-centers Coulomb system. 
These are defined as poles of the meromorphic extension of the Green's functions 
of the separated equations. We then show how to approximate them in 
different semiclassical energy regimes.

These approximations lead to strong evidence that relates
the energies of the resonances far from the real axis (i.e. not-exponentially
close to it w.r.t.~the semiclassical parameter) to that of the closed
hyperbolic trajectories.

Our work is strongly inspired by \cite{AgmonKlein} but we treat a more interesting situation since the scattering by two nuclei is richer than the one by one nucleus. We get similar results as in \cite{AgmonKlein}, except for the expansion of the Green function in partial waves. In \cite{AgmonKlein}, the latter can be justified thanks to a special property of spherical harmonics. We did not succeed in proving it in our context (and this would be an important result). This explains why we did not completely connect our definition of resonances to usual ones.

Compared to other results on resonances, we provide quite precise informations in an usually unpleasant context since our potential (as in \cite{AgmonKlein}) contains Coulomb singularities. Except for some results in Section \ref{sec-apprResKxi}, our main contributions are not of semi-classical nature in contrast to those in \cite{BCD1,BCD2,BCD3,Sj}.

The structure of the paper is as follows.

In Section~\ref{c4}
we introduce the two-centers problem both in its classical and quantum
mechanical formulation. We describe its main properties and the
separation of the differential equation associated to the operator
into radial and angular equations.

In Section~\ref{sec:spec2d} we describe the spectrum of the operator obtained
from the angular differential equation and the properties of its analytic continuation.

In Section~\ref{Ab} we focus on the spectrum of the operator obtained from the
radial differential equation and the analytic continuation of its resolvent.
This is done constructing explicitly two linearly independent solutions with
prescribed asymptotic behaviour. They mimic the incoming and outgoing waves
of scattering theory, in fact we will use them to construct the Jost
functions, and consequently define and analyse the Green's function
and the scattering matrix.
The main results are contained in Theorem~\ref{thm:asympsol} and Theorem~\ref{thm:anaext} and their corollaries. In particular they provide the key ingredients to define the Jost functions and their analytic continuation in Corollary~\ref{cor:formfpm}.
In Theorem~\ref{thm:asympsol} is proven the existence and uniqueness of the 
incoming and outgoing waves for real and complex values of the parameters.
In Theorem~\ref{thm:anaext}, it is shown that these solutions admit an
analytic continuation across the positive real axis into the second Riemann
sheet. 

In Section~\ref{Pw1} we explain how the resolvent of the two-centers system
relates to the angular and radial operators.

In Section~\ref{sec:defres2d} we apply the theory developed for the angular
and radial operators to the objects described in Section~\ref{Pw1}.
Here we define the resonances for the two-centers problem (see
\eqref{eq:defRes2d}) and analyse some of their properties.
The rest of the section is devoted to the computation of approximated
values of the resonances in different semiclassical energy regimes, see
in particular \eqref{eq-Enmvalue3}, \eqref{eq-rescoupling}
and \eqref{eq:AppResAft}.

In Section~\ref{sec:numerical} we use the approximations obtained in
the previous section to compute the resonances and study their
relationship with the structure of the underlying classical systems.
The numerics strongly support the relation between the resonances that
we've found and the classical closed hyperbolic trajectories.

In Section~\ref{conc} we make some additional comments relating our results
for the planar two-centers problem to the three-dimensional one and to the
$n$-centers problem.

In the Appendix~\ref{appendix} we describe how to modify the generalised
Pr\"ufer transformation in the semi-classical limit to get precise
high-energy estimates.
These results are needed for the high-energy approximation obtained in
Section~\ref{sec:hee}.

{\bf Notation.}
In this article $ \bN=\{1,2,3,\ldots\}$, $\bR^{*}:=\bR\setminus\{0\}$.

%
\section{The two-centers system on $L^{2}(\bR^{2})$}\label{c4}
%

\subsection{The two-centers Coulomb system}

We consider the operator in $L^2(\bR^2)$, given by
\beq
\cH := -h^2\Delta +V(q) \qmbox{with}
V(q) := \frac{-Z_1}{|q-s_1|} + \frac{-Z_2}{|q-s_2|},
\Leq{1}
where $h>0$ is a small parameter.

This describes the motion of an electron in the field of two nuclei of
charges $Z_i\in\bR^{*} =\bR\setminus\{0\}$, fixed at positions
$s_1\neq s_2\in\bR^2$, taking into account only the electrostatic
force.
By the unitary realisation $Uf(x) := |\det A|^{-1/2} f(Ax +b)$ of an
affinity of $\bR^2$ we assume that the two centers are at
$s_1 := a := \bsm 1\\0\esm$ and $s_2:=-a$.

\begin{remarks}$ $\\\vspace{-0.4cm}
\begin{compactitem}
\item Notice that if we set $Z_{1} = Z_{2} >0$ in the operator in
  (\ref{1}), we get the Schr\"odinger operator for the simply ionized
  hydrogen molecule ${\rm H}_{2}^{+}$ \cite{bbs66,lots86,SR}, whereas for
  $Z_{1} = -Z_{2}$ it describes an electron moving in the field of a
  proton and an anti-proton \cite{H1RExpEVRel,klaus83}.
  Another example covered by this model is the doubly charged
  helium-hydride molecular ion ${\rm He}\, {\rm H}^{++}$, with
  $Z_1=2Z_2>0$, see \cite{WDR}.
\item Even if \eqref{1} does not directly describe the interactions in molecules, it is related to the study of scattering theory for such systems. In Example 1.3 in \cite{CJK}, the scattering of a heavy particle by a molecule is partially studied and, thanks to a natural physical assumption, the Hamiltonian of the heavy particle is given by \eqref{1} plus an additional potential correction. In the paper \cite{JKW}, scattering cross sections for diatomic molecules are estimated in a semi-classical regime related to the Born-Oppenheimer approximation. A Schr\"odinger operator of the type \eqref{1} enters in the computations as an effective Hamiltonian for the scattering process. 
  \hfill$\Diamond$
\end{compactitem}
\end{remarks}

\subsection{Elliptic coordinates}\label{sec:pec3dintro}
The restriction to the rectangle $M:=(0,\infty)\times(-\pi,\pi)$ of the
map
\beq
  G:\bR^2\to\bR^2\qmbox{,} \bsm\xi \\
  \eta \esm \mapsto \bsm \ch(\xi)\co(\eta) \\
  \sh(\xi) \s(\eta) \esm
\Leq{elliptic}
defines a $C^{\infty}$ diffeomorphism \beq G:M\to G(M) \Leq{def-G}
whose image $G(M)=\bR^{2}\setminus(\bR\times\{0\})$ is dense in
$\bR^{2}$. Moreover it defines a change of coordinates from
$q\in\bR^{2}$ to $(\xi,\eta)\in M$. These new coordinates are called
\emph{elliptic coordinates}.

\begin{figure}[h!]
\begin{center}
 \includegraphics[width=0.4\linewidth]{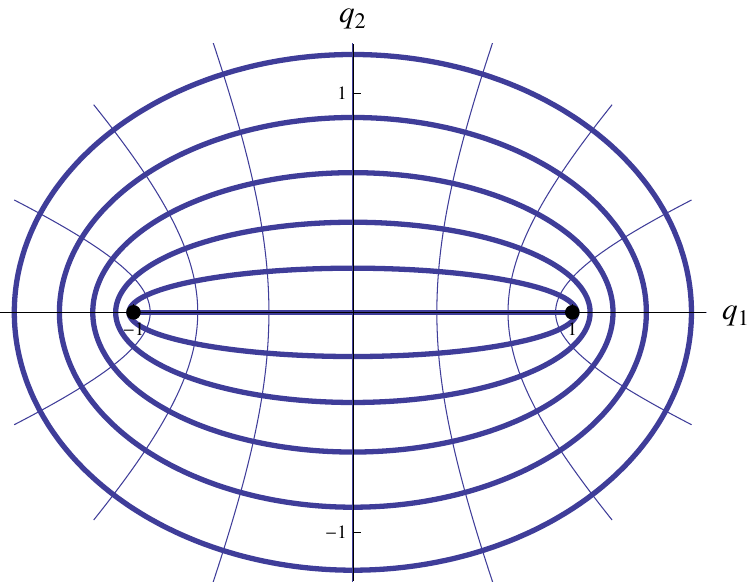}
\caption{\small Elliptic coordinates.}
\label{fig:coordPE}
\end{center}
\end{figure}

\begin{remarks}\label{rmk:coord}
\begin{compactenum}
\item In the $(q_{1},q_{2})$-plane the curves $\xi=c$ are ellipses
  with foci at $\pm a$, while the curves $\eta=c$ are confocal half
  hyperbolas, see Figure \ref{fig:coordPE}.
\item The Jacobian determinant of $G$ equals
  \beq
    F(\xi,\eta) := \det(DG(\xi,\eta)) =
    \sh^{2}(\xi)+\s^{2}(\eta) =
    \ch^2(\xi)-\co^2(\eta).
  \Leq{def-F}

  Thus the coordinate change \eqref{elliptic} is degenerate at the
  points $(\xi,\eta)\in\{0\}\times\{0,\pm\pi\}$ in $\overline{M}$.
  For $\xi=0$ the $\eta$ coordinate parametrizes the $q_{1}$-axis
  interval between the two centers. For $\eta=0$ ($\eta=\pm\pi$) the
  $\xi$ coordinate parametrizes the positive (negative) $q_{1}$-axis
  with $|q_{1}|>1$.  \hfill$\Diamond$
\end{compactenum}
\end{remarks}

\subsection{Classical results}
\label{sec:classical}

The classical analogue of \eqref{1} is described by the Hamiltonian
function on the cotangent bundle $T^{*}Q_2$ of $Q_{2} := \bR^{2}\setminus\{\pm a\}$
relative to the two-center potential given by:
\begin{equation}\label{1c}
  H:T^*Q_2\to \bR\qmbox{,}
  H(p,q) := \frac{|p|^{2}}2 + \frac{-Z_1}{|q-a|} + \frac{-Z_2}{|q+a|}.
\end{equation}

\begin{lemma}[see e.g. \cite{S14}]
Using $G$ defined in \eqref{def-G}, and $Z_\pm := Z_2 \pm Z_1$,
$H$ is transformed by the elliptic coordinates into
\beq
  H\circ {(G^{-1})^*}(p_\xi,p_\eta,\xi,\eta) =
  \frac1{F(\xi,\eta)}\big(H_{1}(p_\xi,\xi) + H_{2}(p_\eta,\eta)\big)
\Leq{eq:sep-midstep0}
where  $(G^{-1})^*:T^*M\to T^*Q_2$ is the cotangential lift of $G^{-1}$,
and
\beq
  H_{1}(p_\xi,\xi) := \frac{p_{\xi}^{2}}2 - Z_{+} \ch(\xi)\qmbox{,}
  H_{2}(p_\eta,\eta) := \frac{p_{\eta}^{2}}2 + Z_{-} \co(\eta).
\Leq{eq:sep-midstep1}
There are two functionally independent constants of motion
$H$ and $L:=H_{1} - \ch^{2}(\xi)H$ with values $E$ and $K$ respectively.
\end{lemma}

Taken together, the constants of motion define a vector-valued
function on the phase space of a Hamiltonian.  We can study the
structure of the preimages of this function (its level sets), in
particular their topology. In the simplest case the level sets are
mutually diffeomorphic manifolds.

\begin{defi} {\bf (see \cite[Section 4.5]{AM})}\label{def:bs}
  Given two manifolds $M, N$, $f\in C^{\infty}(M,N)$ is called
  {\bf locally trivial} at $y_{0}\in N$ if there exists a neighborhood
  $V\subseteq N$ of $y_{0}$ such that $f^{-1}(y)$ is a smooth
  submanifold of $M$ for all $y\in V$ and there there is a map
  $g\in C^{\infty}(f^{-1}(V),f^{-1}(y_{0}))$ such that
  $f\times g:f^{-1}(V)\to V\times f^{-1}(y_{0})$ is a diffeomorphism.

The {\bf bifurcation set} of $f$ is the set
\[
 \cB(f) := \{ y_{0}\in N \mid f
 \mbox{ is not locally trivial at } y_{0}\}.
\]
\end{defi}
Notice that if $f$ is locally trivial, the restriction
$g\rst{f^{-1}(y)}:f^{-1}(y)\to f^{-1}(y_{0})$
is a diffeomorphism for every $y\in V$.
\begin{remark}
The critical points of $f$ lie in
$\cB(f)$ (see \cite[Prop. 4.5.1]{AM}), but the converse is true
only in the case $f$ is proper (i.e. it has compact preimages).
\hfill$\Diamond$
\end{remark}

Define the function on the phase space as follows
{(omitting a projection in the second component)}
\beq
  \cF := \bsm H \\ H_{\xi} \circ G^*\esm : T^{*}Q_2\to \bR^{2},
\Leq{def:enmomMapping}
where $H_{\xi}(p_\xi,\xi) := H_1(p_\xi,\xi) - \cosh^2(\xi) E$.

\begin{theorem}[\cite{S14}]\label{thm:defbifdia}
Let $(Z_{1},Z_{2})\in\bR^{*}\times\bR^{*}$, then the bifurcation
set of \eqref{def:enmomMapping} for positive energies equals
\[
\cB\l(\cF\ri)
\cap\;(\bR_+\times\bR)
 =
\l\{
(E,K)\in\cL \mid
E\geq 0
\text{ and }
K_{+}(E)\leq K \leq K_{-}(E)
\ri\}.
\]
Here
$
\cL := \cL_{0}\cup\cL_{-}^{1}\cup\cL_{-}^{2}\cup\cL_{-}^{3}
       \cup\cL_{+}^{2}\cup\cL_{+}^{3}{\subset \bR^2}
$
with
\beq
\begin{array}{lcl}
\cL_{0} := \{E=0\}
, &\qquad& \cL_{-}^{1} := \{K=Z_{-}-E\}, \\
\cL_{+}^{2} := \{K=-Z_{+}-E\}, &\qquad&
  \cL_{-}^{2} := \{K=-Z_{-}-E\}, \\
\cL_{+}^{3} := \{4EK=Z_{+}^{2}\}, &\qquad&
  \cL_{-}^{3} := \{4EK=Z_{-}^{2}\},
\end{array}
\Leq{eq:bifCurves}
and $K_+$ and $K_-$ are defined by
\[
K_{+}(E) :=
\begin{cases}
- \infty, & E > 0 \\
-(Z_{+} + E), & E \leq \min\l(-\frac{Z_{+}}2, 0\ri) \\
\frac{Z_{+}^{2}}{4E}, & 0 \geq E >  \min\l(-\frac{Z_{+}}2, 0\ri)
\end{cases},
\]
\[
K_{-}(E) :=
\begin{cases}
Z_{-} - E, & E \leq \frac{Z_{-}}2 \\
\frac{Z_{-}^{2}}{4E}, & E >  \frac{Z_{-}}2
\end{cases}.
\]
\end{theorem}

The energies lying on the line $\cL^2_{+}$ are the ones
associated with the closed hyperbolic trajectory bouncing between
the two centers \cite{S14}.

Moreover, for $|Z_{+}| < Z_{-}$ the set of energy parameters included
in the region $\{E\geq 0\}\cap\l\{ (E,K)\in\cL_{+}^{3} \Mid E <
\frac{|Z_{+}|}{2}\ri\}$ and contained between the curves $\cL_{+}^{2}$ and
$\cL_{+}^{1}$ is somewhat special: on the configuration space they are
associated with a family of bounded trajectories trapped near the attracting
center \cite{S14}.

\subsection{Separation in elliptic coordinates}
The importance of the change of coordinate \eqref{def-G} for the
quantum problem is clarified by the following well-known theorem (see
e.g.\ \cite{babhas}).
Here we enlarge the domain of $G$ to $\ov{M}$.

\begin{theorem}\label{thm:sepThm}
Let
$u\in C_{a}(\bR^2) := \l\{
u \in C(\bR^{2}) \;\Big|\; u\rstr_{\bR^{2}\setminus\{\pm a\}}
\text{ is twice continuously differentiable}
\ri\}$.
The eigenvalue equation
\[
\big(-h^{2}\Delta + V(q)\big) u(q) = E u(q), \quad E\in \bR,
\]
transformed to prolate elliptic coordinates, separates with the ansatz
\[
u \circ G(\xi, \eta) = f(\xi) g(\eta)
\]
into the decoupled system of ordinary differential equations
\[
\begin{cases}
\l(-h^2\pa_\xi^2 - Z_+\ch(\xi) - E\ch^2(\xi) + \mu\ri) f(\xi) = 0 \\
\l(-h^2\pa_\eta^2  + Z_-\co(\eta) + E\co^2(\eta) - \mu\ri) g(\eta) = 0,
\end{cases}
\]
where $\mu\in\bC$ is the separation constant,
\[
\begin{aligned}
f &\in \ C^{2}_N([0,\infty))\ :=
  \l\{ h\in C^{2}([0,\infty)) \mid h'(0)=0\ri\},\\
g &\in \ C^{2}_{\rm per}([-\pi,\pi])\ :=
  \l\{h\in C^{2}([-\pi,\pi]) \mid h^{(k)}(-\pi) = h^{(k)}(\pi)
  \text{ for } k=0,1
\ri\}
\end{aligned}
\]
and we have set $Z_{\pm}:=Z_{2}\pm Z_{1}$ and $\pa_{\alpha} =
\frac{\pa}{\pa\alpha}$.
\end{theorem}

\begin{remark}
Without loss we assume $Z_{-}\in[0,\infty)$ and $Z_{+}\in\bR$, $Z_{+}\neq Z_{-}$, i.e. $Z_{2} \geq Z_{1}$.
\hfill$\Diamond$\end{remark}
\begin{remark}

Since $G$ is a diffeomorphism and since
$F$ defined in (\ref{def-F}) equals $\det(DG)$, the
transformation to prolate elliptic coordinates $(\xi, \eta)$ defines a
unitary operator
\[
\cG : L^{2}(\bR^{2}, dq) \to L^{2}(M,d\chi)\qmbox{, with}
d\chi := F(\xi,\eta)\,d\xi \, d\eta.
\]
\hfill$\Diamond$\end{remark}

\proofof{\ref{thm:sepThm}}
We set $r_1 := |q-s_{1}|$, $r_2 := |q-s_{2}|$ and transform to \emph{elliptic coordinates}. We have
\[
r_{2,1}^2 = (q_1\pm1)^2+q_2^2 = \big(\ch(\xi)\pm\co(\eta)\big)^2.
\]
Thus the distances from the centers equal
\[
r_1 = \cosh\xi-\cos \eta \qmbox{and} r_2 = \cosh\xi+\cos\eta.
\]
For $F(\xi,\eta) = \sh^{2}(\xi)+\s^{2}(\eta) = \ch^2(\xi)-\co^2(\eta)$
we obtain
\[
 V\circ G(\xi,\eta) =
  - \frac{Z_1}{ |q-a| } - \frac{Z_2}{ |q+a| }
  = {-\frac{Z_+\ch(\xi)-Z_-\co(\eta)}{F(\xi,\eta)}}
\]
and the Laplacian $\Delta$ acts in
elliptic coordinates as
\beq
  \Delta_{\cG} :=
  \frac{1}{F(\xi,\eta)} \l(\pa_\xi^2 + \pa_\eta^2 \ri).
\eeq
With the ansatz
\[
\widetilde{u}(\xi,\eta) = f(\xi)g(\eta)
\qmbox{ with }
f\in C^{2}_{N}([0,\infty))
\mbox{ and }
g\in C^{2}_{\rm per}([-\pi,\pi])
\]
the first equation separates and we obtain the decoupled system of
ordinary differential equations \beq -h^{2}\pa_\xi^2 f(\xi) +
\l(V_{\xi}(\xi)+\mu\ri)f(\xi) = 0 \qmbox{,} -h^{2}\pa_\eta^2 g(\eta) +
\l(V_{\eta}(\eta)-\mu\ri)g(\eta) = 0 \eeq where $V_{\xi}$ and
$V_{\eta}$ are the multiplication operators for the functions
\beq
  V_{\xi}(\xi) := - Z_+\ch(\xi) - E\ch^2(\xi)
  \qmbox{,}
  V_{\eta}(\eta):= Z_-\co(\eta) + E\co^2(\eta)
\eeq
\qed

\begin{remark}
  Here the separation constant $\mu$ plays the role of the spectral
  parameter in time independent Schr\"odinger equations, and energy
  $E$ the one of a coupling constant.  \hfill$\Diamond$
\end{remark}

\begin{proposition}\label{prop:saExt}
The operator $\cH$ on $L^2(\bR^2)$ defined as in (\ref{1})
is unitarily equivalent to the operator in $L^{2}(M,d\chi)$, given by
\[
\cH_{\cG} := -h^{2} \Delta_{\cG} + V_{\cG}
\qmbox{with}
V_{\cG}(\xi,\eta) := - \frac{Z_{+}\ch(\xi) - Z_{-} \co(\eta)}{F(\xi,\eta)}.
\]
$\cH_\cG$ has form core
$$
\cG\l(C_0^\infty(\bR^2)\ri) = \l\{ f\in C_0^\infty\l(\overline M\ri)
\mid f(\xi,\pi) = f(\xi,-\pi)\mbox{ and } \pa_\xi f|_{\xi=0}=0
\ri\}.
$$
It admits a unique self-adjoint realisation with domain
$\cG(D(\cH))$ with
\beq
D(\cH) := \l\{ u\in L^2(\bR^2) \mid Vu\in
  L^1_{\rm loc}(\bR^2),\; u\in H^1_{\rm loc}(\bR^2),\;
  \cH u \in L^2(\bR^2) \ri\},
\Leq{eq:domain2d}
where $\cH u$ is to be understood in distributional sense.
\end{proposition}
\proof It is well-known that $\cH$ has a self-adjoint realisation on
$L^2(\bR^2)$.  The proof is based on the infinitesimal form
boundedness of $V$ w.r.t.\ $\Delta$ \cite[Theorem 3.2]{Ag82} and the
KLMN Theorem \cite[Theorem X.17]{RSII}.  In this way the operator is
well-defined and has form domain $H^{1}(\bR^{2})$.  Moreover its
domain $D(\cH)$ is given by \eqref{eq:domain2d}, see \cite[Theorem
3.2]{Ag82}.

The domain of the unitarily transformed $\cH_{\cG}=\cG \cH\cG^{-1}$
is then transformed to $\cG(D(\cH))$.

Finally $C_0^\infty(\bR^2)$ is a form core for the quadratic form
associated to $\cH$, therefore it is unitarily transformed to a form
core for the quadratic form associated to $\cH_\cG$. See \cite[Section
VIII.6]{RSI} for the definitions. The form of the operator is given by
Theorem \ref{thm:sepThm}.  \qed

It is natural at this point to move our point of view from the study
of $\cH_{\cG} - E$ on $L^{2}(M,d\chi)$ to the study of the separable
operator
\[
K_{E} := K_{\xi}\otimes\Id+ \Id\otimes K_{\eta}
\]
acting on
$L^{2}(M,d\xi\, d\eta) = L^{2}([0,\infty), d\xi) \otimes
L^{2}([-\pi,\pi],d\eta)$.
Here
\beq\begin{split}
K_{\xi}(h) &:= K_{\xi, E, h} :=
  -h^{2}\pa_\xi^2 - Z_+ \ch(\xi) - E \ch^{2}(\xi),\\
K_{\eta}(h) &:= K_{\eta, E, h} :=
  -h^{2}\pa_\eta^2  + Z_- \co(\eta) + E \co^2(\eta).
\end{split}\Leq{def-Kxieta}

In fact, the separation reduces the problem to the study of two
Sturm-Liouville equations
\beq
  (K_{\xi} + \mu) f(\xi) = 0 \qmbox{ and }
  (K_{\eta} - \mu) g(\eta) = 0.
\Leq{SLSL}
Following the standard convention used in the literature, we will call
the first equation \emph{radial equation} and the second equation
\emph{angular equation}.  For the proper boundary conditions on
$L^{2}([0,\infty), d\xi)$ respectively $L^{2}([-\pi,\pi],d\eta)$ they
define essentially self-adjoint operators.

More specifically the eigenvalue equation of $K_{\eta}(h)$ is in the
class of the so called Hill's equation.  In view of Proposition
\ref{prop:saExt}, we are interested in the $2\pi$-periodic solutions
of the equation, i.e. we look for $g\in L^2([-\pi,\pi],d\eta)$
such that
\[
g(-\pi) = g(\pi) \qmbox{and} g'(-\pi) = g'(\pi).
\]

For $K_{\xi}(h)$ it is clear that $0$ is a regular point, 
we will see later how to treat the singular point $\infty$
(we refer the reader to \cite{We} for additional information concerning
regular and singular points of Sturm-Liouville Problems).
For what concerns the boundary conditions in $0$, as suggested by Proposition
\ref{prop:saExt} we will require \beq f'(0) = 0.  \Leq{eq:BC2dR}
The transformation needed to move from $\cH_\cG - E$ to
\eqref{def-Kxieta} is obviously not unitary, as we are passing from a
semibounded operator to a family of non-semibounded ones.  On the
other hand, their spectra are related, and we will study
$\sigma(\cH_\cG)$ by means of the spectra associated to
\eqref{def-Kxieta}.

%
\section{Spectrum of the angular operator and its analytic continuation}\label{sec:spec2d}
%

We now turn the attention to the second equation in \eqref{SLSL}, the
angular equation.
Let
\beq 
T := T_\eta(Z_-, h, \mu, E) := h^{-2}K_\eta(h) - h^{-2}E,
\Leq{eq:eigeqT} 
with parameters $Z_{-}\in \bR$ and $E\in(0,\infty)$. 
With this definition, $h^2 [T\psi](\eta) = 0$ denotes the eigenvalue equation 
for $K_{\eta}$.

We start considering the simpler case of equal charges ($Z_{-} = 0$). 
Then the eigenvalue equation $[T\psi](\eta) = 0$ is the
Mathieu equation
\beq
  [T\psi](\eta) =
  -\pa_{\eta}^{2}\psi(\eta) - \frac{2\mu - E}{2h^{2}}\psi(\eta)
  + 2 \frac E{4h^{2}} \cos(2\eta)\psi(\eta) = 0
\Leq{eq-kappaeta}
with periodic boundary conditions in $[-\pi,\pi]$.
We apply Floquet theory (see \cite{Ea,MW,MS,TeODE}), using the
fundamental matrix
\beq
  \cF(\lambda, \delta) := \bsm f_{1} & f_{2} \\
  f_{1}' & f_{2}' \esm (\pi ; \lambda, \delta),
  \qquad \lambda:=\frac{2\mu - E}{2h^{2}},\;
  \delta:= \frac E{4h^{2}},
\Leq{eq:fSolMathNot}
built from the fundamental system
of solutions $\eta \mapsto f_{i}(\eta; \lambda, \delta)$, with
\beq
  f_{1}(0; \lambda, \delta) = 1
  = f_{2}'(0; \lambda, \delta) \qmbox{,} f_{2}(0; \lambda, \delta)
  = 0 = f_{1}'(0; \lambda, \delta)
\Leq{eq:fundSolMath}
(henceforth the prime $'$ means the partial derivative
w.r.t.\ the first variable).
The potential $V(\eta) := \cos(2\eta)$ being even, it follows that all
the $2\pi$-periodic solutions must be either $\pi$-periodic or
$\pi$-antiperiodic in $[0,\pi]$ (or $[-\pi,0]$).

The structure of the periodic solutions and their eigenvalues for the
Mathieu equation is well-understood (see \cite[Chapter 2]{MS}): For
each integer $n\geq 0$ one finds two solutions
$\ce_{n}(\bullet;\delta)$ and $\se_{n+1}(\bullet;\delta)$, called
\emph{Mathieu Cosine} and \emph{Mathieu Sine} respectively, that have
exactly $n$ zeroes in $(0,\pi)$ and that are $\pi$-periodic for even
$n$ and $\pi$-antiperiodic for odd $n$, the corresponding eigenvalues
being $\lambda^+_{n}(\delta)$ and $\lambda^-_{n+1}(\delta)$
respectively.  For parameter values $E\in\bR$, $\delta\in(0,\infty)$ the
$\lambda^+_{n}$ and $\lambda^-_{n+1}$ are real and
\[
  \lambda^+_{0} < \lambda^-_{1} < \lambda^+_{1}
  < \lambda^-_{2} < \lambda^+_{2} < \cdots.
\]

The following facts are proved in \cite[Chapter VII.3.3]{Kato},
\cite[Chapter 2.4]{MSW}, \cite[Chapter 2.2]{MS} and \cite{V98}.
\begin{compactenum}
\item The eigenvalues of the Mathieu operators are real-analytic
  functions in $\delta\in\bC$, whose algebraic singularities all lie
  at non-real branch points.

\item \label{item:ordGrow} They can be defined uniquely as functions
  $\lambda^\pm_{n}(\delta)$ of $\delta$ by introducing suitable cuts
  in the $\delta$-plane.  Moreover they admit an expansion in powers
  of $\delta$ with finite convergence radius
  $r_n$ such that $\liminf_{n\to\infty} \frac{r_n}{n^2} \geq C$ for some $C>0$.
\item The number of branch points is countably infinite, and there are
  no finite limit points.
\item\label{item:sepSp0} The operator $T$ corresponding to
  \eqref{eq-kappaeta} can be decomposed according to
\[
  L^{2}([-\pi,\pi]) = L_{0}^{+} \oplus L_{1}^{+}
    \oplus L_{0}^{-} \oplus L_{1}^{-}
\]
where the superscripts $\pm$ denote respectively the sets of even and
odd functions and where the subscripts $0$ and $1$ denote respectively
the sets of functions symmetric and antisymmetric with respect to
$x=\pi/2$.
\item\label{item:sepSp} The restrictions of $T$ to the four subspaces
  $L^{\pm}_{0/1}$ are self-adjoint and have only simple eigenvalues,
  as given by the following scheme:
\begin{align*}
L_{0}^{+} &\;:\;
  \lambda_{n}^{+}, \psi_{n}^{+},
  \qquad n= 0,2,4,6,\ldots; \\
L_{1}^{+} &\;:\; \lambda_{n}^{+}, \psi_{n}^{+},
  \qquad n= 1,3,5,\ldots; \\
L_{0}^{-} &\;:\; \lambda_{n}^{-}, \psi_{n}^{-},
  \qquad n= 1,3,5,\ldots; \\
L_{1}^{-} &\;:\; \lambda_{n}^{-}, \psi_{n}^{-},
  \qquad n= 2,4,6,\ldots.
\end{align*}
\item All the eigenvalues in each of the four groups of the previous
  remark belong to the same analytic function, i.e.\ the eigenvalues
  in the same group lie on the same Riemann surface \cite{MS, V06}.
\item The eigenfunctions $\eta\mapsto\psi_{n}^{\pm}(\eta)$ are
  themselves analytic functions of $x$ and $\delta$. For all $n\in\bN$
  they coincide with the Mathieu Cosine and the Mathieu Sine
  introduced above, namely $\psi_n^+ \equiv \ce_n$ and
  $\psi_{n+1}^- \equiv \se_{n+1}$ ($n\in \bN_0$).
\end{compactenum}
\smallskip

Despite the completeness and the clarity of perturbation theory for
one-parameter analytic families of self-adjoint operators, the
situation is much more intricate and much less complete in presence of
more parameters. On the other hand we can use our restrictions on the
parameters and the special symmetries of the potential to play in our
favor.

For a general value of $Z_-$, the eigenvalue equation is
\beq
[T\psi](\eta) = -\pa_{\eta}^{2}\psi(\eta) + \l(
   \frac{Z_{-}}{h^{2}} \co(\eta)
  + \frac E{2h^{2}} \co(2\eta)
  - \frac{2\mu-E}{2h^{2}}
  \ri)\psi(\eta) = 0,
\Leq{eq:kappaetaZnonzero}
with periodic boundary conditions on $[-\pi,\pi]$ and eigenvalue
  $\mu$.  Let us call
\beq
  \lambda := \frac{2\mu-E}{2h^{2}},
  \qquad \gamma_{1} := \frac{Z_{-}}{h^{2}},
  \qquad \gamma_{2} := \frac E{2h^{2}}.
\Leq{eq:lg1g2}
Notice that the main
difference between \eqref{eq:kappaetaZnonzero} and the Mathieu equation
is that now the period of the potential is no more smaller than the
length of the considered interval.
Thus, in applying Floquet
  theory we do not anymore look for solutions which are (anti-)periodic under
  translation by $\pi$.

\begin{remark}\label{rmk:singlespec}
  By standard Sturm-Liouville theorems (see for instance
  \cite[Theorems 2.3.1 and 3.1.2]{Ea}) we know that for every choice
  of $\gamma_{1}$ and $\gamma_{2}$ the spectrum
  of $K_\eta(h)$
  is discrete, at most doubly degenerate and accumulates only at
  infinity.  Anyhow it follows from \cite[Theorem 7.10]{MW} using a
  change of variable that in this case there cannot be coexistence of
  $2\pi$-periodic eigenfunctions for the same eigenvalue. Thus the
  spectrum is non-degenerate.  \hfill$\Diamond$
\end{remark}

It is proved in \cite{S43} that, for real-valued $E$ and $Z_{-}$, the
eigenvalues of $h^{-2}K_\eta(h)$
form a countably infinite set
$\{\lambda_{n}(\gamma_{1}, \gamma_{2}, h)\}_{n\geq 0}$
of transcendental real analytic (actually entire) functions of the
parameters $\gamma_{1},\gamma_{2}\in\bR$, so that in the
$(\gamma_{1},\gamma_{2},\lambda)$ space the sets
\[
\l\{ \l(\gamma_{1},\gamma_{2},\lambda_{n}(\gamma_{1}, \gamma_{2})\ri)
\mid (\gamma_{1},\gamma_{2})\in\bR^{2}\ri\}
\]
define a countably infinite number of uniquely defined
real-analytic surfaces.

We can apply analytic perturbation theory \cite[Chapter VII]{Kato} to
\[
T(\beta) := T + \beta (1 + \co(2\eta))
\]
where $T$ is defined in \eqref{eq:eigeqT} and $\beta$ is assumed to be defined by $h$ and some real parameter $E_{\rm im}$ as follows
\[
\beta(E_{\rm im},h) := i \frac{E_{\rm im}}{2h^{2}}
\qquad (\text{with } i=\sqrt{-1},\; h\in(0,\infty)).
\]
Therefore $T(\beta)$ is merely \eqref{eq:kappaetaZnonzero}
with complex $E$. It is evident that $T(\beta)$ defines a self-adjoint
analytic family of type (A) in the sense of Kato. Therefore
\cite[Chapter VII]{Kato} each
$\lambda_{n}(\gamma_{1},\gamma_{2},\beta)$ admits an analytic
extension on the complex plane around each real $E$ that can be
expanded as a series in $\beta=i E_{\rm im}/2h^{2}$
with an $n$-dependent convergence radius $\rho_n$.
Remark \ref{rmk:singlespec} concerning the simplicity of the spectrum
and the construction described at points
\ref{item:sepSp0}. and \ref{item:sepSp}. 
on page \pageref{item:sepSp0} is still valid.
Therefore we may continue to regard each eigenvalue as simple
restricted on its proper subspace and consider the lower bound of the
convergence radius in terms of the eigenvalues' spacing in the proper
subspace. These distances are known to be at least of order $n$, in the sense that there exists $C>0$ 
such that $\liminf_{n\to\infty} \frac{n\text{th-distance}}{n} \geq C$, see \cite[Chapter VII.2.4]{Kato}.

In the particular case considered, we can use the ansatz given by
\cite[Theorem 1.1]{MW} to bound the distance between the periodic
solutions with a boundary-value problem. To this end we can use the
discussion of \cite[Section 5]{V95} and apply it to our case to obtain
the following rough estimate, generalizing point (\ref{item:ordGrow})
on page \pageref{item:ordGrow}.

\begin{theorem}\label{thm:CRad}
  Let $E > 2 |Z_{-}|$. Then the convergence radii $\rho^{D,N}_{n}$
  corresponding to  \eqref{eq:kappaetaZnonzero} with Dirichlet
  (resp. Neumann) boundary conditions satisfy
\[
\liminf_{n\to\infty}\;
  \frac{\rho^{D,N}_{n}}{n^{2}} \geq \frac6{13}
\]
\end{theorem}
\proof 
In \cite{V95}, Section 5, it is shown that a result like our Theorem \ref{thm:CRad} holds for the Mathieu equation (see \cite[Theorem 5.1]{V95}).
This is a particular case of a more general theorem on the quadratic growth of the convergence radii for the eigenvalues of a big family of differential equations (see \cite[Theorem 3.4]{V95}).

To apply \cite[Theorem 3.4]{V95} and obtain the theorem for the Mathieu equation, it is enough to check the assumptions and use the estimates obtained there to get the constants in the growth rate.
This check relies on some crude estimates on incomplete elliptic integrals and on the potential that can be used also for our problem.

Indeed, we can replace the estimate $|2\cos(2z)|\leq 2\cosh(2\Im z)$ for the Mathieu potential by a corresponding estimate for $\cos(2x)+\frac{Z_{-}}{2E}\cos(x)$: if $E \gg |2Z_{-}|$, then
\[
\l| \cos(2z)+2\frac{Z_{-}}{E}\cos(z) \ri| \leq 2\cosh(2\Im z).
\]
Then, the constants in the proof of \cite[Theorem 5.1]{V95}, would coincide with the constants obtained for our potential: $R=2\cosh(2\delta)$, $R_0=2$, $U^2 = \frac{\pi^2}{16}+\delta^2$ (notation from \cite[Section 5]{V95}). And choosing $\delta=\frac12$ one can check that the assumptions of \cite[Theorem 3.4]{V95} are satisfied and the growth constant is $\frac6{13}$ also in this case.
\qed

\begin{remark}
As for \cite[Theorem 5.1]{V95}, we used very crude estimates. The constants, and in particular the lower bound for the growth rate, are far from being optimal also in this case and could be improved following the enhancements presented in \cite{V98}.
\end{remark}

\begin{remark}
  We expect that Theorem~\ref{thm:CRad} still holds true 
  for $0 < E \leq 2|Z_-|$.
\hfill$\Diamond$\end{remark}

%
\section{Asymptotic behaviour of solutions of the radial Schr\"odinger equation
and their analytic extensions}\label{Ab}
%

The general estimates that we develop in this section are needed in
order to justify the formal step in the separation of variables and
the construction of the Green's functions. We proceed with a
philosophy close to the one of \cite{AgmonKlein}.

With the substitution $E=k^2$ of its parameter, the radial equation in
\eqref{SLSL} takes the form \beq v''(\xi, k) + h^{-2} \l(k^2
\ch^2(\xi) + Z_+ \ch(\xi) - \mu\ri)v(\xi,k) = 0 \Leq{eq:schro} where
$\xi>0$, $h>0$ and $k\in\bC$ are arbitrary.  Now for $l\in\bN$ we set
$\mu:=\mu_{l}$, the $l$-th eigenvalue of $K_{\eta}$ (counted in
ascending order for real parameters and then extended analytically).
We assume w.l.o.g. that $h=1$, since $h$ can be absorbed in the other
parameters.

We will be interested in the solutions $v_\pm(\xi,k) :=
v_\pm(\xi,k,\mu)$ of \eqref{eq:schro} which decay as $\xi\to\infty$
for $k$ in the upper, resp.\ lower, half-plane $\bC_\pm =
\{k\in\bC\mid \Im(k) \lessgtr 0\}$. We call them, following
\cite{AgmonKlein} ``outgoing'', resp. ``incoming'', and we will make a
specific choice of such a family of solutions by fixing the behaviour
of $v_\pm(\xi, k)$ as $\xi\to\infty$.

We want to construct a phase function that
is an approximate solution of the eikonal equation for the Schr\"odinger equation \eqref{eq:schro},
that is characterized by a particular asymptotic behaviour and that is
analytic in $k$. We would like to consider something of the form
\beq
  \phi(\xi,k) \sim \int_0^\xi
  \sqrt{k^2 \ch^2(t) + Z_+ \ch(t) - \mu}\;
dt,
\Leq{eq:phfunct0}
but this gives a well-defined analytic function only for $|k|^{2} >
|Z_{+}-\mu|$. For our analysis it will be essential that the phase
function is analytic in $k\in\bC\setminus\{0\}$. To construct it we
reconsider the previous ansatz and perform a change of variables. If
we call $\tau = \sh(t)$, the above equation is transformed into
\beq
  \phi(\xi,k) \sim \int_{0}^{\sh(\xi)} \sqrt{k^{2}
  - q(\tau)}\;d\tau
\qmbox{with} q(\tau) :=
  \frac{\mu}{1+\tau^{2}} -
  \frac{Z_{+}}{\sqrt{1+\tau^{2}}}.
\Leq{eq:long:short}
If we call $r = \sinh(\xi)$, we may consider the map $r\mapsto\phi(\arcsinh(r),k)$ 
to be the phase function of a 
long-range potential, asymptotic to $r\mapsto k r$ as $r\to \infty$,
(see \eqref{eq:phineqqsym} for a more precise statement),
plus a short-range perturbation.

\subsection{Decomposition into long and short range}
\label{sec:dec-lr}

To construct the phase function $\phi$, we introduce an appropriate decomposition of the potential $q$ into short and long range parts.

Let $j\in\bN$. We define $l_j, s_j \in (0, \infty) \to \bR$ by \beq
s_{j}(\tau) := q(\tau) - l_{j}(\tau)
\quad\mbox{and}\quad
l_j(\tau) := - \chi(\tau)\frac{Z_+}{\sqrt{1+\tau^{2}}},
\Leq{eq:decompo}
where $\chi(\tau) = 1$ if $Z_+ \geq 0$ and otherwise is defined as follows:
 $\chi \in C_c^\infty((0, \infty); [0, 1])$ such that $\chi(\tau) = 0$ if $\tau \leq j|Z_+|$ and $\chi(\tau) = 1$ if $\tau \geq j|Z_+| + 1$.

Note that $s_{j}(\tau)\in L^{1}((0,\infty))$, $l_{j}\in C^{2}((0,\infty))$,
\[
\sup_{\tau>0} l_{j}(\tau) \leq 1/j
\qmbox{and}
l_{j}(\tau) = - \frac{Z_{+}}{\sqrt{1+\tau^{2}}}
\text{ for } \tau > R_{j},
\]
for $R_j := j|Z_+| + 1$.

Let $\Omega_{j}:=\{ k\in\bC \mid |k|^{2} > 1/j \}$ and $\phi_{j} \in (0, \infty) \times \Omega_{j} \to \bC$, defined by
\beq
\phi_{j}(\xi,k) := \int_{0}^{\sh(\xi)} \sqrt{k^{2}
  - l_{j}(\tau)}\;d\tau.
\Leq{eq:phfunct1}
Here we have taken the principal branch of the square root, i.e.\ the uniquely
determined analytic branch of $\sqrt{z}$ that maps $(0,\infty)$ into
itself. 

Note that $\phi_{j}(\xi, \cdot)$ is analytic in $\Omega_{j}$ and  $\phi_{j}(\cdot, k)\in C^{2}((0,\infty))$. Furthermore, for $k\in\Omega_{j}$, $\phi_{j}(\cdot, k)$ satisfies the eikonal equation
\beq
|\pa_{\xi}\psi(\xi)|^2 = k^2 - l_{j}(\sh(\xi))
\Leq{eq:pheik}
on $(0, \infty)$.

\begin{theorem}\label{thm:defPhFun}
Let 
\[
D:=\{ (\xi, k) \in (0, \infty)\times\bC\setminus\{0\} \mid \sh(\xi) \geq | k^{-2} Z_+| + 1 \}.
\]
There exist a function $\phi:D\to\bC$ satisfying the following properties:
\begin{enumerate}
\item\label{eq:phineq1} For all $(\xi, k)\in D$, $\phi(\xi,-k) = -\phi(\xi, k)$.
\item\label{eq:phfunct} For all $j\in\bN$, the restriction of $\phi-\phi_{j}$ to $(R_j, \infty)\times\Omega_{j}$ doesn't depend on $\xi$ and is an analytic function of $k$.
\item For all $k\in\bC\setminus\{0\}$, $\phi(\xi, \cdot)$ is analytic on each $\Omega_{j}$, for $j\in\bN$ such that $\sh(\xi) > R_j$.
\item For all $k\in\bC\setminus\{0\}$, $\phi(\cdot, k)\in C^2((0,\infty))$ and satisfies the eikonal equation \eqref{eq:pheik} on $(R_j, \infty)$ where $j $ is the integer part of $|k^2|^{-1}$.
\end{enumerate}
\end{theorem}

The theorem follows from the construction above with the same proof as \cite[Proposition 2.1]{AgmonKlein}.


\begin{remark}
The phase function $\phi$ defined in the previous theorem is
not unique. This is, however, immaterial for our purposes. In fact, our main concern
is to have a controlled behaviour, as $\xi\to\infty$ (see Proposition \ref{pr:phfun})
and good analyticity properties in order to identify the two (unique)
waves $v_\pm$ for a wide range of parameters.
\end{remark}

Henceforth we will refer to the $\phi(\xi,k)$ defined in Theorem
\ref{thm:defPhFun} as a \emph{global phase function}.

\begin{proposition}\label{pr:phfun}
The global phase function $\phi(\xi,k)$ has the asymptotic behaviour given by
\beq\hspace*{-1.1cm}
\phi(\xi,k) = k\sh(\xi) + \frac{Z_+}{2k}\xi + {\cal O}(1)
= \frac k2 e^\xi \big(1 + o(1)\big)
\qmbox{as} \xi\to\infty.
\Leq{eq:phineqqsym}
\end{proposition}
\begin{remark}\label{rem:simplifyyourlife}
  In the proposition the term $s(\tau) := \frac{\mu}{1+\tau^{2}}$ has
  been dropped out. In fact it belongs to the short range component
  $s_{j}$ of \eqref{eq:decompo} and choosing in \eqref{eq:decompo} a
  different decomposition of $q(\tau)$ into a short-range and long-range 
  part, keeping $l(\xi)$ fixed near infinity, modifies
  $\phi(\xi,k)$ by an analytic function of $k$ alone.
  \hfill$\Diamond$
\end{remark}
\proof Without losing generality we can suppose $|k| > |Z_{+}|$ and
consider the simplified phase function
\beq
  \phi(\xi, k) := \int_0^\xi
  \sqrt{k^2 \ch^2(t) + Z_+ \ch(t)}\; dt = k\int_{_0}^\xi \ch(t)
  \sqrt{1+\frac{Z_+}{k^2\ch(t)}}\; dt
\Leq{eq:simplPhFun}
as $\xi\to\infty$:
\begin{align*}
\phi(\xi, k) = {}&
k\int_{0}^\xi \ch(t) \l(1 + \frac{Z_+}{2k^2 \ch(t)}
+ {\cal O}\l(k^{-2}\cosh^{-2}(t)\ri) \ri)\;dt \\
= {}& k \sh(\xi) + \frac{Z_+}{2k} \xi
+ {\cal O}(1),
\end{align*}
Writing $\sh(\xi)= (e^x - e^{-x})/2$ and collecting the growing term
we have the thesis.
\qed

The Liouville-Green Theorem \cite[Corollary 2.2.1]{Ea89} guarantees
that for each $k\in\bC$
there exist two linearly independent solutions of \eqref{eq:schro} whose asymptotics as $\xi\to\infty$ is given by
\[
y_{1,2} (\xi) =
\textstyle{\frac1{\sqrt{\phi'(\xi,k)}}}
\exp\big({\pm i \phi(\xi,k)}\big)
\big(1 + o(1)\big) \quad\text{ for } \xi\to\infty.
\]
In particular, it follows from the asymptotic estimate of Proposition
\ref{pr:phfun} that \eqref{eq:schro} must be in the Limit Point Case
at infinity
(more precisely Case I of \cite[Theorem 2.1]{BEMP})
if we set $r(x) := \cosh^{2}(x)$, $p :=1$ and $\lambda:=k^{2}$. In
what follows we investigate the regularity of the solutions with
respect to $\xi$ and $k$.

\begin{theorem}[Outgoing and incoming
    solutions]\label{thm:asympsol}
For each $k\in\bC\z$, equation \eqref{eq:schro} has
unique solutions $v_{\pm}(\xi,k)$ verifying the asymptotic
relation
\beq
v_{\pm}(\xi,k) =
\sqrt{2} e^{-\frac{\xi}{2}}\exp \big(\pm i \phi(\xi,k) \big)
\big( 1 + o(1)\big)
\qmbox{ as} \xi\to\infty.
\Leq{eq:vpmasymp}
\eqref{eq:vpmasymp} holds uniformly in any truncated cone
\[
\Lambda_{\pm}(\eta,\delta)\, :=
\{ k\in\bC\z \mid \eta \leq \arg(\pm k)
\leq \pi-\eta,\; |k|\geq\delta\}
\qmbox{with}\eta\geq0,\ \delta >0.
\]
The family of solutions $k\mapsto v_{\pm}(\xi , k)$ defined by \eqref{eq:vpmasymp} is analytic in
the half planes $k\in\bC_{\pm}$ pointwise in $\xi$, and extends continuously to $k\in\overline{\bC}_{\pm}\z$.

\end{theorem}
\begin{figure}[h!]
\begin{center}
 \includegraphics[width=0.4\linewidth]{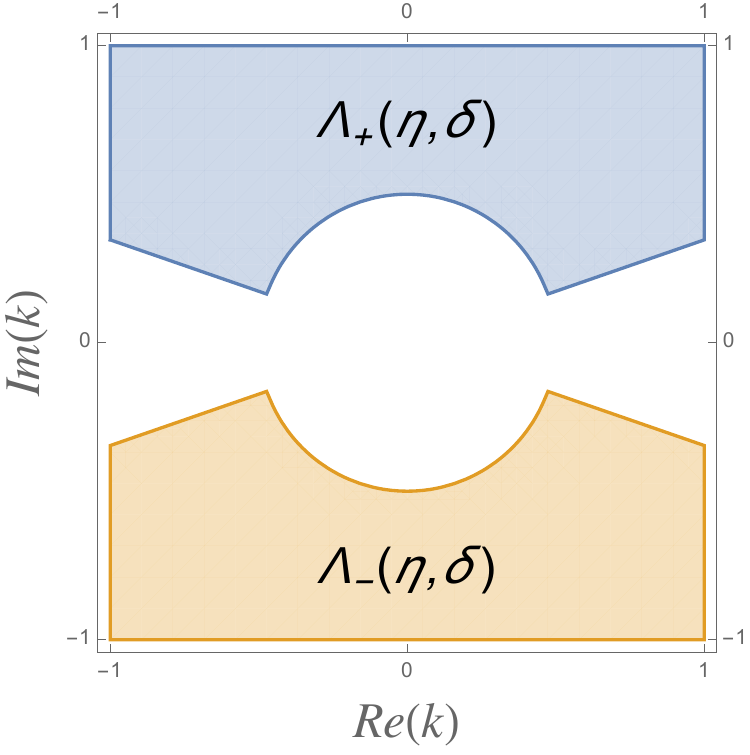}
\caption{\small Cones $\Lambda_\pm$ for $\eta = 1/3$ and $\delta = 1/2$.}
\label{fig:dpm}
\end{center}
\end{figure}

\begin{remark}
  \eqref{eq:phineq1} and the uniqueness of Theorem \ref{thm:asympsol}
  imply that $v_{+}(\xi,k) = v_{-}(\xi,-k)$. In particular it suffices
  to consider $v_{+}$.  \hfill$\Diamond$
\end{remark}
\proof In view of Theorem \ref{thm:defPhFun} and the subsequent
remark, we can reduce the proof to the case where the phase function
$\phi$ is given by \eqref{eq:simplPhFun} for $\xi > 0$ and $|k|^{2} >
|Z_{+}|$. We call $\phi$ a \emph{local phase function}.
Let
\beq
V_{\pm}(\xi,k) := \l(\frac{k}{\pa_\xi\phi(\xi,k)}\ri)^{\frac12} e^{\pm
  i \phi(\xi,k)}
\Leq{eq:appsol}
define the approximate solutions of \eqref{eq:schro}.

For $|k| \geq \delta$ the function $V_{\pm}$ satisfies the comparison
equation
\beq
  V''_{\pm}(\xi,k) + \l(k^{2} \ch^{2}(\xi) + Z_{+}\ch(\xi)
  +\eh S\phi(\xi,k) \ri) V_\pm(\xi,k) = 0
\Leq{eq:approx}
where $S\phi$ denotes the Schwarzian derivative
\beq
  S\phi =
  \frac{\phi'''}{\phi'}-\frac32\l(\frac{\phi''}{\phi'}\ri)^{2}
\Leq{eq:schder}
w.r.t.\ $\xi$.  For $k\in\Lambda_{\pm}(\eta,\delta)$
we consider the inhomogeneous Volterra Integral Equation
\cite{Tricomi}
\beq
  v_{\pm}(\xi, k) = V_{\pm}(\xi,k) -
  \int_{\xi}^{\infty} K_{k}(\xi,t) F_{k}(t) v_{\pm}(t,k)\; dt
\Leq{eq:VIE}
where $F_{k}(t) = \frac12 S\phi(t,k) + \mu$ is the
function that expresses the difference between the Schr\"odinger
equation \eqref{eq:schro} and the comparison equation
\eqref{eq:approx} and $K(\xi,t)$ is the Green's function associated
with equation \eqref{eq:appsol}:
\beq
  K(\xi,t) = W(V_{-}, V_{+})^{-1}
  \l\{ V_{+}(\xi)V_{-}(t) - V_{+}(t)V_{-}(\xi)\ri\}
\Leq{eq:kernel}
(the parameter $k$ being suppressed),
with Wronskian $W(V_{-},V_{+}) := V_{-}V_{+}' - V_{-}'V_{+} = 2 i k$.

To give \eqref{eq:VIE} meaning we need to check if the definition
makes sense and a solution can be found.

We
explicitly compute $S\phi$ and thus $F$ using
  \eqref{eq:schder},
obtaining
\[
S\phi\,(\xi) = \frac{10 k^4 - Z_+^2 - 2k^4 \ch(2\xi)+ Z_+ \sech(\xi)\l(12 k^2 + 5 Z_+ \sech(\xi)\ri)
}{8\l(Z_+ + k^2 \ch(\xi)\ri)^2}
\]
and thus, for real $\xi$ and for every $k\in\Lambda_+(\eta,\delta)$,
we have
\beq
  \lim_{\xi\to\infty}|F(\xi)| = {\textstyle\frac 18} + \mu
  \qmbox{ and }
  C_{F}:=\sup_{\xi\in(0,\infty)}|F(\xi)|
  <\infty.
\Leq{eq:estSphixi}
Of course $C_{F}$ depends on $Z_{+}$, $\mu$ and $k$, thus on $\eta$
and $\delta$.  Moreover from \eqref{eq:appsol} and
\eqref{eq:phineqqsym}, writing $k\in\Lambda_+(\eta,\delta)$ as $k =
k_r + i k_i$ ($k_r,\, k_i$ real), we get
\beq\hspace*{-1cm}
\l| V_\pm (\xi,k) \ri|
= \sqrt{2} e^{-\frac \xi2}  (1+o(1)) \l| e^{i k
\l(\frac{\phi(\xi,k)}{k}\ri)
} \ri|
\leq C_{V} e^{-\frac \xi2} \exp\big({-\textstyle
    \frac{k_i}2 e^{\xi}(1+o(1)\big)
},
\Leq{eq:solBound}
where $\ds C_{V}(k) := \sup_{\xi\in(0,\infty)} \l(e^{\xi/2}|k/\phi'(\xi,k)| \ri)
< \infty$
by \eqref{eq:simplPhFun}.
Therefore for $0<\xi\leq t< \infty$ we have
\begin{align*}\hspace*{-0.9cm}
\l| K(\xi,t) \ri| = {}&
  \l|\frac1{2ik}\sqrt{\frac{k^2}{\phi'(t,k)\phi'(\xi,k)}}
  \l( e^{i(\phi(\xi,k)-\phi(t,k))} - e^{i(\phi(t,k)-\phi(\xi,k))}\ri)\ri| \\
  \leq {}& \frac{C_{V}^{2}}2 e^{-\frac{\xi+t}2} C_{K}
  \l|\textstyle\exp\l(- i k \int_{\xi}^{t} \ch(\tau)
  \sqrt{1+\frac{Z_+}{k^2\ch(\tau)}}\,d\tau \ri) \ri|,
  \eqnumtag\label{eq:KBound}
\end{align*}
where
\[
C_{K}(k) := \ds\sup_{t,\xi \in\bR^{+}} \l| 1 -
  \textstyle\exp\l(2 i k \int_{\xi}^{t} \ch(\tau)
  \sqrt{1+\frac{Z_+}{k^2\ch(\tau)}}\,d\tau\ri) \ri| 
  \leq 2.
\]

It follows from \eqref{eq:estSphixi}, \eqref{eq:solBound} and
\eqref{eq:KBound} that the Volterra Integral Equation \eqref{eq:VIE}
is well-defined as a mapping from the function space
\beq
\cC_{\pm}(\eta,\delta):=\Big\{f\in C^2
  \big((0,\infty)\times \Lambda_{\pm}(\eta,\delta)\big) \Big|\,
  \forall k\in\Lambda_{\pm}(\eta,\delta),  \|f\|_{k}:=
\!\sup_{x\in(0,\infty)}\!\l|f(x,k)\, e^{\mp i \phi(x,k)
}\ri| < \infty \Big\}
\Leq{def:ccplus}
to itself.
In particular, being $V_{\pm} \in \cC_{\pm}(\eta,\delta)$ we can apply
the Picard iteration procedure to find a solution of the equation and
prove its existence. We claim that the solution must be
unique. Suppose that there exists two solutions
$v_{+},\tv_+\in\cC_{+}$ of \eqref{eq:VIE}, then
\beq\hspace*{-0.2cm}
\psi(\xi,k) := v_{+}(\xi,k) - \tv_+(\xi,k)
= -\int_{\xi}^{\infty} K(\xi,t) F(t)\psi(t,k)\; dt.
\Leq{eq:defdiff}
At this stage, it is not obvious that the r.h.s. of \eqref{eq:VIE} is a contraction, that would allow us to conclude the proof in a standard way. 
In the rest of the proof we show that for appropriate initial values this is indeed the case, therefore proving the unicity and the uniformity of the estimates.
The previous estimates applied to \eqref{eq:defdiff} give
\begin{align*}
\l|\psi(\xi,k)\ri|={}& \l| \int_{\xi}^{\infty} K(\xi,t)
  F(t)\psi(t,k)\; dt \ri|
\leq \int_{\xi}^{\infty} \l| K(\xi,t) F(t)\psi(t,k)\ri| \; dt \\
\leq{}& \frac{C_{K} C_{F} C_{\psi}}{2}
        \l| \sqrt{\frac{k}{\phi'(\xi,k)}} \ri| \l| e^{i\phi(\xi,k)} \ri|
        \int_{\xi}^{\infty}
         \l| \sqrt{\frac{k}{\phi'(t,k)}} \ri|
        \;dt \\
\leq{}& \frac{C_{K} C_{F} C_{\psi} C_{V}}{2}
        e^{-\frac{\xi}{2}}
        \l|e^{i\phi(\xi,k)}\ri|
        \int_{\xi}^{\infty}
        \sqrt{2} e^{-\frac{t}{2}} (1+o(1))
        \;dt \\
\leq{}& \frac{C_{K} C_{F} C_{\psi} C_{V} C_{I}}{2}
        e^{-\frac{\xi}{2}}
        \l|e^{i\phi(\xi,k)}\ri|
        \int_{\xi}^{\infty}
        e^{-\frac{t}{2}}
        \;dt
= C_{\psi} C_{\rm tot} e^{-\xi} \l|e^{i\phi(\xi,k)}\ri|
\eqnumtag\label{eq:estdiff1}
\end{align*}
where $C_{\psi}(k) := \|\psi\|_{k}$,
$
C_{I}:= \sup_{\xi\in(0,\infty)} \sqrt{2} \l((1+e^{-2\xi})
\sqrt{1+\frac{Z_+}{k^{2}\ch(\xi)}}\ri)^{-\frac12}
$
and
$
C_{\rm tot} := C_{K} C_{F} C_{V} C_{I}
$.
Using equations \eqref{eq:defdiff} and \eqref{eq:estdiff1} we can
reiterate the procedure, in fact defining
\[
\psi_{1}(\xi,k)  := \int_{\xi}^{\infty} K(\xi,t) F(t)\psi(t,k)\; dt
\qmbox{and}
\psi_{n}(\xi,k)  := \int_{\xi}^{\infty} K(\xi,t) F(t)\psi_{n-1}(t,k)\; dt,
\]
one can prove by induction that
\beq
\l|\psi(\xi,k)\ri| = \l|\psi_{n}(\xi,k)\ri|
\leq \frac{\;C^{n}_{\rm tot}\;e^{-n\xi}}{(2n-1)(2n-3)\cdots3\cdot1}
  \l|e^{i\phi(\xi,k)}\ri|
\leq C_{\psi} \frac{C^{n}_{\rm tot}\;e^{-n\xi}}{n!}
  \l|e^{i\phi(\xi,k)}\ri|
\Leq{eq:diffboundexp}
uniformly in $k\in\Lambda_{+}(\eta,\delta)$ and for all $n\in\bN$.
The convergence of
\beq
  \sum_{n=1}^{\infty} C_{\psi} \frac{C^{n}_{\rm tot}}{n!}
    e^{-n\xi} \l|e^{i\phi(\xi,k)}\ri|
  = C_{\psi}\l|e^{i\phi(\xi,k)}\ri|
    \l( e^{C_{\rm tot} e^{-\xi}} -1\ri)
\Leq{eq:convVolterra}
implies that $\l|\psi(\xi,k)\ri| = 0$, i.e. $\tv_+ = v_{+}$.

The same inequality implies that after some iterates the homogeneous
integral equation is a contraction, and coupled with the bounds on
$V_+$ it implies that \eqref{eq:VIE} has a unique fixed point.  This
proves the existence and uniqueness of the solution.  In fact if we
define
\[
v_{0,+}(\xi,k)  :=V_{+}(\xi,k)\qmbox{,}
v_{n,+}(\xi,k) := - \int_{\xi}^{\infty} K(\xi,t) F(t)v_{n-1,+}(t,k)\; dt,
\]
the Picard iteration converges to $v_{+} = \sum_{n=0}^\infty v_{n,+}$,
and the series converges absolutely uniformly in
$k\in\Lambda_+(\eta,\delta)$ with
$
\l|v_{+}(\xi,k) \ri|
\leq
\l| V_{+}(\xi,k) \ri| e^{C e^{-\xi}}
$
for some positive constant $C$. Therefore one has
\[
v_{+}(\xi,k) = V_{+}(\xi,k) (1 + o(1)) \qmbox{ as } \xi\to\infty
\]
and \eqref{eq:vpmasymp} holds.

The fact that all the bounds are valid for $k\in\bR$ completes the
proof.
\qed

\begin{remark}
  It is possible to compute an explicit bound like
  \eqref{eq:diffboundexp} using the fact that
\[
\l|v_{n,+}(\xi)\ri| \leq C_V e^{-\frac{\xi}{2}} \l|e^{i\phi(\xi,k)}\ri|
  \frac{C_{\rm tot}^n e^{-n\xi}}{2^n n!} .
\]
In particular the dependence on $\mu$, the parameter of the
  short-range potential in \eqref{eq:long:short}, appears in the
  constant $C_{\rm tot}$. In view of the previous estimates it can be
  bounded by $|\mu| {\cal O}(1)$. Therefore we can be more precise
  and estimate
\beq
  v_{\pm}(\xi,k) = \sqrt{2} e^{-\frac{\xi}{2}}
  e^{\pm i \phi(\xi,k)} \big( 1 + M_{\pm}(\xi,k,\mu) \big)
  \qmbox{ as } \xi\to\infty,
\Leq{eq:vpmasymp1}
where for some constant $C\neq0$ we have
$M_{\pm}(\xi,k,\mu) = e^{C |\mu|e^{-\xi}}o(1)$.
\hfill$\Diamond$\end{remark}

\begin{remark}\label{remark:anyotherfamily}
  Let $w$ be any other family of solutions of
  \eqref{eq:schro}, analytic in $k\in \bC\z$ and satisfying for
  $k\in\Lambda_+(\eta,\delta)$ the estimate $w(\xi,k) =
  o(1)$ as $\xi\to\infty$.  Then
\[
w(\xi,k) = \gamma(k)v_+(\xi,k),
\]
where $\gamma(k)$ is a nowhere-vanishing analytic function of
$k\in\Lambda_+(\eta,\delta)$.  \hfill$\Diamond$\end{remark}

\begin{remark}\label{rmk:ansol}
  In case $Z_{+} = 0$, the solutions of $\eqref{eq:schro}$ are given
  by linear combinations of the modified Mathieu functions (${\rm Mc}$
  and ${\rm Sc}$) \cite[\S16.6]{BMP3}. In particular, if we
  look at their asymptotic behaviour, we find out that up to a constant
  factor
  \beq
  v_{+}(\xi,k) = {\rm Mc}\l(\mu-\frac{k^{2}}2,
  \frac{k^{2}}{4},\xi\ri)
  \Leq{eq:knownSol}
  where ${\rm Mc}(a,q,x)$ is the modified Mathieu cosine,
  i.e.\ the solution of
\[
y''(x) - (a - 2q \ch(2x))y(x) = 0
\]
that decays for $\sqrt{q}\in\bC_{+}$. It is well-known \cite[Chapter
2]{MS} that the function in the RHS of \eqref{eq:knownSol} admits an
analytic continuation through the positive real axis on the negative
complex plane for $-\pi/2 \leq \arg(k) \leq \pi/2$ and that for
$x\to\infty$ and $k\in\bC_{+}$ it has the following asymptotic
behaviour \cite{BMP3, MS}
\[
{\rm Mc}\l(\mu-\frac{k^{2}}2, \frac{k^{2}}{4},x\ri) =
e^{-\frac x2} \exp\big({i {\textstyle \frac{k}{2}}
  e^x (1+o(1))}\big)\big(1+o(1)\big),
\]
in line with
the estimates \eqref{eq:phineqqsym} and \eqref{eq:vpmasymp},
valid for all $Z_{+}$.
\hfill$\Diamond$
\end{remark}

For what follows we will need to work in a slightly different
setting. If we perform the change of variable defined by
  $\xi   \mapsto {\Log}(x+1)$ (with the principal branch
  ${\rm Log}$ of the logarithm), for
  $\tilde v(x,k):= v(\Log(x+1),k)$
  ~Equation \eqref{eq:schro}
  takes the form
\beq\label{eq:schrlog}
\hspace*{-0.4cm}\begin{split}
\l((x+1) {\tilde v}'(x,k)\ri)' &+ h^{-2}\,q(x,k,Z_{+},\mu)\,{\tilde v}(x,k) = 0
\qquad\mbox{with}\\
q(x,k,Z_{+},\mu) := &
\frac{k^{2}}4 \l(x+1 + 2(x+1)^{-1} + (x+1)^{-3} \ri)
 + \frac{Z_{+}}2 \l( 1 + (x+1)^{-2} \ri) - \frac{\mu}{x+1}.
\end{split}
\eeq
where $x >0$, $h >0$ and $k\in\bC\z$. As before we assume $h=1$ for
the moment.

\begin{remark}\label{remark:newequationvpm}
  In this case Theorem \ref{thm:asympsol} and Remark
  \ref{remark:anyotherfamily} is still valid and in accord with the
  Liouville-Green Theorem we have two unique solutions that as
  $x\to\infty$ are asymptotic to
\beq\hspace*{-0.9cm}\begin{split}
{\tilde v}_{\pm}(x,k) = {}& \frac{1}{\sqrt{x+1}} e^{\pm i \Psi(x,k)}(1+o(1)) \\
={}& \frac{1}{\sqrt{x+1}} \textstyle \exp\l(\pm i \l( \frac{k}{2}x +  \frac{Z_{+}}{2k}\log(x+1) +  \frac{k}{2} \ri)\ri) \\
&\cdot\textstyle\exp\l(\pm i \l(\frac{Z_{+}^{2}}{4k^{3}} (x+1)^{-1} + {\cal O}\l((x+1)^{-2}\ri)\ri)\ri)(1+o(1))
\end{split}\Leq{eq:vlogpmasymp}
where $\Psi(x,k) = \phi(\Log(x+1),k)$. The asymptotic
behaviour \eqref{eq:vlogpmasymp} holds uniformly for $k$ in any sector
$\Lambda_{\pm}(\eta,\delta) = \{ k\in\bC \mid \eta \leq \arg(\pm k)
\leq \pi-\eta,\; |k|\geq\delta\}$ with $\eta\geq0$ and $\delta >0$.
The family of solutions defined by \eqref{eq:vlogpmasymp} is analytic
in $k\in\bC_\pm\z$ and extends continuously to
$k\in\overline{\bC}_\pm\z$.
\hfill$\Diamond$\end{remark}

\begin{remark}
  From now on we write with an abuse of notation $\phi(x,k)$ in place
  of $\Psi(x,k)$.
  \hfill$\Diamond$
\end{remark}

\noindent
Before presenting Theorem \ref{thm:anaext}, the main result of
this section, we need the following lemma.

\begin{lemma}\label{lemma:uniqueextsol}
  Let $\cK$ be a compact set in $\bC\z$. Then for any $-\pi < \theta <
  \pi$, there is a constant $A_\theta$ such that any solution of
  Equation \eqref{eq:schrlog} verifies the estimate
\beq
\l|\tv(x,k)\ri|\leq A_\theta\; (|c|+|c'|)\frac{1}{\sqrt{x}} e^{|\Im\phi(x,k)|}
\Leq{eq:cpxEst}
for $x\in e^{i\theta}[0,\infty)$ and $k\in\cK$, where
$c = \tv(0,k),\  c'=\tv'(0,k)$
are the initial data at $x=0$.
\end{lemma}
\proof
We start proving \eqref{eq:cpxEst} in the case $\eta \leq |\arg k|
\leq \pi-\eta$ for any $\eta \geq 0$ and $\theta = 0$
(i.e. $x\in(0,\infty)$).  All the constants that we are going to use
without an explicit definition are defined as previously.  Using the
approximate solutions given by \eqref{eq:appsol} defined by
$\cV_\pm(x,k) := V_\pm(\log(x+1),k)$, we determine $a_+$
and $a_-$ from the initial data requiring
\beq
c  = a_+ \cV_+(0,k) + a_- \cV_-(0,k),
\qmbox{,}
c' = a_+ \cV'_+(0,k) + a_- \cV'_-(0,k).
\Leq{eq:apamInitialData}

Then $\tv(x,k)$ satisfies the Volterra Integral Equation
\beq\hspace{-1cm}
\tv(x,k) = a_+ \cV_+(x,k) + a_- \cV_-(x,k) +
\int_0^x \cK(x,t)\cF(t) \tv(t,k)\;\frac{dt}{t+1}
\Leq{eq:VIE1}
where $\cK(x,t):=K(\Log(x+1),\Log(t+1))$ and $\cF(t):=F(\Log(t+1))$
are defined from the respective function
\eqref{eq:kernel} and \eqref{eq:VIE}.  Notice similarly as in the
previous theorem that for $0\leq t \leq x$ there exist constants
$C_0(\eta,\delta)$ and $C_\cV$ such that we have
\begin{align*}
|\cK(x,t)|
                   \leq{}&\frac{C_0(\eta,\delta)}2
                       \l|\frac{1}{\phi'(x,k)}\ri|
                       \l|\frac{1}{\phi'(t,k)}\ri|
                       \exp\big({|\Im(\phi(x,k)-\phi(t,k))|}\big) \\
                   \leq{}& \frac{C_\cV^2 C_0(\eta,\delta)}{2}
                       \frac{1}{\sqrt{(x+1)(t+1)}}
                       \exp\big({|\Im(\phi(x,k)-\phi(t,k))|}\big).
                       \eqnumtag\label{eq:Kbound1}
\end{align*}

Define now
\beq
\cV(x,k) = \sqrt{2} (|a_+| + |a_-|) \frac{1}{\sqrt{x+1}}
  \exp\big({|\Im \phi(x,k)|}\big).
\Leq{eq:Vnorm}
The sequence
\[
\tv_{0}(x,k)  := a_+ \cV_+(x,k) + a_- \cV_-(x,k) \qmbox{,}
\tv_{n}(x,k)  := \int_{0}^{x} \cK(x,t) \cF(t)\tv_{n-1}(t,k)\;
\frac{dt}{t+1},
\]
is uniformly convergent. In fact, suppressing the dependence of the
constant on $\eta$ and $\delta$, we have
$|\tv_0(x,k)|\leq C_\cV \cV(x,k)$
and, using the transformed version of \eqref{eq:Kbound1},
it follows by induction that
\beq
|v_n(x,k)| \leq \frac1{n!} \cV(x,k) L^n(x),
\Leq{eq:AppVIE1}
where
\[
L(x) := C_0 \int_0^x \l|\frac{1}{\phi'(t,k)}\ri|
  |\cF(t)| \;\frac{dt}{t+1}
= C_0\, C_\cV \int_0^x \frac{1}{\sqrt{t+1}}
  |\cF(t)| \;\frac{dt}{t+1}
\leq C_0 C_\cV C_\cF \frac{1}{\sqrt{x+1}}
\]
is uniformly bounded for $x\in(0,\infty)$. Therefore $\sum_{n=0}^\infty
\tv_n(x,k)$ converges uniformly and absolutely and coincides with the
given solution $\tv(x,k)$ of \eqref{eq:VIE1} for
$\eta \leq |\arg k|\leq\pi-\eta$, $\eta\geq0$.
In particular being $a_\pm$ bounded in
terms of the initial data $c$ and $c'$, we obtain \eqref{eq:cpxEst}
for real values of $x$.

At this point it is enough to notice that as soon as we do not cross
the branch cut of the logarithm, all the inequalities and the
equations written up to this point are valid, therefore the result
holds replacing $x$ with $e^{i\theta}x$ for every
$-\pi < \theta < \pi$.
\qed

\subsection{Analytic continuation}
\label{ssec:anacont}

We are ready to prove that the functions $v_\pm$ can be analytically
extended in $k$ up to the positive real axis. To this end we consider
the transformed form $\tv_{\pm}$.
\begin{remark}
  The potential $q$ defined in \eqref{eq:schrlog} is analytic in $\bC\setminus(-\infty,-1]$. 
  Therefore its analyticity in the cone
\beq
\Sigma_{\alpha,\beta} := \l\{z\in\bC\setminus\{0\} \mid -\alpha < \arg z < \beta\ri\}
\Leq{def:anastrip}
for all $\alpha,\beta\in[0,\pi)$ is clear.
\hfill$\Diamond$\end{remark}

\begin{theorem}\label{thm:anaext}
  Let ${\tilde v}_\pm(x,k)$ be defined as in Remark
  \ref{remark:newequationvpm}. Then ${\tilde v}_+(x,k)$ admits an
  analytic continuation in $k$ through the positive real $k$-axis into
  the region
\[
\l\{k\in\bC\z \mid -\beta < \arg k < \beta \ri\},
\]
${\tilde v}_-(x,k)$ admits an analytic continuation into
\[
\l\{k\in\bC\z \mid -\alpha < \arg k < \alpha \ri\},
\]
for any $\alpha,\beta\in[0,\pi)$ and both verify the asymptotic
relation \eqref{eq:vpmasymp}
\beq
{\tilde v}_{\pm}(x,k) = \frac{1}{\sqrt{x}}
  e^{\pm i \phi(x,k)} \l( 1 + o(1) \ri)
\qmbox{as} x\to\infty \mbox{ in } \Sigma_{\alpha,\beta},
\Leq{eq:vpmasympcone}
where \eqref{eq:vpmasympcone} holds locally uniformly in $k$ and
uniformly in $x$.
Furthermore an analytic continuation of ${\tilde v}_+(x,k)$ and
${\tilde v}_-(x,k)$ through the negative real axis is defined via
\beq
{\tilde v}_+(x,k) = {\tilde v}_-(x,-k).
\Leq{eq:anaextvpm}
\end{theorem}
\begin{figure}[h!]
\begin{center}
 \includegraphics[width=0.4\linewidth]{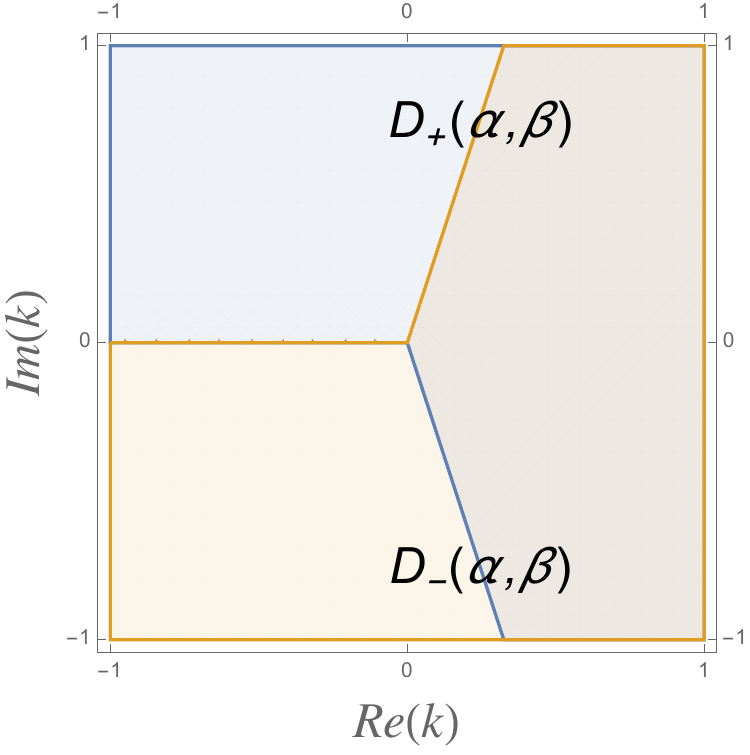}
\caption{\small Domains $D_\pm$ for $\alpha = 2\pi/3$ and $\beta = 2\pi/5$.}
\label{fig:dpm}
\end{center}
\end{figure}
\begin{remark}
  If $\alpha+\beta > \pi$, the analytically continued function
  ${\tilde v}_\pm(x,k)$ may be double-valued for $k\in\bC_{\mp}$.
  By an abuse of notation we denote the corresponding, possibly
  not simply-connected, domain by
\beq
  D_\pm(\alpha,\beta) := \l\{k\in\bC\z \mid -\beta
    < \arg(\pm k) < \pi + \alpha \ri\}.
\Leq{eq:dom2d}
See Figure \ref{fig:dpm}.
\end{remark}
\proof
It is well-known \cite[Chapter 3.7]{CL55} that, as solutions of the
linear differential equation \eqref{eq:schrlog} with analytic
coefficients, ${\tilde v}_\pm(x,k)$ admit an analytic
continuation in $x$ into the region $\Sigma_{\alpha,\beta}$.  The main
point of this proof is to use this information to obtain the
analyticity in $k$ via dilation. More in details we will
imitate the strategy of \cite[Theorem 2.6]{AgmonKlein}, refining the
crude bound of Theorem \ref{lemma:uniqueextsol} by using the
Phragmen-Lindel\"of principle.  This allows us to identify the dilated
solutions with a decaying solution of the dilated equation. In view of
Lemma \ref{lemma:uniqueextsol},
(up to multiplication with a function only depending on $k$)
this solution is uniquely defined by the asymptotic behaviour
as $x$ goes to infinity.

Let us consider ${\tilde v}_+(z,k)$ along a ray
$\Gamma:=\{z\in \bC\setminus\{0\} \mid \arg z = \gamma\}$ with
$0<\gamma<\beta$.  Then for $x > 0$ and $k\in\bC_{+}\z$, the function
\beq
\omega(x,k,\gamma) := {\tilde v}_+(e^{i\gamma}x,k)
\Leq{eq:wxikgamma}
satisfies the equation
\beq
\l((e^{i\gamma}x+1) \omega'(x,k)\ri)'
  + \frac{e^{2i\gamma}}{h^{2}}\,
  q(e^{i\gamma}x,k,Z_{+},\mu) \, \omega(x,k) = 0
\Leq{eq:wgammaeq}
with $q$ from \eqref{eq:schrlog}.
Moreover the initial data
\beq
\omega(0,k,\gamma) = {\tilde v}_+(0,k),
\quad \omega'(0,k,\gamma) = e^{i\gamma} {\tilde v}_+'(0,k),
\Leq{eq:initdata}
are analytic in $k\in\bC_{+}\z$.

To obtain an analytic continuation of ${\tilde v}_+(x, k)$ into
the lower half-plane, first observe that by the Liouville-Green
Theorem and Remark \ref{remark:newequationvpm}, Equation
\eqref{eq:wgammaeq} has a unique solution $\omega_+(x,k,\gamma)$ in
the cone
$-\gamma < \arg k < \pi - \gamma$ characterized by the asymptotic
relation
\beq
\omega_+(x,k,\gamma) =
\frac{1}{\sqrt{e^{i\gamma}x}} e^{i\phi(e^{i\gamma}x, k)} (1 + o(1))
\qmbox{as} x\to\infty.
\Leq{eq:asympW}
We claim that in fact
\beq
\omega_+(x,k,\gamma) = \omega(x, k, \gamma)
\qquad\mbox{for } x\in(0,\infty), \quad 0 < \arg k < \pi-\gamma.
\Leq{eq:asyconicide}
Then $\omega_+(0,k,\gamma)$ and $\omega_+'(0,k,\gamma)$ provide the
analytic continuation of the initial data for ${\tilde v}_+(x,k)$
into the region $-\gamma < \arg k < 0$, implying that $\tv_+(x,k)$
can be continued analytically into the lower half-plane.

To prove \eqref{eq:asyconicide}, we observe that
$x\mapsto \tv_+(x,k)$ is of exponential type for
$x\in\Sigma_{\alpha,\beta}$ and decays exponentially for $\Im(k) >0$.
Then it follows from the Phragmen-Lindel\"of principle
\cite[VI.4]{Conway}, applied to
\beq
g(x,k) := \sqrt{x} \exp\big(-i\phi(x,k)\big) {\tilde v}_+(x,k)
\Leq{eq:defg}
that for fixed $\Im(k) >0$ the function ${\tilde v}_+(x,k)$
decays exponentially as $x\to\infty$ in a small cone containing
$(0,\infty)$.

Therefore Remark \ref{remark:newequationvpm} and Remark
\ref{remark:anyotherfamily} applied to the dilated function
$\omega_+(x,k,\widetilde{\gamma})$ for some small
$\widetilde{\gamma}>0$ imply that $\omega_+(x,k,\widetilde{\gamma})$
is a multiple of $\omega(x,k,\widetilde{\gamma})$. This means moreover
that it decays at a rate given by the expected function
\[
\frac{1}{\sqrt{e^{i\widetilde{\gamma}}x}}
  \exp\big(i\phi(e^{i\widetilde{\gamma}}x,k)\big).
\]

We can repeat this procedure a finite number of times and deduce that
for fixed $k$ the analytic function $g(x,k)$ is uniformly
bounded as $x\to\infty$ within an angle $-\epsilon < \arg x <
\gamma+\epsilon$ for some $\epsilon >0$. Since by
  \eqref{eq:vlogpmasymp}
\[
\lim_{x\to\infty} g(x,k)=1,
\]
it follows from Montel's theorem \cite[VII.2]{Conway} that this limit
is assumed uniformly as $x\to\infty$ in $0 \leq \arg x \leq
\gamma$. This proves \eqref{eq:asyconicide}. Since
$\gamma\in(0,\beta)$ was arbitrary, we obtain an analytic continuation
of ${\tilde v}_+(x,k)$ to $-\beta < \arg k <\pi$. It remains to
prove \eqref{eq:vpmasympcone}.

For $-\alpha < \gamma < \beta$ we can apply Lemma
\ref{lemma:uniqueextsol} to the dilated function $\omega(x,k,\gamma)$
to have
\beq
g(x,k) = {\cal O}(1) \qmbox{as} x\to\infty
\mbox{ within } \Sigma_{\alpha,\beta}.
\Leq{eq:gbound}
We already know from \eqref{eq:asyconicide} that $g(x,k) \to 1$ as
$x\to\infty$ along any ray such that $0 < \eta \leq \arg(k x) \leq \pi
- \eta$ for some $\eta \geq 0$. Therefore we have that also
  locally uniformly in $k\in\bC\z$, $-\beta < \arg k < \pi$
\[
g(x,k) = {\cal O}(1) \qmbox{as} x\to\infty
\mbox{ within } \Sigma_{\alpha,\beta}
\]
and $g(x,k)$ is uniformly bounded along the boundary rays of
$\Sigma_{\alpha,\beta}$. That $g(x,k)$ is uniformly bounded in
$x\in\Sigma_{\alpha,\beta}$ is now a consequence of the
Phragmen-Lindel\"of Principle. The fact that $g(x,k)$ tends to $1$
as $x\to\infty$ since it does so along some ray contained in its
interior, completes the proof of the theorem.
\qed

\begin{remark}
  The analytical extension of $\tv(x,k) = v(\Log(x+1),k)$ gives in turn
  the extension of $v(\xi,k)$.
  \hfill$\Diamond$
\end{remark}

%
\subsection{Generalised eigenfunctions,
  Green's function and the scattering matrix}\label{sec:scatt2d}
%

We are now ready to construct the main elements for the partial wave
expansion required to give a definition of the resonances of our
operator.

We considered in the previous section the {\em outgoing}
  respectively {\em incoming} solutions as the solutions meeting a
``regular'' boundary condition at infinity.  Because of the fact
  that the boundary conditions are at infinity it requires some work
to prove that they can be analytically extended to the second Riemann
sheet across the positive real axis.

This is much simpler for the solution $\tv_0(x,k)$ of
\eqref{eq:schrlog} (or the corresponding $v_0(\xi,k)$ of
\eqref{eq:schro}) that is {\em regular} in $0$ in the sense of
the boundary conditions derived from \eqref{eq:BC2dR}, i.e.
\beq
\tv_0(0,k) = 1\qmbox{,}\tv_0'(0,k) = 0.
\Leq{eq:regsol}
Being the solution of a boundary problem with analytic coefficients
and analytic initial conditions, the following theorem follows as a
corollary of the standard theory of complex ordinary differential
equations (see \cite[Chapter 1.8]{CL55}).

\begin{theorem}[The regular solution]\label{thm:regsol2d}
  The unique solution $\tv_0(x,k)$ of \eqref{eq:schrlog} defined by the
  condition \eqref{eq:regsol} is analytic in the cone
  $x\in\Sigma_{\alpha, \beta}$, $k\in\bC\z$ defined in
    \eqref{def:anastrip} and satisfies
  \beq
    \tv_0(x,k) = \tv_0(x,-k).
  \Leq{eq:vpmrelneg}
\end{theorem}

\begin{remark}
  Working with \eqref{eq:schrlog} or \eqref{eq:schro} is
  equivalent. We will use each time the representation that makes the
  proofs and the computations easier. Therefore in what follows we do
  not continue to remark that the properties are equivalent. It is
  always possible to understand in which setting we are working,
  looking at the name of the functions and the variables.

From now on, we will always assume that the Wronskian is defined in its
generalised form given by
\[
W_x(f,g) := p(x)\l( f(x)g'(x) - f'(x)g(x) \ri),
\]
where the notation comes from \eqref{eq-hsleq}.
\end{remark}
We are finally ready to introduce the basic elements for scattering
theory on the half-line. We call \emph{Jost functions} associated to
the radial equation \eqref{eq:schrlog} and our choice of phase
function $\phi(x,k)$ the Wronskians
\beq
f_\pm(k) :=
W\big(\tv_\pm(\bullet ,k),\tv_0(\bullet ,k)\big).
\Leq{eq:Wro2d1}
They connect the regular solution to the incoming and outgoing ones
via the identity
\beq
W(\tv_-,\tv_+) \tv_0 = f_+ \tv_- - f_- \tv_+,
\qmbox{with}
W(\tv_+,\tv_-) = 2ik,
\Leq{eq:WroEq}
that follows expanding explicitly the Wronskian and using the
asymptotic behaviour of the solutions in their domain of
analyticity. In particular this implies the following corollary of
Theorem \ref{thm:regsol2d} and Theorem \ref{thm:anaext}.

\begin{corollary}\label{cor:formfpm}
  The Jost functions $f_\pm(k)$ are analytic in $k\in D_\pm
  (\alpha,\beta)$ defined in \eqref{eq:dom2d} and verify
\beq
f_\pm(k) = \pm (2ik) \lim_{x\to\infty} e^{i\gamma/2}\sqrt{x}
\textstyle\exp\l(\pm i\phi(e^{i\gamma} x,k)\ri) \tv_0(e^{i\gamma} x,k),
\Leq{eq:udeterminefpm}
where $\gamma\in(-\alpha,\beta)$ satisfies $\gamma \gtrless - \arg(k)$
according to the choice of sign of \eqref{eq:udeterminefpm}.
\end{corollary}

It will be convenient for what follows to change the normalisation
$\tv_0(0,k) = 1$ to one at ``infinity'' in the sense of Corollary
\ref{cor:formfpm}. Namely if $f_+(k)\neq 0$, we define the
\emph{generalised eigenfunction} of the radial equation
\eqref{eq:schrlog} and our choice of phase function $\phi(x,k)$ the
function
\beq
e(x,k) := f_+(k)^{-1} \tv_0(x,k).
\Leq{eq:GenEf}
With this notation we introduce for $k\in\Sigma_{\alpha,\beta}$ with
$f_+(k)\neq 0$ the \emph{radial Green's function}
\beq
G(x,x';k) := e(x_<,k)\tv_+(x_>,k),
\Leq{eq:RGFunc2d}
where for $x,x' > 0$, $x_{<} := \min\{x,x'\}$ and
$x_{>} := \max\{x,x'\}$.
$G(x,x';k)$ is a fundamental solution of the radial Schr\"odinger
equation \eqref{eq:schrlog}.

\begin{remark} We now consider the spectral parameter $\mu$ appearing
  in Equation \eqref{eq:schro} as a perturbation of the operator
  $K_{\xi}$ defined in \eqref{def-Kxieta}. Consequently we will write
\[
K_\xi(Z_+,\mu) := K_{\xi} + \mu
\]
for the perturbed operator.
\hfill$\Diamond$\end{remark}

\begin{remark}
  Notice that eventual zeros of $f_+(k)$ for $k\in\bC_+\z$ correspond
  to eigenvalues of the operator.
\hfill$\Diamond$\end{remark}

In view of Theorem \ref{thm:anaext} and \ref{thm:regsol2d},
$G(x,x';k)$ possesses a meromorphic continuation in $k$ into the
possibly two-sheeted domain, projecting to $D_+(\alpha,\beta)$
defined by \eqref{eq:dom2d}.

Finally we introduce the so-called \emph{scattering matrix element}
\beq
s(k) = \frac{f_-(k)}{f_+(k)}
\Leq{eq:defsmel}
which in view of Corollary \ref{cor:formfpm} is a meromorphic function
of $k$
over $D_{+}(\alpha,\beta)\cap D_{-}(\alpha,\beta)$.

\begin{lemma}\label{lemma:relationsgfgesme}
Let $x,x' >0$ and $-\beta < \arg(k) < \alpha$.
\begin{compactenum}
\item The radial Green's function and the radial generalised
  eigenfunctions satisfy the functional relation
  \beq G(x,x';k) -
    G(x,x';-k) = -2ik\, e(x_{<},k)e(x_{>},-k).
  \Leq{eq:funcRel}
\item The scattering matrix element satisfies the following relation
\beq
s(-k) = s(k)^{-1}.
\Leq{eq:smelrel}
\item The scattering matrix elements and the radial generalised
  eigenfunctions satisfy the functional relation
\beq
s(k)e(x,-k) = e(x,k).
\Leq{eq:funcRelsmge}
\end{compactenum}
\end{lemma}
\proof
From \eqref{eq:anaextvpm} and \eqref{eq:vpmrelneg} we have that
\beq
f_+(-k) =W\big(\tv_+(\bullet ,-k),\tv_0(\bullet ,-k)\big) =
W\big(\tv_-(\bullet ,k),\tv_0(\bullet ,k)\big) =f_-(k)
\Leq{eq:fpmrel}
for $k\in D_+(\alpha,\beta)\cap D_-(\alpha,\beta)$. Therefore, using
\eqref{eq:WroEq} and the definitions of the radial Green's function
and the radial generalised eigenfunctions, we get
\beqno
\lefteqn{G(x,x';k) - G(x,x';-k) =
  e(x_<,k)\tv_+(x_>,k)-e(x_<,-k)\tv_+(x_>,-k)}\\
&=&\tv_0(x_<,k)
  \big(f_-(-k)^{-1} \tv_+(x_>,k)-f_+(-k)^{-1} \tv_+(x_>,-k)\big)\\
&=&\tv_0(x_<,k)f_-(-k)^{-1}f_+(-k)^{-1}
\big(f_+(-k) \tv_+(x_>,k)-f_-(-k) \tv_+(x_>,-k)\big)\\
&=&\tv_0(x_<,k)f_+(k)^{-1}f_+(-k)^{-1}
\big(f_+(-k) \tv_-(x_>,-k)-f_-(-k) \tv_+(x_>,-k)\big)\\
&=&-2ik\, e(x_<,k)f_+(-k)^{-1}\tv_0(x_>,-k)
=-2ik\, e(x_<,k)e(x_>,-k).
\eeqno

The second part and the third part follows as a direct application of
\eqref{eq:fpmrel} to the definition of the scattering matrix elements.
\qed

A first consequence of Lemma \ref{lemma:relationsgfgesme} is that it
is enough to discuss the scattering matrix elements in the angle
$-\beta < \arg(k) < \alpha$.

With the above definitions we can discuss the notion of eigenvalues
for the radial non-selfadjoint Schr\"odinger operator $K_\xi(Z_+,\mu)$
in $L^2((0,\infty),\ch^2(\xi)d\xi)$. We define
\beq
\cE_{Z_+,\mu} :=
\Big\{
k\in\overline{\bC}_+\z\mid  f_+(k) = 0,
 e^{-\xi/2} e^{i\phi(\xi,k)}\in L^2((0,\infty),\ch^2(\xi)d\xi)
\Big\}.
\Leq{def:evrso}

If $k\in\cE_{Z_+,\mu}$, we call $k$ an \emph{eigenvalue} of this
\emph{quadratic eigenvalue problem}. All other zeros of the Jost
function $f_+(k)$ are called resonances of $K_\xi(Z_+,\mu)$ and we
denote them by
\beq
\cR_{Z_+,\mu}:=
\l\{
k\in D_+(\alpha,\beta)\setminus\cE_{Z_+,\mu} \mid f_+(k) = 0
\ri\}.
\Leq{eq:defRes}

\begin{remarks}\label{rem:plesGfSm}
\begin{compactenum}
\item The condition $\xi\mapsto e^{ -\frac{\xi}2} e^{i\phi(\xi,k)}\in
  L^2((0,\infty),\ch^2(\xi)d\xi)$ is automatically fulfilled when
  $k\in\bC_+\z$, independently of $\mu$.
\item There cannot be real positive $k\in\cE_{Z_+,\mu}$.  In fact, if
  there would exist $k\in(0,\infty)$ in $\cE_{Z_+,\mu}$, then by Theorem
  \ref{thm:anaext} we would have $v_+(\xi,k)\in
  L^2((0,\infty),\ch^2(\xi)d\xi)$, but it is evident from the asymptotic
  behaviour of $v_{+}$ that this is impossible.  On the other hand, we
  cannot exclude a priori the presence of real $k$ in $\cR_{Z_+,\mu}$.
\item Two Jost functions cannot vanish simultaneously in $-\beta <
  \arg(k) < \alpha$, otherwise $\tv_+$ and $\tv_-$ (or $v_+$ and
  $v_-$) would be linearly dependent in contradiction with their
  asymptotic behaviour. Therefore the points of
  $\cE_{Z_+,\mu}\cup\cR_{Z_+,\mu}$ contained in $-\beta < \arg(k) <
  \alpha$ are in one to one correspondence with all the poles of the
  scattering matrix elements $s(k)$.

  In view of the definitions \eqref{eq:GenEf} and \eqref{eq:RGFunc2d},
  the set $\cE_{Z_+,\mu} \cup \cR_{Z_+,\mu}$ can be identified with
  the set of poles of the radial Green's function $G(\xi,\xi';k)$ or    
  with the set of
  poles of the generalised radial eigenfunctions $e(x,k)$.
\item The set $\cR_{Z_+,\mu}$ of resonances does not depend on the
  choice of the phase function which determines the Jost functions
  $f_\pm(k)$, the generalised radial eigenfunctions and the scattering
  matrix elements.  \hfill$\Diamond$
\end{compactenum}
\end{remarks}

%
\section{Formal partial wave expansion of the Green's
  function}\label{Pw1}
%
For real $E$ we know from Remark \ref{rmk:singlespec}  that the spectrum of
$K_{\eta}=K_{\eta}(E,Z_-,h)$ consists of an infinite number of
simple eigenvalues
\[
\mu_0(E)<\mu_1(E)<\mu_2(E)<\mu_3(E)<\ldots
\]
tending to infinity, where in the notation of Remark \ref{rmk:singlespec}
we have
$
  \mu_n := \lambda_n + \gamma_2.
$
These extend to analytic functions of $E$
in some neighborhood of the real line. We shall denote by $\vv_{n,E}$
the eigenfunctions
\[
K_{\eta}(E)\vv_{n,E}(\eta) = \mu_{n}(E)\vv_{n,E}(\eta), \quad
n\in\bN_0, 
\]
normalised by
\[
\|\vv_{n,E}\|^2 = \int_{-\pi}^{+\pi}|\vv_{n,E}(\eta)|^2\;d\eta=1
\]
for $E\in(0,\infty)$ and then extended analytically. We choose $\vv_{n,E}$
real for $E$ real.

Define
\beq
K := F\; \cH_{\cG}
\Leq{def:K2d}
with $\cH_{\cG}$ from Proposition  \ref{prop:saExt} and
$F$ from \eqref{def-F}.
Instead of solving $(\cH_{\cG} - E)u = f$ in
$L^{2}(M, F(\xi,\eta)d\xi d\eta)$ for
$E\in\bC\setminus\sigma(\cH_{\cG})$, we look at the solutions of
\beq
\big( K - F(\xi,\eta)E\big)u (\xi,\eta)= F(\xi,\eta) f(\xi,\eta).
\Leq{22-2d}
We already know (see \eqref{def-Kxieta}) that
 \[
\big(K - F(\xi,\eta)E\big) u(\xi,\eta) =
  K_{E} \,u(\xi,\eta) = (K_{\xi}+K_{\eta})u(\xi,\eta).
 \]

Now, using the completeness of the orthonormal base
$\l\{\vv_{n,E}\ri\}_{n\in\bN}$
for $E\in\bR$, $u$ possesses the expansion
\beq
u(\xi,\eta) = \sum_{n\in\bN_0}u_n(\xi,\eta)
\qmbox{with} u_n(\xi,\eta) := \vv_{n,E}(\eta)\psi_{n,E}(\xi),
\Leq{33-2d}
where
\[
\psi_{n,E}(\xi) =
\int_{-\pi}^{+\pi}\vv_{n,E}(\eta)u(\xi,\eta)\;d\eta.
\]
This expansion extends to complex values of $E$ by analyticity
(note that no complex conjugate is involved, since $\vv_{n,E}$ is chosen
real for $E\in\bR$).
Analogously we get
\beq
F(\xi,\eta)f(\xi,\eta) = \sum_{n\in\bN_0}\vv_{n,E}(\eta)g_{n,E}(\xi)
\qmbox{with}
g_{n,E}(\xi) := \int_{-\pi}^{+\pi}\vv_{n,E}(\eta)(Ff)(\xi,\eta)\;d\eta.
\Leq{44-2d}

Substituting (\ref{33-2d}) and (\ref{44-2d}) into (\ref{22-2d}) one
gets
\[
\l(K_{\xi}+K_{\eta}\ri) \sum_{n\in\bN_0}
u_n(\xi,\eta) = \sum_{n\in\bN_0}\vv_{n,E}(\eta)g_{n,E}(\xi)
\]
or equivalently
\beq
\sum_{n\in\bN_0}\vv_{n,E}(\eta)
\big(\l(K_{\xi}(E)+\mu_{n}(E)\ri)\psi_{n,E}(\xi)-g_{n,E}(\xi)\big)=0.
\Leq{55-2d}

\begin{remark}\label{rmk:resKxi}
  (\ref{55-2d}) extends to complex points $E\not\in\sigma(H)$, where
  $K_{\xi}(E)+\mu_{n}(E)$ possesses an inverse $R_{n}(E)$ by means of
  the Green's function defined in \eqref{eq:RGFunc2d}.
  \hfill$\Diamond$
\end{remark}
\beq
\psi_{n,E}(\xi) = R_n(E)g_{n,E}(\xi)
= \int_{(0,\infty)}G_n(\xi,\tilde{\xi};E)\int_{-\pi}^{+\pi}\vv_{n,E}
(\te)(Ff)(\tilde{\xi},\te) \;d\te\;d\tilde{\xi},
\Leq{88-2d}
using (\ref{55-2d}). Combining (\ref{88-2d}) and (\ref{33-2d}) we
obtain
\[u(\xi,\eta) = \sum_{n\in\bN_0}\vv_{n,E}(\eta)\iint_{M_{0}}
  G_n(\xi,\tilde{\xi};
E)\vv_{n,E}(\te)(Ff)(\tilde{\xi},\te)\;d\tilde{\xi}
\,d\te\]
and we read off the partial wave expansion for the Green's function
\beq\hspace*{-1cm}
G(\xi,\eta;\tilde{\xi},\te;E) =
\sum_{n\in\bN_0}\vv_{n,E}(\eta)\vv_{n,E}(\te)G_n(\xi,
\tilde{\xi};E)(\cosh^2\tilde{\xi}-\cos^2\te).
\Leq{99-2d}

It would be of great interest to be able to prove that the sum
converges in the sense of distributions in the product space
$D'(M)\otimes D'(M)$. Then we could use our results on the
analytic continuation of the $G_n$ and of the angular eigenfunctions
to give a meromorphic continuation of the
$G(\xi,\eta;\tilde{\xi},\tilde{\eta};E)$ in $E$ to the second Riemann
sheet (or $k\in\bC_-$).

Anyhow, for each fixed $N\in\bN$, we can consider the
restriction $K_N$ of the operator $K$ to the subspace
\beq
\Upsilon_N(E) := \bigoplus_{n=0}^{N} \Phi_n(E)\otimes
  L^2((0,\infty),\ch^2(\xi)d\xi)\\
\subset L^2([-\pi,\pi],d\eta)\otimes L^2((0,\infty),\ch^2(\xi)d\xi)
\Leq{eq:subspace}
where $\Phi_n(E)$ is the subspace spanned by $\vv_{n,E}$. The relative
Green's function
\[\hspace*{-1cm}
G_N(\xi,\eta;\tilde{\xi},\te;E) =
\sum_{n=0}^N\vv_{n,E}(\eta)\vv_{n,E}(\te)G_n(\xi,
\tilde{\xi};E)(\cosh^2\tilde{\xi}-\cos^2\te)
\]
is the truncated sum obtained from \eqref{99-2d}.
Being a finite sum of well-defined terms, it is convergent. 
Moreover it follows from the
results of the previous sections that it possesses a meromorphic
continuation in $E$ to the second Riemann sheet.

\section{Resonances for the two-centers problem}\label{sec:defres2d}

With the expansion of Section~\ref{Pw1} and the theory developed in the previous sections, we are finally ready to define the resonances for the two-centers problem and analyse some of their properties. This is done in Section \ref{sec:defres2sssc}.

The rest of the section is then devoted to asymptotically locate these resonances.
In particular in Section \ref{sec-apprResKxi} we show that the resonances can be computed as roots of some explicit asymptotic equation, and in the subsequent sections we explicitly solve this equation in different semiclassical energy regimes.

\subsection{Definition of the resonances}\label{sec:defres2sssc}
The operator $K_{\eta}$ defined by \eqref{eq:kappaetaZnonzero} has
discrete spectrum $\mu_{n}(k^{2})$ admitting an analytic continuation
in $k^{2} := E$ in some neighborhood of the real axis. At the same
time for each $\mu$, the resolvent of the operator $K_{\xi}(\mu,Z_+)$
(see Remark \ref{rmk:resKxi}) can be extended in terms of $k$ to the
negative complex plane, having there a discrete set of poles
$k_{m}(\mu)$.

With the definitions given in Section \ref{sec:scatt2d} we set
\beq
\hspace*{-2mm}
\cE_{n} :=
\big\{
k\in\overline{\bC}_+\z \mid  f_+(k,\mu_{n}(k^{2})) = 0,
e^{-\xi/2} e^{i\phi(\xi,k,\mu_{n}(k^{2}))}\in L^2((0,\infty),\ch^2(\xi)d\xi)
\big\}.
\Leq{def:evrso2d}

If $k\in\cE_n$ (for some $n\in\bN_0$), we call $k$ an \emph{eigenvalue}
of the quadratic eigenvalue problem for $K=K(Z_{-},Z_+)$ defined
  in \eqref{def:K2d}. All other zeros of the Jost function
$f_+(k,\mu_{n}(k))$ are called \emph{resonances} of $K(Z_{-},Z_+)$
and we denote them by
\beq
\cR_{n}:=
\l\{
k \in D_+(\alpha,\beta)\setminus\cE_{n} \mid f_+(k,\mu_{n}(k^{2})) = 0
\ri\}.
\Leq{eq:defRes2d}
\begin{proposition}
  The sets $\cE_{n}$ and $\cR_{n}$ are made by an at most countable
  number of elements $k_m\in D_+(\alpha,\beta)$ ($m\in I \subseteq
  \bN$) of finite multiplicity such that
  $f_+(k_m,\mu_{n}(k_m^{2})) = 0$.
\end{proposition}
\proof $f_+(k)$ and $\mu_n(k^2)$ being non-constant analytic functions of $k$, the
statement is clear. \qed

\begin{remark}
  Notice that if $k^2$ is an eigenvalue of the full operator $K$ (or
  its restriction $K_N$), then it must be an eigenvalue of
  $K_\xi(Z_+,\mu_n)$ for some $\mu_n(k^2)$ (i.e. an element of
  $\cE_n$).
\hfill$\Diamond$\end{remark}
\begin{remark}
By definition
$\cE_{n}\cap \cR_{n} = \emptyset$.  Furthermore, it is clear looking
at the asymptotic behaviour \eqref{eq:phineqqsym}
of the phase function 
that it is impossible that $k\in\cE_n$ and $k\in\cR_{n'}$ for $n\neq n'$.
\hfill$\Diamond$\end{remark}
Relying on the previous discussion and on Remark \ref{rem:plesGfSm}.2
we can switch from the $k^2$ plane to the $k$ plane and refer to
\beq
\cE^N := \bigcup_{n=0}^N \cE_n, \qquad
\cR^N := \bigcup_{n=0}^N \cR_n
\Leq{eq:EVResSetN}
as the sets of \emph{eigenvalues} and \emph{resonances} of
$K_N$. Moreover, in view of Remark \ref{rem:plesGfSm}.2, the points of
$\cE^N\cup\cR^N$ contained in $D_+(\alpha,\beta)\cap
D_-(\alpha,\beta)$ are in one-to-one correspondence with the poles of
the scattering matrix elements $s_n(k):=s(k,\mu_n)$ and with the poles
of the Green's functions
$G_n(\xi,\widetilde\xi;k):=G(\xi,\widetilde\xi;k,\mu_n(k))$ for
$n\in\{0,\ldots,N\}$.

\begin{remark}
If we suppose that \eqref{99-2d} is convergent, we can refer to
\beq
\cE := \bigcup_{n=0}^\infty \cE_n, \qquad
\cR := \bigcup_{n=0}^\infty \cR_n
\Leq{eq:EVResSet}
as the sets of \emph{eigenvalues} and \emph{resonances} of $K$. As for
the restricted operator, in view of Remark \ref{rem:plesGfSm}.2, the
points of $\cE\cup\cR$ contained in $D_+(\alpha,\beta)\cap
D_-(\alpha,\beta)$ are in one-to-one correspondence with the poles of
the scattering matrix elements $s_n(k)$ and with the poles of
the Green's functions $G_n(\xi,\widetilde\xi;k)$.
\hfill$\Diamond$\end{remark}

%
\subsection{Computation of the resonances of
  $K_{\xi}$} \label{sec-apprResKxi}
%

Consider the equation
\beq\hspace*{-1cm}
0 = K_{\xi}(h) \psi(\xi)= -h^{2}\pa_\xi^2\psi(\xi)
  - Z_+ \ch(\xi)\psi(\xi) -  E \ch^{2}(\xi)\psi(\xi)
\Leq{eq-kxi}
with the condition $\psi'(0) = 0$. The potential
\[
V(\xi; Z_{+}, E) := - Z_+\ch(\xi) - E\ch^{2}(\xi)
\]
has a Taylor expansion around $\xi=0$ given by
\begin{align*}
V(\xi; Z_{+}, E) &= - \frac{Z_+}2 \l(e^{\xi} + e^{-\xi}\ri)
  - \frac{E}{4} \l(e^{\xi} + e^{-\xi}\ri)^{2} \\
&= -Z_{+} - E - \l(E + \frac{Z_+}2\ri) \xi^{2} + \cO(\xi^{4})
= A - \omega^{2}\xi^{2} + \cO(\xi^{4}),
\end{align*}
where $A := -Z_{+} - E$ and $\omega=\sqrt{E + \frac{Z_+}2}$.

Let now $E +\frac{Z_+}2 > 0$. We would like to apply the theory
developed in \cite{BCD1,BCD2,BCD3} and \cite{Sj} to get the resonances from the
eigenvalues
\[
e_{n}(h) = h(2n+1)\omega \qquad (n\in\bN_{0})
\]
of the harmonic oscillator
\[
H_{osc} = -h^{2}\pa_{\xi}^{2} + \omega^{2}\xi^{2},
\]
according to
\[
A_{n} (h,E,Z_{+})= -Z_{+} - E - ih(2n+1)\omega + \cO(h^{3/2}).
\]

\begin{remark}
  \cite{BCD1,BCD2,BCD3} and \cite{Sj} are not directly
    applicable, as there it is \emph{essential} to assume that the
  potential is bounded, and this is clearly false in \eqref{eq-kxi}.
\hfill$\Diamond$\end{remark}

The problem stressed by the previous remark can be solved. With the
change of variable given by $y := \sh(\xi):(0,\infty)\to(0,\infty)$ we
change the measure from $\ch^{2}(\xi)\;d\xi$ to $\sqrt{y^{2}+1}\; dy$.
At the same time the differential equation of $K_{\xi}(Z_{+},\mu)$
takes the form
\[\hspace*{-0.5cm}
- h^{2}(y^{2}+1)\pa_{y}^{2}u(y) - h^{2}y \pa_{y} u(y)
  + \l(\mu - k^{2}(y^{2}+1) - Z_{+}\sqrt{y^{2}+1}\ri)u(y) = 0.
\]
Note that $\mu$ will correspond to an eigenvalue of the angular
equation $K_{\eta}$, and as such it will be an analytic function of
$E$.  Moreover it will be real for real values of $E$ (see Section
\ref{sec:spec2d}).

With the ansatz
\[
u(y) := \frac1{\sqrt[4]{y^{2}+1}} v(y)
\]
we can rewrite the differential equation in Liouville normal form as
\beq
\frac{y^{2}+1}{\sqrt[4]{y^{2}+1}}\l(
-h^{2 } \pa_{y}^{2}v(y) +
V(k,Z_{+},\mu,h; y)v(y)
\ri) = 0
\Leq{eq:diffeqResNew}
where
\[
V(k,Z_{+},\mu,h; y):=-k^{2} - \frac{Z_{+}}{\sqrt{y^{2}+1}}
  + \frac{\mu}{1+y^{2}} - \frac{y^{2} - 2}{4(y^{2}+1)^{2}}h^{2}.
\]
This potential $V$ has the following properties:
\begin{compactitem}
\item it is smooth in $(0,\infty)$;
\item it is bounded;
\item it is analytic in a cone centered at the positive real axis;
\item it has a non-degenerate global maximum at $y=0$;
\item around the maximum $V$ can be expanded in Taylor series as
\[
V(k,Z_{+},\mu,h; y) = A - \omega^{2}y^{2} + \cO(y^{4}),
\]
where $A := -Z_{+} - k^{2} +\mu - \frac{h^{2}}{2}$ and
$\omega=\sqrt{ \mu + \frac54h^{2} - \frac{Z_+}2}$.
\end{compactitem}
Therefore it satisfies the assumptions of \cite{BCD1,BCD2,BCD3} and
\cite{Sj}, there a resonance is an exact zero of some symbol in the semi-classical parameter, and we are left to compute the leading terms of this symbol. This allows us to approximate the resonances with the eigenvalues of the harmonic oscillator according to
\beq
A_{n} (h,E,Z_{+},\mu)= -Z_{+} - k^{2} + \mu - ih(2n+1)\omega
  + \cO(h^{3/2}).
\Leq{eq-Andef}
This given, we have a solution of \eqref{eq:diffeqResNew} if $v$ is
identically $0$ or if $A_{n} = 0$. In summary,
\begin{proposition}
	For any given $Z_+$ and $\mu$, the resonances of $K_{\xi}(Z_{+},\mu)$ are asymptotically given by the zeroes of a symbol $A_{n} (h,E,Z_{+},\mu)$ whose expansion as $h \to 0$ is provided by \eqref{eq-Andef}.
\end{proposition}

From this
formula one can have a first very rough approximation of the resonances $E_n =
k^2_n$ in orders of $\Re(\mu) \gg 0$ and $h$ small but constant as follows
\beq
\Im E_n =  (2n+1)h\sqrt{\Re\mu} +\Im\mu+ {\cal O}
  \l((\Re\mu)^{-1/2}\ri) \qmbox{,}
\Re E_n = \sqrt{\Re\mu - Z_+ + {\cal O}\l((\Re\mu)^{-1/2}\ri)}.
\Leq{eq:bigEAppBigmu}

\begin{remark}\label{rmk:app1}
  The approximation \eqref{eq-Andef} identifies the resonances
  generated by the top of the potential (at $\xi = 0$) and these
  corresponds to the resonances generated by the classical closed
  hyperbolic trajectory bouncing between the two centers (see Remark
  \ref{rmk:coord}.1).
\hfill$\Diamond$\end{remark}

\begin{remark}\label{rmk:app2}
  In \cite{S14} it is proven that for $Z_+ < 0$, $|Z_+|
  < Z_-$, there is for small energies a region of the phase-space
  characterized by closed orbits related to a local minimum of the
  potential. We expect in this case the appearance of some
  shape resonances at exponentially small distance in $h$ from the
  real axis (see \cite{HaSi1,HaSi2} and \cite[Chapter 20]{H-S}). We
  plan to study the existence and the distribution of these other
  resonances in a future work.
\hfill$\Diamond$\end{remark}

%
\subsection[Resonances for $Z_{-} = 0$]{Eigenvalues asymptotics and
  resonant regions for $Z_{-} = 0$ near the bottom of the
  spectrum} \label{sec-app2dz-0}
%

As we did previously, before studying the general system, let us have
a look to the simplest case $Z_{-} = 0$.  With a proper renaming of
the constants and the notation of \eqref{eq:fSolMathNot}, in
\cite[Section 2.331]{MS} it is proved that
\begin{theorem}
  For $\delta\rightarrow+\infty$ and $n\in\bN_{0}$, the
  eigenvalues $\lambda^\pm_n$ of the Mathieu equation written in the
  form $-y''(x)+ (2\delta\cos(2x) - \lambda) y(z)=0$ are
\[
\lambda^+_{n}(\delta) = -2\delta + (4n+2)\sqrt{\delta} + \cO(1)\qmbox{,}
\lambda^-_{n+1}(\delta) = -2\delta + (4n+2)\sqrt{\delta} + \cO(1).
\]
\end{theorem}
Thus we have as a direct consequence the following theorem.
\begin{corollary}\label{thm-muZ-null}
In the limit $h\searrow 0$ and for every $E>0$ we have
\[
\mu^{+}_{n} (h, E, 0)= (2n+1) \sqrt{E}\; h + \cO(h^{2})\qmbox{,}
\mu^{-}_{n} (h, E, 0)= (2n+1) \sqrt{E}\; h + \cO(h^{2}).
\]
where $\mu_{n}^\pm$ are the eigenvalues described in Section \ref{Pw1}
reindexed using the parity separation described by item \ref{item:sepSp}.
of our 'fact sheet' in Section \ref{sec:spec2d} on page \pageref{item:sepSp0}.
\end{corollary}

We can use this result in combination with
(\ref{eq-Andef}) to obtain the following proposition.

\begin{proposition} The resonances in the set $\cR_n$ (see \eqref{eq:defRes2d}) are given asymptotically as $h\to 0$ by the solutions of the following equation
\[
-A_{n}(h,E,Z_{+},\mu^{+}_{m}(h,E,0)) = 0.
\]
\end{proposition}

Neglecting the error terms and writing $E = k^2$ we have
\beq\hspace*{-0.9cm}
k^2 + Z_{+} -(2n+1) k h + ih (2m+1) \sqrt{(2n+1) k h
  + \frac{5h^2}4 -\frac{Z_+}{2}} = 0.
\Leq{eq-Enmvalue3}

%
\subsection[Resonances for $Z_{-}> 0$]{Eigenvalues asymptotics and
  resonant regions for $Z_{-} > 0$ near the bottom of the
  spectrum} \label{sec-app2dz-n0}
%
Notice that we can always define $Z_{-}$ in such a way that it is
non-negative. In presence of the $Z_{-}$ term the equation
$K_{\eta}\psi(\eta) = 0$ assumes the form
\beq
0 =-h^{2}\pa_\eta^2\psi(\eta)  + \l(-\mu + Z_- \co(\eta) +
E \co^2(\eta)\ri)\psi(\eta),
\Leq{eq-kappaetanonzero}
with periodic boundary conditions on $[-\pi, \pi]$.

\begin{remark} 
  In view of \eqref{eq-kappaetanonzero}, we have, for all normalized $\psi$ in the domain of $K_\eta$, 
  \[(\psi , K_\eta ^0\psi)-Z_-\leq (\psi , K_\eta \psi)\leq (\psi , K_\eta ^0\psi)+Z_-.\] 
  By the min-max principle (see \cite[p. 75]{RSIV}), we get, for all $n$, 
  \[|\mu _n(h, E, Z_-)-\mu _n(h, E, 0)|\leq Z_-,\] 
  where the behaviour of $\mu_n(h, E, 0)$ is given by Corollary \ref{thm-muZ-null}.
  \hfill $\Diamond$
\end{remark}

To obtain better estimates for the spectrum in orders of small $h$ we
use the $\epsilon$-quasimodes \cite{AK99, La93}. If $A$ is a
self-adjoint operator on $D(A)$ in a Hilbert space $\cH$, then for
$\epsilon>0$ one calls a pair
\[
\big(\widetilde\psi,\widetilde E\big) \in D(A)\times\bR,
\qmbox{with}
\big\|\widetilde\psi \big\|=1 \mbox{ and }
\big\| \big( A-\widetilde E \big) \widetilde\psi\big\| \leq \epsilon
\]
an \emph{$\epsilon$-quasimode} (so with this notation an eigenfunction
$\psi$ with eigenvalue $E$ is a $0$-quasimode).

The existence of an $\epsilon$-quasimode
$\big(\widetilde\psi,\widetilde E\big)$
implies that the distance between $\widetilde E$ and the spectrum of
$A$ fulfils
\beq
\dist \l(\sigma(A), \widetilde E \ri) \leq \epsilon.
\Leq{eq-distspec}
In particular there exists an eigenvalue $E$ of $A$ in the interval
$[\widetilde E -\varepsilon, \widetilde E + \varepsilon]$
if we know that in that interval the spectrum is discrete.

In our case we want to replace $A$ with an operator of the form
\beq
P_h :=- h^2 \frac {d^2}{dx^2} + V(x)
\Leq{eq-Phdefsinglemin}
with periodic boundary conditions on $L^2([-\pi,\pi])$
with $2\pi$-periodic $V\in C\big([-\pi,\pi],\bR^+\big)$, so that
\[
V(x)=\frac {x^2}4+ W(x)\qmbox{and}W(x) =\cO(x^{m_0})
\mbox{ for }m_0\in\bN\setminus \{1,2\}.
\]
Let $\chi\in C^2_0(\bR,[0,1])$ have support in $[-\pi,\pi]$ and
equal one on $[-\pi/2,\pi/2]$. We choose the positive constant
$c_n^h$ so that
\[
\psi_n^h\in L^2\big([-\pi,\pi]\big)\subseteq L^2(\bR)
\qmbox{,}
\psi_n^h(x):=c_n^h\, \chi(x)\, D^h_n(x)\, \exp\big(\!-x^2/(2h)\big)
\]
is of $L^2$ norm one.\\
It is a well-known fact that on $L^2(\bR)$ for
$\tilde P_h:=- h^2 \frac {d^2}{dx^2} + x^2/4$
\[
\tilde P_1 \tilde D^{1}_n = E^1_n \tilde D^1_n
\]
with $E^1_n := n + \frac 12$, $D^1_n$ the normalised Hermite
Polynomials
\beq
D^1_{n}(x) := \frac{(-1)^{n}}{n! \sqrt{2\pi}}\;
            e^{\frac{x^{2}}{4}}\frac{d^{n}}{dx^{n}}
            e^{-\frac{x^{2}}{2}},
\qquad n\in\bN_{0},
\Leq{def-fd}
and the Hermite functions
${\tilde D}^{1}_n(x):=D^{1}_n(x)e^{-\frac{x^{2}}{2}}$.
It thus follows from $L^2$ dilation that
\beq
\tilde P_h  \tilde D^{h}_n = E^h_n \tilde D^h_n
\qquad \mbox{ with } E^h_n := h E^1_n\mbox{ and }
\tilde D^h_n := h^{-\frac14} \tilde D^1_n
\l(h^{-\frac12} x\ri).
\Leq{eq-propPCF}
\begin{lemma}\label{lem-ohmhalfquasimode:new}
 $(\psi^h_n, E^h_n)\ (n\in\bN_0)$ are ${\cal O}(h^{m_0/2})$--quasimodes
 for $P_h$.
\end{lemma}
\proof
$\bullet$ For any polynomial $p\in \bC[x]$ the function $x\mapsto
p(x)\exp(-x^2/h)$ is of order $\cO\big(\exp(-x^2/(2h)\big)$ for
$h\searrow0$, uniformly in $|x|\in [\pi/2,\infty)$.  Thus
\[
\int_{\pi/2}^\infty |p(x)| \exp(-x^2/h)\,dx=\cO(h^\ell)
\qmbox{and}
\int_{-\infty}^{-\pi/2}\hspace*{-2mm}
|p(x)| \exp(-x^2/h)\,dx=\cO(h^\ell)
\quad(\ell\in\bN).
\]
$\bullet$
By compactness of the support of $\chi\in C^2_0(\bR,[0,1])$,
$\chi$, $\chi'$ and $\chi''$ are bounded.\\
$\bullet$ The first two remarks imply that
$\|\psi^h_n-\tilde D^h_n\|=\cO(h^\ell)\ (\ell\in\bN)$.
Since the scaled Hermite function
has norm $\|\tilde D^h_n\|=1$, the normalisation constant equals
$c_n^h=1+\cO(h^\ell)\ (\ell\in\bN)$.  More generally, regarding that
the derivatives of $\tilde D^h_n$ are of the form $x\mapsto
p(x)\exp(-x^2/h)$,
$\big\|\frac{d^r}{dx^r}(\psi^h_n-\tilde D^h_n)\big\|=\cO(h^\ell)
\quad (r,\ell\in\bN_0)$.\\
$\bullet$
So for the case $W=0$ in \eqref{eq-Phdefsinglemin},
$(\psi^h_n, E^h_n)$ are ${\cal O}(h^\ell)$-quasimodes for $P_h$ \
$(n,\ell\in\bN_0)$.\\
$\bullet$ We are thus left to prove that
$\|W\; \psi^h_n\| = {\cal O}\l(h^{\frac{m_0}2}\ri)$.
This, however, follows by a splitting of
the $L^2$ integral, regarding that $W(x)=\cO(x^{m_0})$ uniformly on
the interval $[-\pi/2,\pi/2]$, where $\psi^h_n= c_n^h\tilde D^h_n$,
and that $W$ is bounded on $[-\pi,\pi]$.
\hfill $\Box$

\begin{figure}[h]
\begin{center}
 \includegraphics[width=0.5\textwidth]{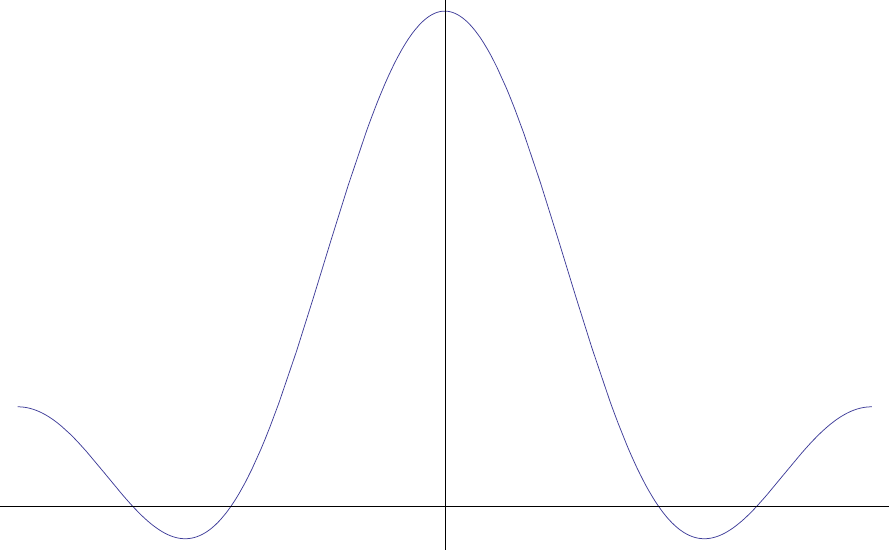}
 \caption{Shape of $Z_- \co(\eta) + E \co^2(\eta)$ in $[-\pi,\pi]$.}
 \label{fig-potshape}
\end{center}
\end{figure}

The potential $\eta\mapsto Z_- \co(\eta) + E \co^2(\eta)$ has in
general two non-degenerate minima at the points $\pm \eta_*$
  with
\[
\eta_* :=  \arccos\l( - \frac{Z_-}{2 E} \ri) \in [\pi/2, \pi],
\]
where the potential reaches
the value $-\frac{Z_-^2}{4 E}$ (see Figure \ref{fig-potshape}).

We construct our quasimodes to be concentrated near one of the
minima. Let the intervals $\Delta_o^+$ and $\Delta_i^+$ be two open
neighborhoods of the rightmost minima such that
$\overline{\Delta_i^+}\subset\Delta_o^+$ and $\Delta_o^+$ is contained
in the positive axis and is strictly separated from $0$. Fix $\chi_+
\in C_0^\infty(\bR)$ such that $\chi_+ = 1$ in $\Delta_i^+$ and
$\chi_+ = 0$ in $\bR\setminus\Delta_o^+$.

\begin{lemma}
  Let $P_h$ be as in (\ref{eq-Phdefsinglemin}) but with $V(x) := \frac
  {(x-x_*)^2}4 + W(x)$ and $W(x) := \sum_{m=m_0}^\infty a_m (x-x_*)^m$
  ($m_0 > 2$) entire of order $1$ and finite type. Define
\[
    \psi^h_n(x) := h^{-\frac14} D_n\l(h^{-\frac12}(x-x_*)\ri)\chi_+(x) =
D^h_n(x-x_*)\chi_+(x),
\]
where $\chi_+$ is the characteristic function defined in the previous
paragraph.  Then $(\psi^h_n(x),E_n^h)$ is an ${\cal
  O}\l(h^{3/2}\ri)$-quasimode for $P_h$.
\end{lemma}
\noindent
{\it Proof.}
Applying the operator to $\psi^h_n$ we have
\begin{align*}
 P_h \psi^h_n ={}& - h^2 {\psi^h_n}''+ \frac{(x-x_*)^2}4\psi^h_n
   + W\;\psi^h_n\\
  ={}& \l( -h^2 {D^h_n}'' + \frac{(x-x_*)^2}4 D^h_n+ W\; D^h_n\ri)\chi_+
      -h^2 \l( 2 h^{-\frac12} {D^h_n}'\chi_+' + D^h_n \chi''\ri)\\
  \overset{(\ref{eq-Phdefsinglemin})}{=}&
      E^h_n \psi^h_n + W\;\psi^h_n
      -h^2 \l( 2 h^{-\frac12} {D^h_n}'\chi_+' + D^h_n \chi''\ri).
\end{align*}

For what concerns $W\;\psi^h_n$ we can apply
Lemma \ref{lem-ohmhalfquasimode:new}, obtaining
\[
\|W\cdot \psi^h_n\| = {\cal O}\l(h^{\frac{m_0}2}\ri).
\]

We need now to take care of the last error term. For this last term
the inequality
\[
\l| h^2 \l( 2 h^{-\frac12} {D^h_n}'\chi_+' + D^h_n \chi''\ri) \ri|
\leq h c_1 e^{-\frac{c_2}h}
\]
holds with proper $c_1, c_2 > 0$ (that depend only on $n$ and
$\chi_+$).  Thus this term integrated on $[a,b]$ will give an error
that can be bounded with any polynomial order of decay, in particular
we can choose it to be
\[
\l\| h^2 \l( 2 h^{-\frac12} {D^h_n}'\chi_+' + D^h_n \chi''\ri)
\ri\| = {\cal O}\l(h^{\frac{m_0}2}\ri).
\hfill\tag*{$\Box$}
\]

We need now to transform our equation into something like $V(x)$ in
the previous theorem. We already know the two minima $\pm \eta_*$. If
we expand $V(\eta) := Z_- \co(\eta) + E \co^2(\eta)$ in the neighborhood
of those minima we obtain
\beq
  V(x) = - \frac{Z_-^2}{4E}
  + E \l( 1 - \frac{Z_-^2}{4 E^2} \ri)
    (\eta \pm \eta_*)^2 + W(\eta \pm \eta_*)
\Leq{eq-expvpm}
for a suitable entire $W$ with $m_0 = 3$ and of order
$1$ and finite type.

To simplify a bit the notation let us call
\[
A:= - \frac{Z_-^2}{4E} \;,
\qquad
B := \sqrt{ E \l( 1 - \frac{Z_-^2}{4 E^2} \ri) }.
\]

We focus for the moment only the localisation near the rightmost
minima, i.e.  we choose $(\eta - \eta_*)$. With the unitary
transformation $\cZ$ defined by change of variable
\[
 z(\eta) := \sqrt{2B}(\eta-\eta_*),
\]
the eigenvalue equation (\ref{eq-kappaetanonzero}) is transformed into
\beq
0 = K_{z}\psi(z) := 2B \l( -h^{2}\pa_\eta^2\psi(z) + \l(\widetilde\mu +
\frac{z^2}4 + \widetilde W(z) \ri)\psi(z) \ri),
\Leq{eq-kappaetatransformed}
where
$\widetilde\mu = \frac1{2B} ( - \mu + A)$
and $\widetilde W$ is entire with $m_0 = 3$ and of order $1$ and
finite type.
If in the spirit of the previous lemmas we define
  \[
    \widetilde \psi^h_{n}(z) := D^h_n(z)\chi(z),
    \qquad
    \widetilde\mu_n^h := A + 2B \l(n + \eh \ri) h,
  \]
where $\chi(z)$ is the transformed of the cut-off localised in the
neighborhood of $\eta_*$, then the couple
$(\widetilde\psi^h_{n}, \widetilde\mu_n^h)$ is an
${\cal O}(h^{3/2})$-quasimode for $K_{z}$ and thus if
\[
\psi^h_{n}(\eta) := \l( \cZ^{-1} \widetilde \psi^h_{n} \cZ \ri) (\eta),
\]
the couple $(\psi^h_{n\pm}, \widetilde\mu_n^h)$ defines an
${\cal O}(h^{3/2})$-quasimode for
$K_\eta$.

Exactly the same happens if we look near the other minimum, i.e. if we
choose $(\eta+\eta_*)$. In other words in the limit of $h\searrow 0$
the spectrum of $K_\eta$ consists of pairs $\mu_n^-(h)$, $\mu_n^+(h)$
with the same asymptotics $\widetilde\mu_n^h$ in the limit.  We have
proved the following.

\begin{theorem}\label{thm:low-lying-app}
Let $E > \frac{Z_-}2 >0$. Define
\beq
\widetilde\mu_n^h := -\frac{Z_-^2}{4E} +
                        \sqrt{ E \l( 1-\frac{Z_-^2}{4 E^2}\ri) }
                        \l(2n + 1 \ri) h.
\Leq{eq-bigmApp2d}
There exists an eigenvalue $\mu_n^h$ of $K_\eta$ and a constant $c$
such that $\l| \widetilde\mu_n^h - \mu_n^h \ri| = \cO(h^{3/2})$.
Moreover, the interval $\l[\widetilde\mu_n^h-2 c h^{3/2},
\;\widetilde\mu_n^h + 2 c h^{3/2} \ri]$ contains at least two
eigenvalues of $K_\eta$.
\end{theorem}

\begin{remark}
  It can be proved by standard methods involving the IMS formula
  \cite[Chapter 3.1]{CFKS} and Agmon estimates \cite{Ag82} that the
  distance between the eigenvalues in each pair is of the order
  $\exp(-C/h)$ with $C\in(0,\infty)$.
\hfill$\Diamond$\end{remark}

We can use this result in combination with
(\ref{eq-Andef}).

\begin{proposition} The resonances in the set $\cR_n\cap\{\Re E >\frac{Z_-}2 >0\}$ (see \eqref{eq:defRes2d}) are given asymptotically as $h\to 0$ by the solutions of the following equation
\beq
A_{n}(h,E,Z_{+},\mu^{+}_{m}(h,E))=0.
\Leq{eq-rescoupling}
\end{proposition}

Neglecting the error terms, the resonances for $\Re E >\frac{Z_-}2 >0$ are given by the solutions of
\[
-E  - Z_{+} -\textstyle{\frac{Z_-^2}{4E} +
			\sqrt{ E -\frac{Z_-^2}{4 E} }
			\l(2m + 1 \ri)} h
+ i h (2n+1) \sqrt{ \textstyle{\sqrt{ E -\frac{Z_-^2}{4 E} }
			\l(2m + 1 \ri) h
			-\frac{Z_-^2}{4E} - \frac{Z_+}{2} }}
 = 0.
\]

\begin{remark}
  For $Z_- = 0$ we recover \eqref{eq-Enmvalue3} of the previous
  section. On the other hand, in Section \ref{sec-app2dz-0} the
  approximation error is of order ${\cal O}(h^2)$ instead of ${\cal
    O}(h^{3/2})$.
\hfill$\Diamond$\end{remark}

For $0 < E < \frac{Z_{-}}2$ the bottom of the potential is reached at
$\pi$ and thus we have to expand the potential around this other
point. It turns out that in this case the eigenvalues are approximated
by
\beq
\widehat\mu_n^h := E - Z_- +
			\sqrt{ \textstyle{\frac{Z_-}{2}} - E }
			\l(2n + 1 \ri) h.
\Leq{eq-smallmApp2d}

\begin{proposition} 
	The resonances in the set $\cR_n\cap\{0 < \Re E < \frac{Z_{-}}2\}$ (see \eqref{eq:defRes2d}) are given asymptotically as $h\to 0$ by the solutions of the following equation
\[
A_{n}(h,E,Z_{+},\widehat\mu^{+}_{m}(h,E))=0.
\]
\end{proposition}

\begin{remark}
  This approach gives good results if we stay localised near the
  bottom of the potential: in this case we can find an approximation
  for the eigenvalue up to an order of any integer power of $h$.

  The deficiency of this approach lies in the fact that we have no
  control on the relative error between $n$ and $h$. We need therefore
  to find a different approximation scheme that keeps track of the
  mutual relation between the parameters.
\hfill$\Diamond$\end{remark}

%
\subsection{High energy estimates}\label{sec:hee}
%
We consider the potential in the form
$V(x) = E \cos^{2}(x) + Z_{-} \cos(x)$. Substituting this value in the
formulae given in Theorem \ref{thm-hmu2} we have
\[
\int_{-\pi}^{\pi} V(x) \,dx = E\pi \qmbox{and}
\int_{-\pi}^{\pi} V^{2}(x)  \,dx = \frac{3 E^{2} \pi}{4} + \pi Z_{-}^{2}
\]
and thus the eigenvalues $\mu_{2m+1}$ and $\mu_{2m+2}$ can be
represented as
\beq
\sqrt{\mu} =
(m+1)h + \frac{E}{4(m+1)h} + \frac{Z_{-}^{2}
  - \frac{E^{2}}{4}}{16 (m+1)^{3}h^{3}}
  + {\cal O}\l( \frac1{m^{5} h^{5}}\ri) + o\l(\frac {1}{m^{3}h}\ri).
\Leq{eq-munewlastapprox}
Therefore we can estimate $\mu_{2m+1}$ and $\mu_{2m+2}$ with
\beq
\mu =  (m+1)^2 h^2 + \frac E2 + \l( Z_-^2 + \frac{E^2}4 \ri)
\frac1{8(m+1)^2h^2}  + {\cal O}\l( \frac1{m^{4} h^{4}}\ri)
+ o\l(\frac1{m^{2}}\ri).
\Leq{eq-muestres}

With this result, we can compute the resonances $E_{n, 2m+1}$ and $E_{n, 2m+2}$.

\begin{proposition} The resonances in the set $\cR_n$ (see \eqref{eq:defRes2d}) are given by the solutions of the following equation
\beq
A_{n}(h,E,Z_{+},\mu_{2m+1}(h,E))=0,
\Leq{eq:AppResAft}
asymptotically as $h\to 0$ and $m\to\infty$ with $mh$ large.
\end{proposition}

More explicitly, for fixed $n$ and up to errors of orders
\[
 h^{\frac32},\qquad
 (mh)^{-4}
 \qmbox{and}
 m^{-2},
\]
we can approximate the resonant energies as solutions of
\[
-\frac{E}2 - Z_+ + (m+1)^2 h^2
  + \frac{ Z_-^2 + \frac{E^2}4 }{8(m+1)^2h^2}
  + i(2n+1)h\sqrt{\textstyle{(m+1)^2 h^2
    + \frac {E-Z_+}2 + \frac{Z_-^2 + \frac{E^2}4}{8(m+1)^2h^2}}
  } = 0.
\]

\begin{remark}
  We cannot hide the term $(m+1)^2 h^2$ inside the error term of order
  $h^{3/2}$ because we want to analyze the asymptotic behaviour for $m
  \geq C/h$ ($C\in (0,\infty)$) and that term is rather big compared with
  $h$.  \hfill$\Diamond$\end{remark}

%
\section{Numerical investigations}\label{sec:numerical}
%

In the previous sections we have explicitly written three implicit
equations to approximate the value of the resonances in terms of the
atomic numbers $n$ and $m$ (and of course of the parameters $h$,
$Z_{+}$ and $Z_{-}$).
In this section we investigate the qualitative structure of
the resonances using the approximations given by
\eqref{eq-rescoupling} and \eqref{eq:AppResAft}.

In view of Remarks \ref{rmk:app1} and \ref{rmk:app2} we know that at
least for certain values of the charges $Z_i$ we are not describing
all the resonances of the system.  On the other hand the additional
resonances should appear only for small $\Re(E)$. Therefore we are going to
consider $\Re(E)$ big enough to be sure that we are analysing an
energy region in which all the resonances should be generated by the
classical closed hyperbolic trajectory between the centers.

In this case equation \eqref{eq:bigEAppBigmu} implies that
$\Re(\mu_m)$ must be big and thus it is evident from
\eqref{eq-bigmApp2d}, \eqref{eq-smallmApp2d} and \eqref{eq-muestres}
that $m$ must be big.  The quasimode approximation obtained in Section
\ref{sec-app2dz-0} and \ref{sec-app2dz-n0} is valid only for
small values of $m$ and $h$, therefore these resonances are
automatically excluded from the analysis.

\begin{figure}[h!]
\subfigure[\small Case $Z_+ = 2$.]{
	\makebox[\textwidth][c]{\includegraphics[width=1.\textwidth]{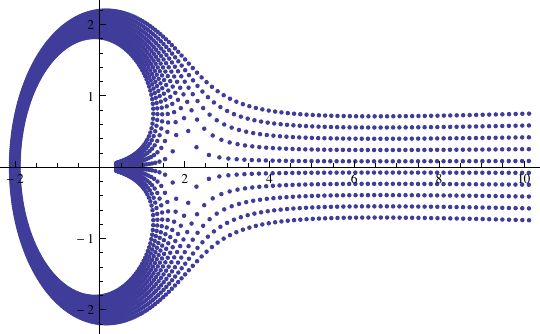}}
	\label{fig:LlSolPos}
}

\subfigure[\small Case $Z_+ = -2$.]{
	\makebox[\textwidth][c]{\includegraphics[width=1.\textwidth]{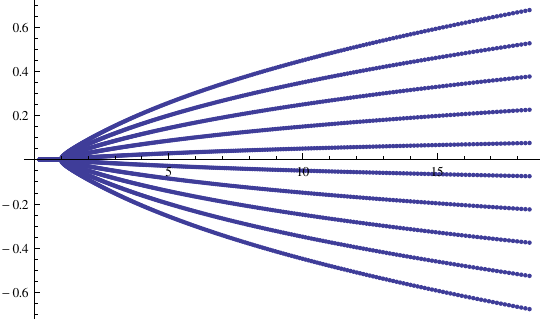}}
	\label{fig:LlSolNeg}
}
\caption{\small Solutions of \eqref{eq-rescoupling} with $Z_- = 0$,
  $h=0.01$, $n=0,\ldots,4$, $m=1, \dots, 250$.}
\label{fig:llSol}
\end{figure}

\begin{figure}[h!]
\subfigure[\small Resonances in $L_+$.]{
	\def\svgwidth{1.\linewidth}
	\makebox[\textwidth][c]{
	\begingroup
	  \makeatletter
	  \providecommand\color[2][]{%
	    \errmessage{(Inkscape) Color is used for the text in Inkscape, but the package 'color.sty' is not loaded}
	    \renewcommand\color[2][]{}%
	  }
	  \providecommand\transparent[1]{%
	    \errmessage{(Inkscape) Transparency is used (non-zero) for the text in Inkscape, but the package 'transparent.sty' is not loaded}
	    \renewcommand\transparent[1]{}%
	  }
	  \providecommand\rotatebox[2]{#2}
	  \ifx\svgwidth\undefined
	    \setlength{\unitlength}{600pt}
	  \else
	    \setlength{\unitlength}{\svgwidth}
	  \fi
	  \global\let\svgwidth\undefined
	  \makeatother
	  \begin{picture}(1,0.61803333)%
	    \put(0,0){\includegraphics[width=\unitlength]{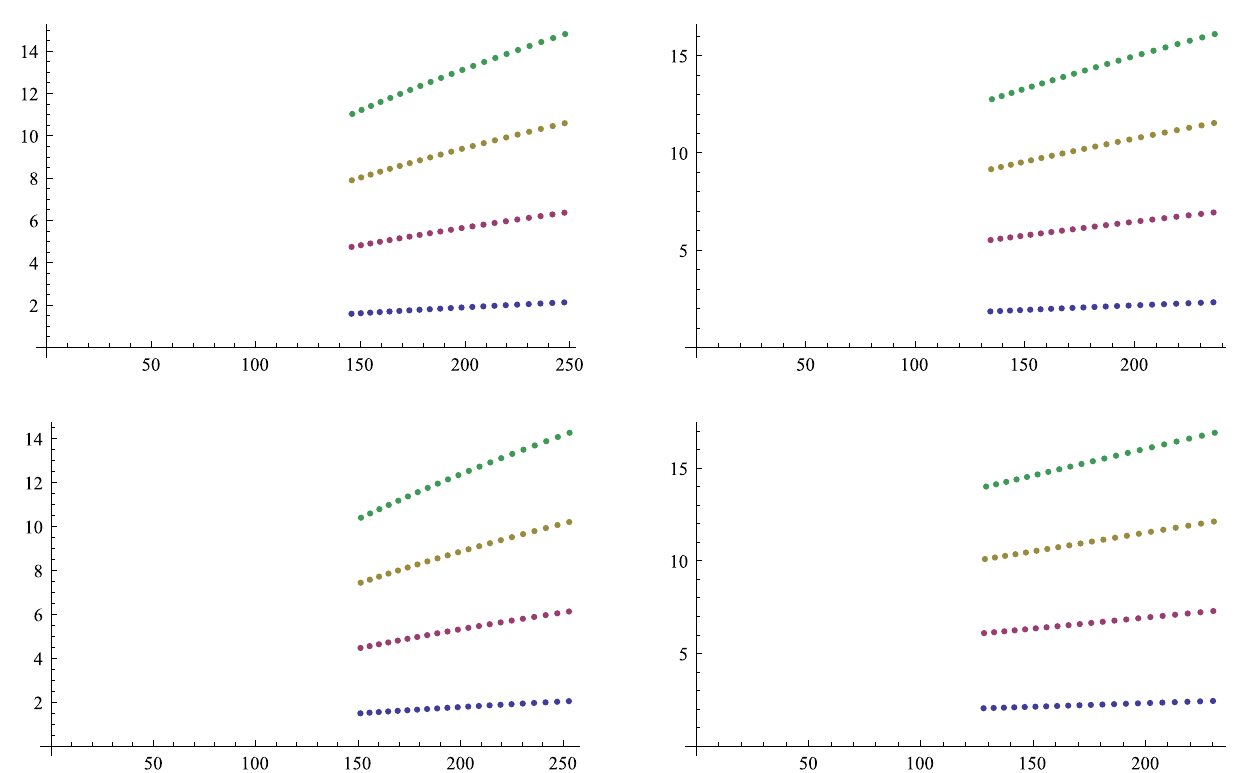}}%
	    \put(0.13693143,0.59801836){\color[rgb]{0,0,0}\makebox(0,0)[lb]{\smash{$(Z_+,Z_-) = (2,4)$}}}%
	    \put(0.66972819,0.59801836){\color[rgb]{0,0,0}\makebox(0,0)[lb]{\smash{$(Z_+,Z_-) = (-2,4)$}}}%
	    \put(0.13693143,0.28068495){\color[rgb]{0,0,0}\makebox(0,0)[lb]{\smash{$(Z_+,Z_-) = (4,2)$}}}%
	    \put(0.66972819,0.28068495){\color[rgb]{0,0,0}\makebox(0,0)[lb]{\smash{$(Z_+,Z_-) = (-4,2)$}}}%
	  \end{picture}%
	\endgroup
	}
	\label{fig:HESolBig}
}

\subfigure[\small Resonances in $S_+$.]{
	\def\svgwidth{1.\linewidth}
	\makebox[\textwidth][c]{
	\begingroup
	  \makeatletter
	  \providecommand\color[2][]{%
	    \errmessage{(Inkscape) Color is used for the text in Inkscape, but the package 'color.sty' is not loaded}
	    \renewcommand\color[2][]{}%
	  }
	  \providecommand\transparent[1]{%
	    \errmessage{(Inkscape) Transparency is used (non-zero) for the text in Inkscape, but the package 'transparent.sty' is not loaded}
	    \renewcommand\transparent[1]{}%
	  }
	  \providecommand\rotatebox[2]{#2}
	  \ifx\svgwidth\undefined
	    \setlength{\unitlength}{600pt}
	  \else
	    \setlength{\unitlength}{\svgwidth}
	  \fi
	  \global\let\svgwidth\undefined
	  \makeatother
	  \begin{picture}(1,0.61803333)%
	    \put(0,0){\includegraphics[width=\unitlength]{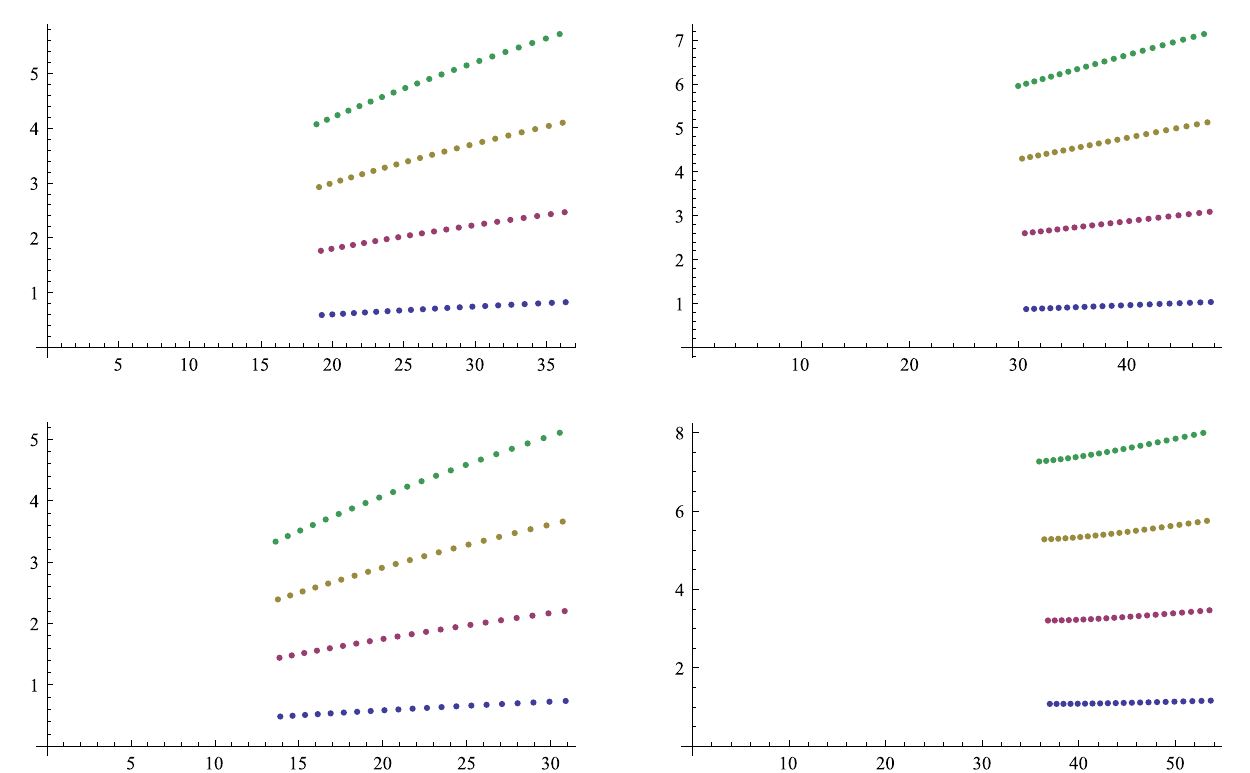}}%
	    \put(0.13211418,0.59775745){\color[rgb]{0,0,0}\makebox(0,0)[lb]{\smash{$(Z_+,Z_-)=(2,4)$}}}%
	    \put(0.64824254,0.59775745){\color[rgb]{0,0,0}\makebox(0,0)[lb]{\smash{$(Z_+,Z_-)=(-2,4)$}}}%
	    \put(0.13211418,0.2830908){\color[rgb]{0,0,0}\makebox(0,0)[lb]{\smash{$(Z_+,Z_-)=(4,2)$}}}%
	    \put(0.64824254,0.2830908){\color[rgb]{0,0,0}\makebox(0,0)[lb]{\smash{$(Z_+,Z_-)=(-4,2)$}}}%
	  \end{picture}%
	\endgroup
	}
	\label{fig:HESolSma}
}
\caption{\small Resonances for $h = 0.05$ and $C = 10$.}
\label{fig:HESol}
\end{figure}

Figure \ref{fig:llSol}\subref{fig:LlSolPos} and
\ref{fig:llSol}\subref{fig:LlSolNeg}
show all the approximated resonances obtained from
\eqref{eq-rescoupling} setting $Z_- = 0$. We plotted all the values
including the one in
regions of energies where we have no control on the error. In these
pictures we can observe an interesting behaviour. In particular
for big values of $m$ we recover the structure shown by the resonances
approximated with \eqref{eq:AppResAft}: see Figure
\ref{fig:HESol}\subref{fig:HESolBig} and Figure
\ref{fig:HESol}\subref{fig:HESolSma}.

The physically interesting resonances are the ones close to the
real axis, this because they can be measured in experiments. Thus to
keep $\Im(E)$ as small as possible we will consider small values of
$n$ (see \eqref{eq:bigEAppBigmu}).

\begin{remark}
Unless differently specified, in the plots we consider $n=0,1,2,3$ and 
$m \in\l\{ \lceil C/h\rceil + k \mid k = 0,1,2,\ldots,20\ri\}$. 
The values of $Z_+$, $Z_-$, $h$ and $C$ will be specified in the 
title or in the caption of the plots.
For practical reasons we plot the resonances in the plane $(\Re(E), -\Im(E))$.
\hfill$\Diamond$\end{remark}

Equation \eqref{eq:AppResAft} has two couples of solutions $(S_+,
L_+)$ and $(S_-, L_-)$, specular w.r.t. the real axis. They correspond
respectively to the resonances and the anti-resonances, i.e. the
resonances defined inverting the roles of the incoming and outgoing
waves $v_\pm$ in the construction of Section \ref{sec:scatt2d}.

We restrict our analysis to the resonances $(S_+, L_+)$.  The two sets
$S_+, L_+\in\bC_-$ characterise two different energy regions, this
meaning that the resonances in $S_+$ have relatively small real part
if compared to the resonances in $L_+$ (see Figure
\ref{fig:HESol}\subref{fig:HESolBig} and Figure
\ref{fig:HESol}\subref{fig:HESolSma}).

The structure that we find is extremely regular.  The first question
that arises is if we are really computing the resonances associated
with energy values on the critical line $\cL_+^2$, associated to the
hyperbolic closed orbits described in \cite{KK,S14}
and summarised in Section~\ref{sec:classical}.

\begin{figure}[h!]
\makebox[\textwidth][c]{
	\setcounter{subfigure}{0}
	\subfigure[\small $(Z_+, Z_-) = (2,4)$, $C=4$, $m=400,\ldots,430$.]{
		\includegraphics[width=0.55\linewidth]{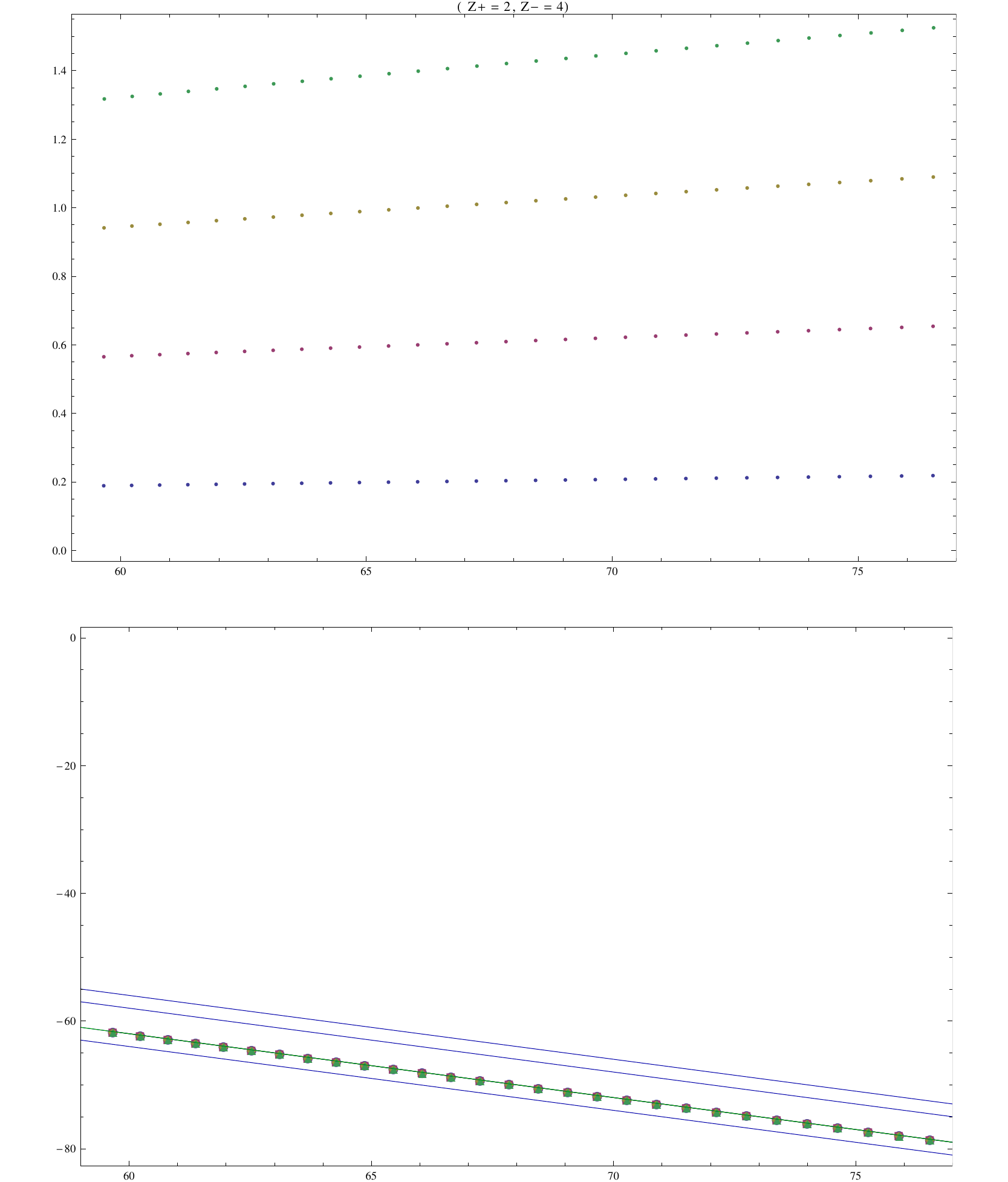}
	}
	\subfigure[\small $(Z_+, Z_-) = (-2,4)$, $C=7$, $m=700,\ldots,730$.]{
		\includegraphics[width=0.55\linewidth]{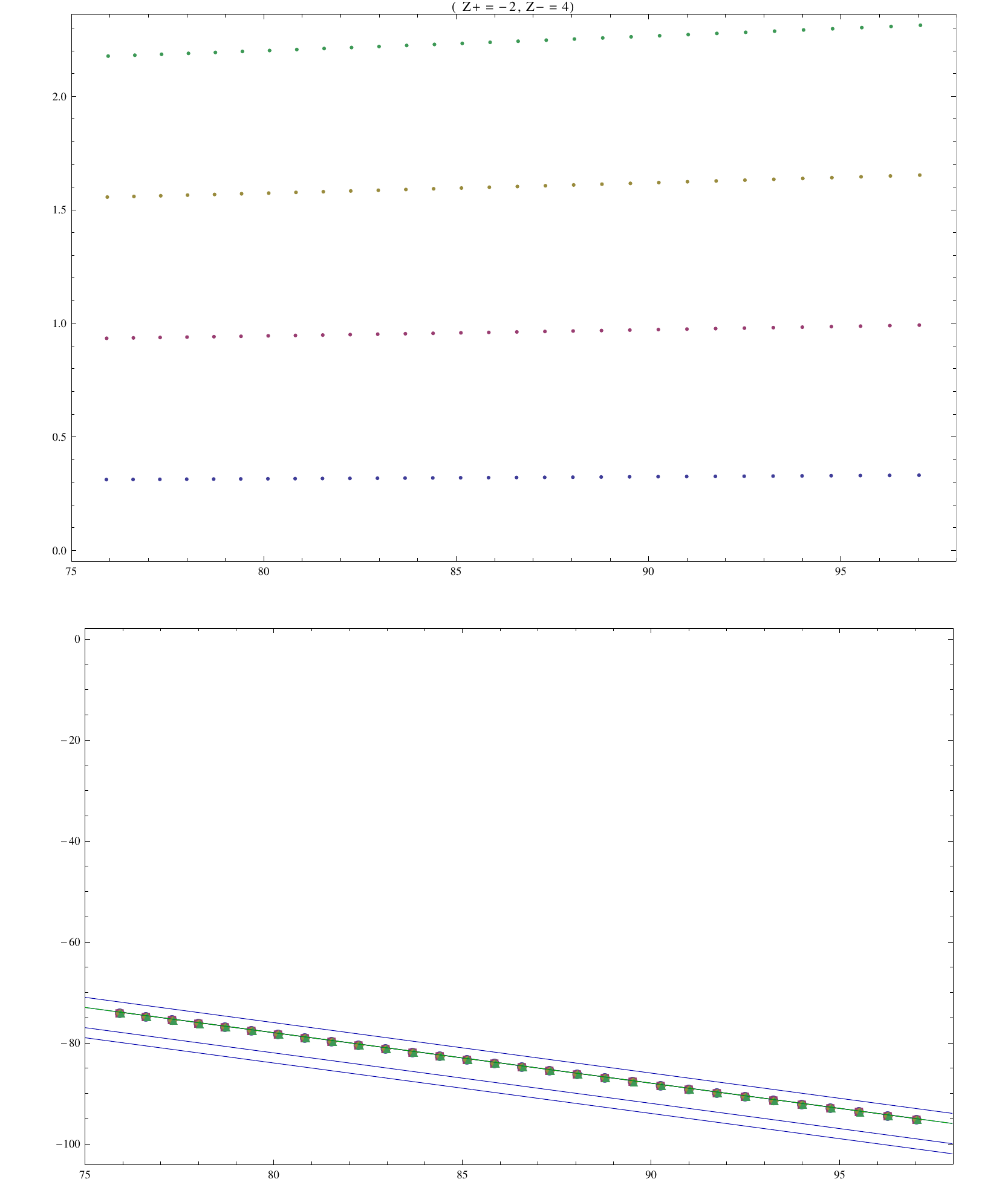}
	}
}\vspace{0.2cm}

\makebox[\textwidth][c]{
	\subfigure[\small $(Z_+, Z_-) = (4,2)$, $C=4$, $m=400,\ldots,430$.]{
		\includegraphics[width=0.55\linewidth]{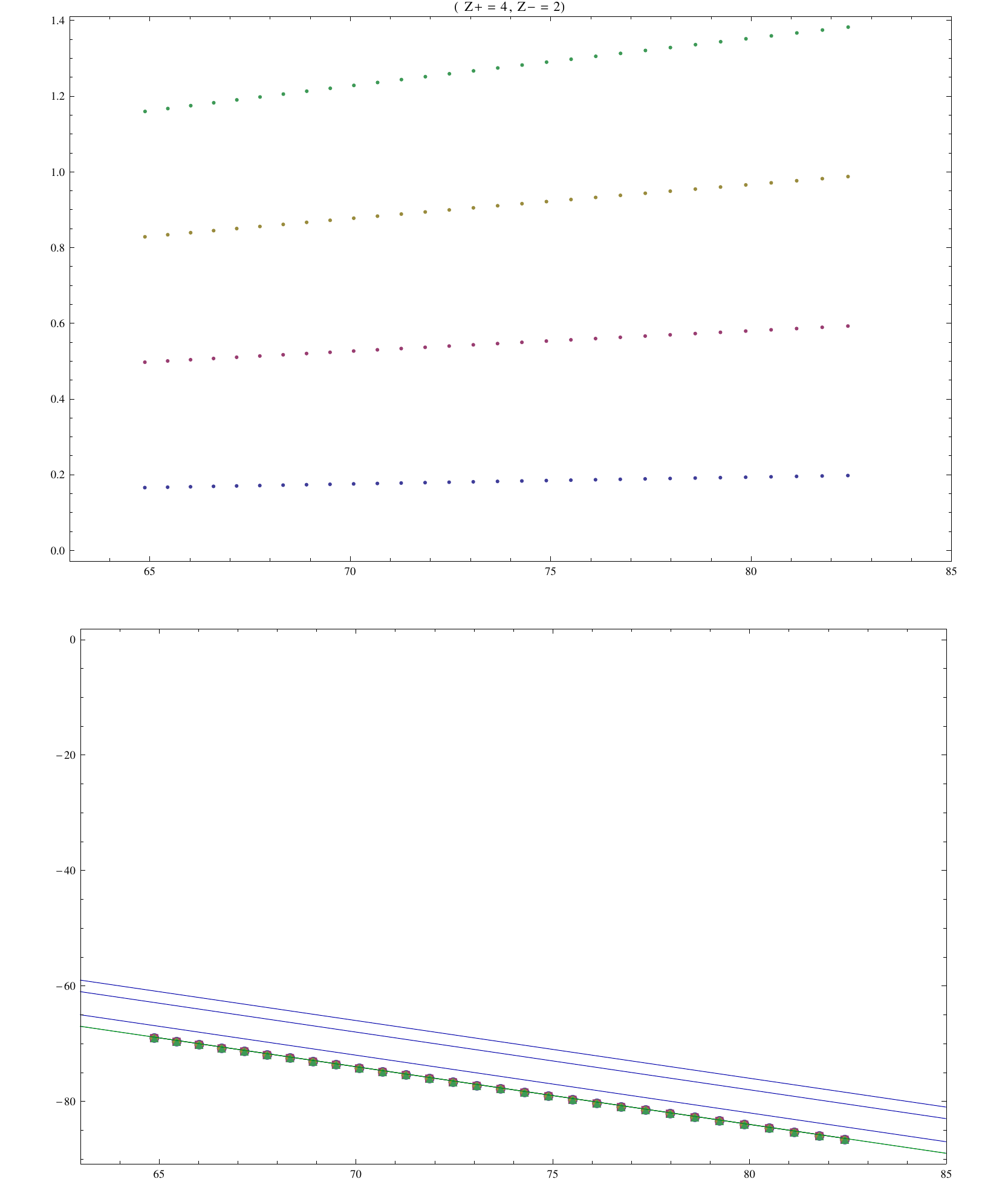}
	}
	\subfigure[\small $(Z_+, Z_-) = (-4,2)$, $C=7$, $m=700,\ldots,730$.]{
		\includegraphics[width=0.55\linewidth]{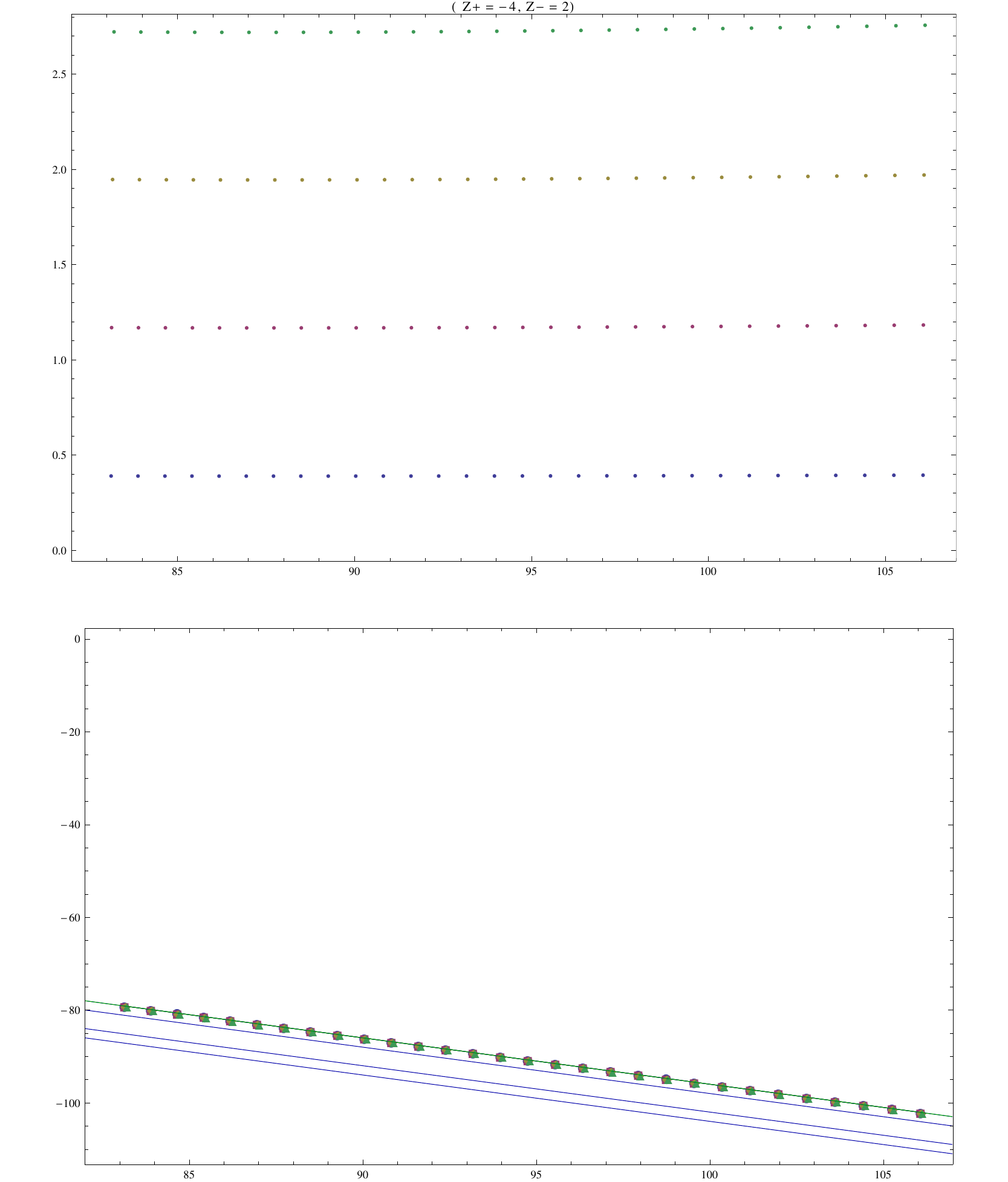}
	}
}
\caption{\small Comparison of the resonances in $L_+$ for $h = 0.001$
  (plot above) and their projection on the bifurcation diagram (plot
  below).}
\label{fig:CompResBD2d}
\end{figure}

For each computed resonance $E_{n,m}$ we can use the approximation
obtained in \eqref{eq-muestres} to estimate the associated constant of
motion $K_{n,m}$. We can thus superimpose the points $(\Re(E),
\Re(K))$ to the bifurcation diagram and visualize how they are
related.  As shown in Figure \ref{fig:CompResBD2d}, the energy
parameters appear to lay exactly upon $\cL_+^2$, giving a strong hint
on the correctness of the result.

A related question
  regards the order of growth of the resonances $E_{n,m}$ in $n$ and
  $m$.  For large energies there is only one bounded trajectory, which
  is closed and hyperbolic.  In the corresponding case for
  pseudo-differential operators the real respectively imaginary parts
  of the resonances in the complex plane are known to be related to
  the action resp.\ Lyapunov spectrum of the the closed trajectory
  (see \cite{gab} for the physics perspective and \cite{GeSj} for a
  mathematical
  proof). \\
  For a two-centers system it is known that the Lyapunov exponent of
  the bounded orbit of energy $E$ diverges like $\ell(E) = \sqrt{E}
  \ln(E)$ (see \cite[Proposition 5.6]{KK}).  As these closed
  trajectories collide with the two centers, where the Coulombic potential
  diverges, these results are not applicable.
  However it is reasonable to normalize the real and imaginary part
of the resonances in $L_+$ (or $S_+$) dividing them by
$\ell(\Re(E))$. In this way it is possible to investigate, at least
qualitatively, the above prediction.

\begin{figure}[h!]
\subfigure[\small Plot for for $E\in L_+$.]{
	\makebox[\textwidth][c]{\includegraphics[width=1.\linewidth]{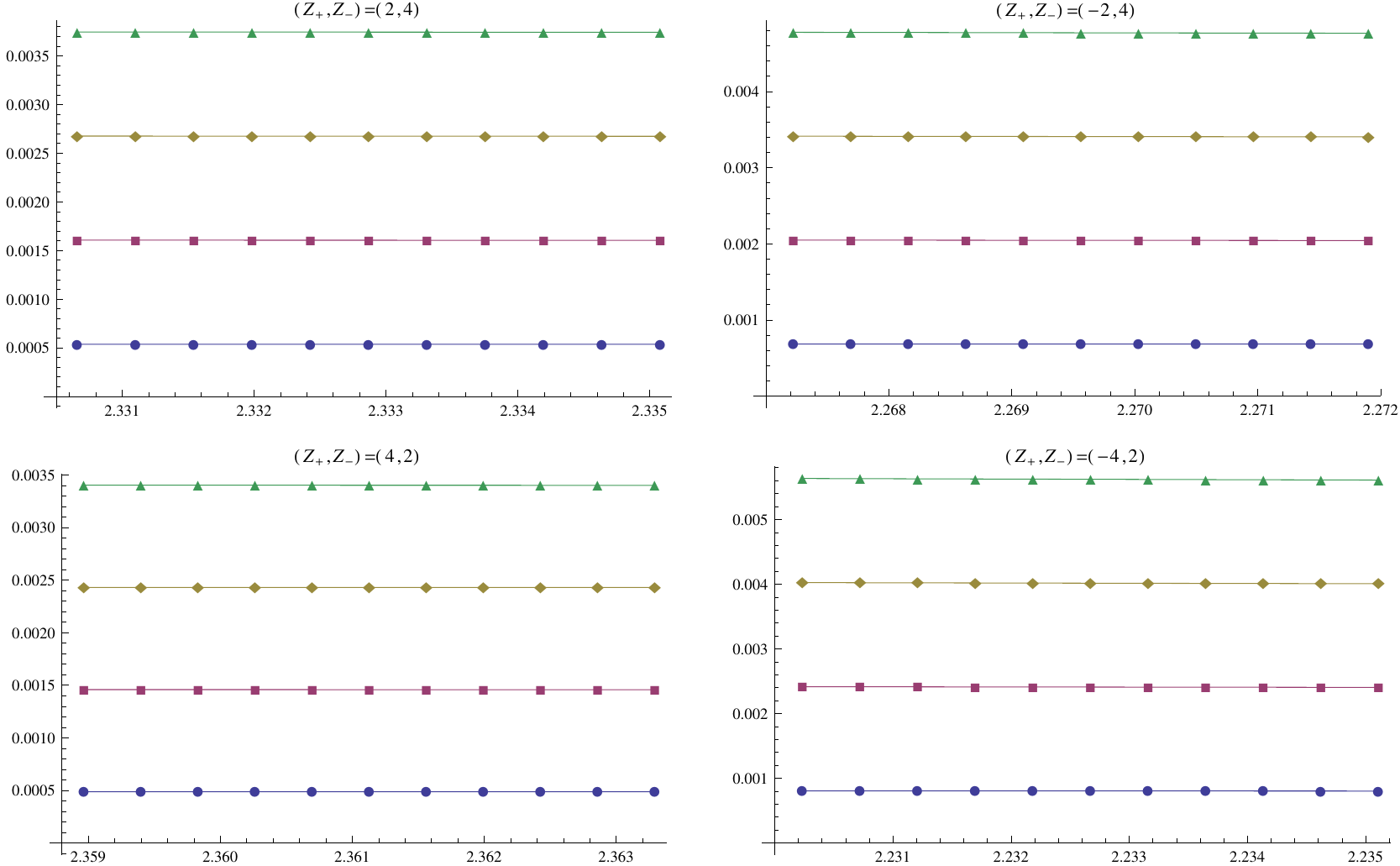}}
	\label{fig:HERESolBig}
}\vspace{0.5cm}

\subfigure[\small Plot for $E\in S_+$.]{
	\makebox[\textwidth][c]{\includegraphics[width=1.\linewidth]{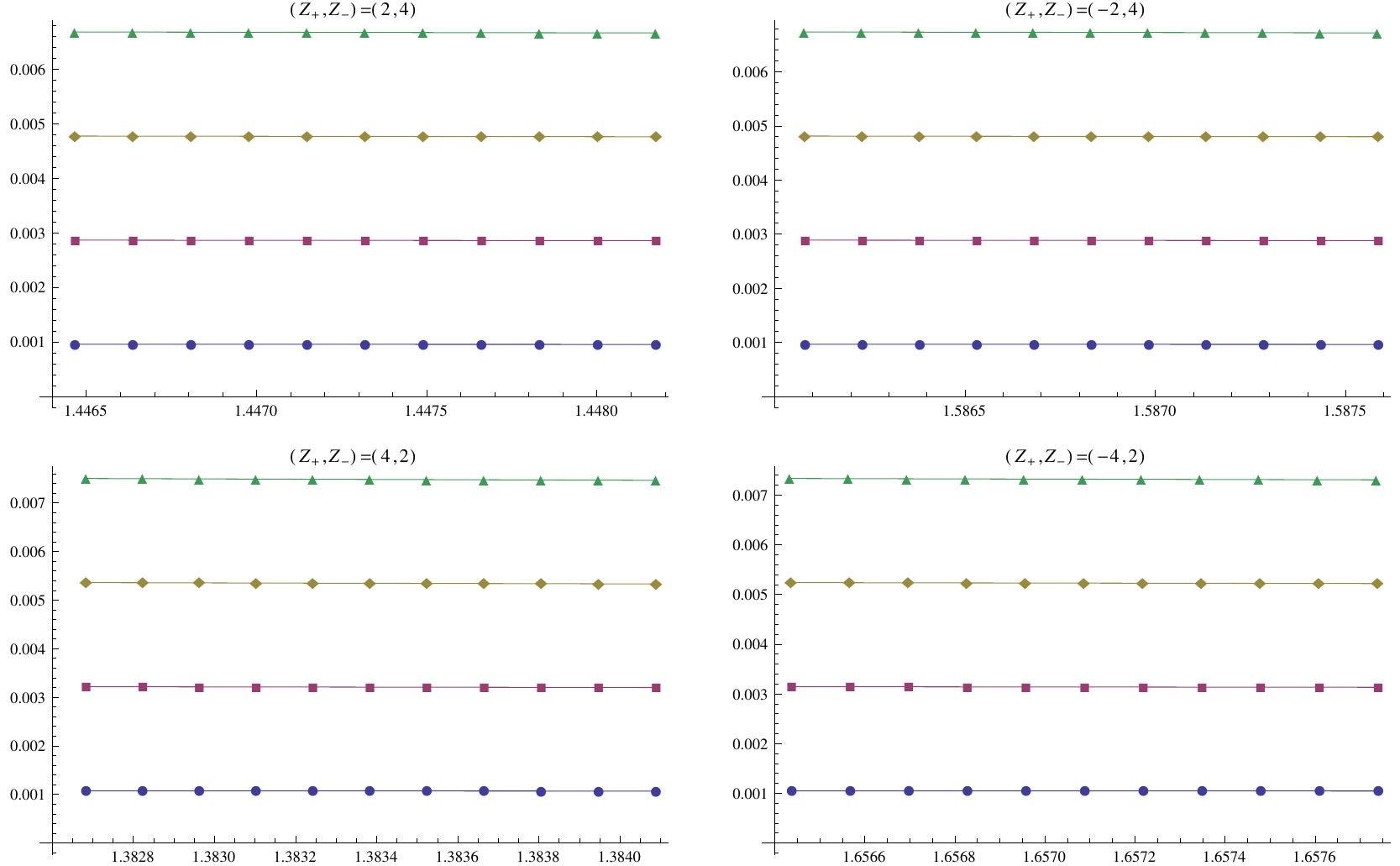}}
	\label{fig:HERESolSma}
}
\caption{\small Renormalised resonances $E/\ell(\Re(E))$ with parameters $h = 0.001$, $C = 9$ and $m = 9000, \ldots, 9010$.}
\label{fig:HERESol}
\end{figure}

The numerics confirm the expected behaviour. It is evident from Figure
\ref{fig:HERESol}\subref{fig:HERESolBig} and
\ref{fig:HERESol}\subref{fig:HERESolSma} that the renormalised
resonances look like distributed on a regular lattice of points with
(almost perfectly) aligned and equispaced real and imaginary parts.

Notice moreover that the vertical spacing of the imaginary parts is $d
= {\cal O}(h)$ and the distance between the real axis and the
resonances with smaller imaginary part is approximately $d/2$, as
expected from the harmonic oscillator perturbation used to approximate
the resonances.

\section{The two-center problem in 3D and the $n$-center problem}
\label{conc}

In \cite[Chapters 3 and 5]{thesis} it is shown that the three-dimensional
two-centers system is not essentially different from the planar
one. In particular all the results obtained for the planar problem
and presented in this paper can be carried almost identical.

However two major difficulties arises. There is a non-trivial
effect coming from the angular momentum that makes the resonances
set more complex and potentially more degenerate. And the numerical
approximations that we get in the planar setting fail to hold
due to the presence of singularities produced by the angular
momentum.

Another important related problem is the study of resonances for
the $n$-centers system.
The classical model for $n\geq 3$ still presents hyperbolic bounded
trajectories \cite{KK, Kn}.
In this case however they form a Cantor set in the phase space.
Moreover the non-trapping condition fails to hold,
thus in the quantum case one expects the resonances to be present and to be
distributed in some complicated way.  There are only few known
examples presenting a similar structure that have been investigated
rigorously (see \cite{FractalNoZw} and \cite{FractalSjZw}).  They
suggests that the resonances are present and
their density near the real energy axis scales with a fractal power of $h$.
The results obtained in this paper strongly support the idea that the
resonances should be present and be strictly related with the underlying
classical hyperbolic structure.

Anyhow for $n\geq 3$, mainly due to the lack of separability, the
singularities of the potential have to be treated by semiclassical
techniques, as in \cite{CJK}.
A lot of work and new ideas are required to properly extend results like
\cite{FractalNoZw,FractalSjZw} to the $n$-center problem.

\section*{Acknowledgements}
The authors are grateful to Hermann
Schulz-Baldes for the interesting and useful discussions.
We thank Paul Abbott for the interesting references and the anonymous referee for the detailed suggestions.

We acknowledge partial support by the FIRB-project RBFR08UH60 (MIUR, Italy).
M. Seri was partially supported by the EPSRC grant EP/J016829/1.

\appendix
\numberwithin{equation}{section}

%
\section{Generalised Pr\"ufer transformation in the semi-classical
  limit}\label{appendix}
%
The method for establishing estimates is based on a modification of
the Pr\"ufer variables described in \cite[Chapter 4.1]{Ea}. Consider a
Sturm-Liouville differential equation on $[x_{1},x_{2}]$ of the form
\beq
  (C(x)y'(x))' + D(x)y(x) = 0
\Leq{eq-slea}
in which $C$ and $D$ are real-valued, not necessarily periodic,
differentiable and with piecewise continuous derivatives.  Suppose
also that $C(x)$ and $D(x)$ are positive and define $R(x) :=
\sqrt{C(x)D(x)}$. If $y$ is a non-trivial real-valued solution of
(\ref{eq-slea}), we can write
\beq
  R(x)y(x) = \rho(x) \sin(\theta(x)),
  \quad
  C(x)y'(x) = \rho(x) \cos(\theta(x)),
\Leq{eq-mpea} where
\[
\rho(x) := \sqrt{R^{2}(x)y^{2}(x) + C^{2}(x)y'^{2}(x)}\qmbox{,}
\theta(x) := \arctan\l( \frac{R(x)y(x)}{C(x)y'(x)}\ri).
\]
Up to now $\theta(x)$ is defined as a continuous function of $x$ only
up to a multiple of $2\pi$.  To solve this problem we select a point
$a_{0}\in[x_{1},x_{2}]$ and we stipulate that
$-\pi\leq\theta(a_{0})<\pi$.  Moreover, if $y(a_{0})\geq0$, we have by
(\ref{eq-slea}) that
\beq
  0 \leq \theta(a_{0}) < \pi.
\Leq{eq-ta0ea}
\begin{lemma}\label{lem-tpropea}
  With the above definitions
\beq
\theta'(x) = \l( \frac{D(x)}{C(x)} \ri)^{1/2}
  + \frac 14 \frac{\l(C(x)D(x)\ri)'}{C(x)D(x)} \sin(2\theta(x)).
\Leq{eq-t'ea}
Let $a_{1}\in(a_{0},x_{2}]$.
If $y(x)$ has $N$ zeroes in $(a_{0}, a_{1}]$ and $y(a_{0})\geq0$, then
\beq
  N\pi \leq \theta(a_{1}) < (N+1)\pi.
\Leq{eq-tboundea}

\end{lemma}
\proof
The theorem is proved in \cite[Chapter 4.1]{Ea}.
\qed

We want to apply (\ref{eq-mpea}) to equation
(\ref{eq-kappaetanonzero}).  In particular we
apply the transform to
\beq
  h^{2}\big(p(x)y'(x)\big)' + \big(\mu - V_{1}(x)\big) y(x) = 0
\Leq{eq-hsleq}
where $p$ and $V_{1}$
have period $2\pi$. Since we are concerned with the limit
$\mu\rightarrow\infty$ (parametrically depending on $h$), we can
consider $\mu$ large enough to have $\mu - V_{1}(x) > 0$ in
$[-\pi,\pi]$.  In the new case (\ref{eq-hsleq}) the two functions
$\theta$ and $\rho$ depend on $\mu$ and $h$ as well as $x$, and we
write
$\theta_{h}(x,\mu)$.
Then (\ref{eq-t'ea}) becomes
\beq
\theta'_{h}(x,\mu) =  \frac1{h} \sqrt{ \frac{\mu-V_{1}(x)}{p(x)} }
  +\frac14 \frac{\mu\, p'(x) - (p(x) V_{1}(x))'}{(\mu - V_{1}(x)) p(x)}
  \sin(2 \theta_{h}(x,\mu) ).
\Leq{eq-t'hmu}

A first consequence of \eqref{eq-t'hmu} is that as
$\mu\rightarrow\infty$
\beq
  \theta'_{h}(x,\mu) = \frac{\mu^{\frac12}}{h}
    \sqrt{ \frac{1-\widetilde{V}_{1}(x)}{p(x)} } + {\cal O}(1),
\Leq{eq-t'firstcons}
where $\widetilde{V}_{1}(x) := V_{1}(x)/\mu$.
Moreover, if $y(x)$ has period $2\pi$ we have
\beq
\theta_{h}(\pi,\mu) - \theta_{h}(-\pi,\mu) = 2 k \pi
\Leq{eq-typerea}
for an integer $k$.

\begin{lemma}\label{lemma-asymptsin2t}
  For $f\in L^1([-\pi,\pi])$ and $c\in\bR\setminus \{0\}$ let
$\theta_{h}(x,\mu)$ satisfy \eqref{eq-t'hmu}. Then
\[
\int_{-\pi}^{\pi} f(x) \sin\l( c\, \theta_{h}(x,\mu) \ri) dx
  \longrightarrow 0
\]
as $\mu\rightarrow\infty$ (and/or $h\!\!\searrow\!\!0$). The same
result holds with $\sin\l( c\, \theta_{h}(x,\mu) \ri)$ replaced by
$\cos\l( c\, \theta_{h}(x,\mu) \ri)$.
\end{lemma}
\proof
To keep the equations compact we drop the $\mu$ dependence of
$\theta_h(x,\mu)$ in the rest of the proof. Fix any $\epsilon > 0$.
Let $g:[-\pi,\pi]\to\bR$ be a continuously differentiable function
such that
\[
\int_{-\pi}^{\pi} \l| f(x) - g(x) \ri| dx < \epsilon.
\]
Then
\beq
\l| \int_{-\pi}^{\pi} f(x) \sin\l( c\, \theta_{h}(x) \ri) dx \ri|
< \epsilon + \l| \int_{-\pi}^{\pi} g(x)
  \sin\l( c\, \theta_{h}(x) \ri) dx \ri|.
\Leq{lem-rl1}
Define
\[
G(x) := g(x) \sqrt{\frac{p(x)}{1 - \widetilde{V}_{1}(x)}}.
\]
Then by \eqref{eq-t'firstcons}
\beqno
\lefteqn{
\int_{-\pi}^{\pi} g(x) \sin\l( c\, \theta_{h}(x) \ri)\, dx =
  \frac{h}{\mu^{\frac12}} \int_{-\pi}^{\pi} G(x)
    \sin\l( c\, \theta_{h}(x) \ri) \theta'_{h}(x)\, dx
  + {\cal O}\l(\frac{h}{\mu^{\frac12}}\ri)} \\
&=  \frac{h}{c\, \mu^{\frac12}} \Big( \l[ G(x)
  \cos\l( c\, \theta_{h}(x) \ri) \ri]_{-\pi}^{\pi}
  - \int_{-\pi}^{\pi} G'(x) \cos\l( c\, \theta_{h}(x) \ri) dx \Big)
  + {\cal O}\l(\frac{h}{\mu^{\frac12}}\ri).
\eeqno
Hence
\[
\l| \int_{-\pi}^{\pi} g(x) \sin\l( c\, \theta_{h}(x) \ri) dx \ri|
\leq \frac{h}{\mu^{\frac12}} K(g)
< \epsilon
\]
if $\mu$ is large enough, $K(g)$ being a number independent of
$\mu$. The lemma follows by the genericity of $\epsilon$ and
\eqref{lem-rl1}.
\qed

For $\mu\to\infty$, the first term on the right hand side of
\eqref{eq-t'hmu} can be rewritten expanding the square root as
\[
\textstyle\frac1h \sqrt{\frac{\mu-V_{1}(x)}{p(x)}} =
\frac{\mu^{\frac12}}{h\sqrt{p(x)}} \l(1- \frac{V_{1}(x)}{2\mu}
  + {\cal O}\l(\mu^{-2}\ri) \ri) =  \frac{\mu^{\frac12}}{h\, \sqrt{p(x)}}
  - \frac{V_{1}(x)}{2 h \mu^{\frac12}\,\sqrt{p(x)}}
  + {\cal O}\l(\frac{1}{h\mu^{\frac32}}\ri).
\]

Then, in the case $p(x) = 1$,
\beq
\theta'_{h}(x,\mu) = \frac1{h} \sqrt{ \mu-V_{1}(x) } -
       \frac14 \frac{ V_{1}'(x) }{ \mu - V_{1}(x) }
         \sin(2 \theta_{h}(x,\mu) ),
\Leq{eq-t'our}
and asymptotically as $\mu\to\infty$ the first term on the right hand
side becomes
\beq
\frac1h \sqrt{\mu-V_{1}(x)}
=  \frac{\mu^{\frac12}}{h} - \frac{V_{1}(x)}{2 h \mu^{\frac12}}
  + {\cal O}\l(\frac{1}{h\mu^{\frac32}}\ri).
\Leq{eq-t'expand}

Let $\mu_{n}$ ($n\in\bN$) denote the eigenvalues of the
Sturm-Liouville periodic problem
\eqref{eq-hsleq} in ascending order (the potential being denoted
  by $V$ instead of $V_1$).
By standard theory of Sturm-Liouville problems (see \cite[Theorems
2.3.1 and 3.1.2]{Ea}) the spectrum is pure point,
  and the $\mu_{n}$ are at most doubly degenerate and accumulate at
infinity.

\begin{theorem}\label{thm-hmu1}
  Let $p(x)=1$. Then as $m\to\infty$, $\mu_{2m+1}$ and $\mu_{2m+2}$
  both satisfy
\[
\sqrt{\mu} = (m+1)h + \frac{\int_{-\pi}^{\pi} V(x)\, dx}{4\pi(m+1)h}
  + o\l(\frac1{mh}\ri).
\]
\end{theorem}
\proof
Fix an $\epsilon > 0$. Let $V_{1}$ be a continuously differentiable
function with period $2\pi$ such that
\beq
V_{1}(x) \geq V(x)
\qmbox{and}
\int_{-\pi}^{\pi }V_{1}(x)\,  dx \leq \epsilon + \int_{-\pi}^{\pi} V(x)\,  dx.
\Leq{eq-thmVhyp}

Let $\mu_{1,n}$ denote the eigenvalue in the periodic problem
associated with $V_{1}(x)$ (and with $p(x)=1$) and $\psi_{1,n}$ its
eigenfunction. Then by \cite[Theorem 2.2.2]{Ea} and the first eq.\ in
\eqref{eq-thmVhyp} we have
\[
\mu_{1,n} \geq \mu_{n}.
\]
We can assume that $\psi_{1,n}(-\pi)\geq 0$ and we apply the modified
Pr\"ufer transformation to $y(x) = \psi_{1,2m+1}(x)$ with $a_{0} =
-\pi$ in \eqref{eq-ta0ea}. Now, from \eqref{eq-ta0ea} and
\eqref{eq-typerea} we have
\[
2k\pi \leq \theta(\pi,\mu_{1,2m+1}) < (2k+1)\pi
\]
for some integer $k$. From the standard theory of Sturm-Liouville
problems (see aforementioned reference) we know that $\psi_{1,2m+1}$
has $2(m+1)$ zeroes in $(-\pi,\pi]$, hence by \eqref{eq-tboundea} with
$a_{1}=\pi$ we have $2k = 2(m+1)$ and thus
\beq
\theta_{h}(\pi,\mu_{1,2m+1})-\theta(-\pi,\mu_{1,2m+1}) = 2(m+1)\pi.
\Leq{eq-tdiff}

Integrating \eqref{eq-t'our} with $\mu=\mu_{1,2m+1}$ over $[-\pi,\pi]$
we obtain
\beq\hspace*{-1.5cm}
2(m+1)\pi = \int_{-\pi}^{\pi}
    \frac1{h} \sqrt{ \mu-V_{1}(x) }\, dx
  - \frac14 \int_{-\pi}^{\pi} \frac{V_{1}'(x) }{ \mu - V_{1}(x) }
    \sin(2 \theta_{h}(x,\mu) ) \, dx.
\Leq{eq-goodapprox}
By Lemma \ref{lemma-asymptsin2t} the rightmost term is $o(\mu^{-1})$
as $m\to\infty$ (becoming $o(h/m^{2})$ in \eqref{eq-mu1approxin} and
thus being suppressed from the equation). For the first integral on
the right we can use the binomial expansion as in
\eqref{eq-t'expand}. Thus \eqref{eq-goodapprox} gives
\[
2(m+1)\pi = \frac{\mu^{\frac12}}{h} 2\pi
  - \frac{\int_{-\pi}^{\pi}V_{1}(x) \, dx}{2 h \mu^{\frac12}}
  + {\cal O}\l(\frac{1}{h\mu^{\frac32}}\ri)
\]
that is
\[
\mu - (m+1)h \sqrt{\mu}
    - \frac{1}{4\pi} \int_{-\pi}^{\pi} V_{1}(x)\, dx
    + {\cal O}\l(\frac{1}{\mu}\ri) = 0.
\]
Solving for $\mu$ one gets
\[
\sqrt{\mu} = \frac12 \l( (m+1)h + \sqrt{(m+1)^{2}h^{2}
  + \frac{1}{\pi}\int_{-\pi}^{\pi}V_{1}(x)\,dx + {\cal O}(\mu^{-1})} \ri).
\]
Extracting $(m+1)h$ and using once more the binomial expansion one gets
\beq
\sqrt{\mu_{2m+1}} = (m+1)h
  + \frac{\int_{-\pi}^{\pi} V_{1}(x) \, dx}{4\pi(m+1)h}
  + {\cal O}\l(\frac1{m^{2}h^{2}}\ri).
\Leq{eq-mu1approxin}
Hence by \eqref{eq-thmVhyp} and by the fact that $\epsilon$ is
arbitrarily small
\[
\sqrt{\mu_{2m+1}} \leq (m+1)h + \frac{\int_{-\pi}^{\pi} V(x) \, dx}{4\pi(m+1)h} + o\l(\frac1{mh}\ri).
\]
The opposite inequality can be proved in the same way. The result for
$\mu_{2m+1}$ holds in the same form using the fact that its
eigenfunction must have $2(m+1)$ zeroes.
\qed

So far we have not used any differentiability-related property of
$V$. Using the differentiability,
we can make the previous estimate much more precise for $m$
  large.
\begin{theorem}\label{thm-hmu2}
  Let $p(x)=1$, let
  $r\in\bN$, and let $\frac{d^{r}}{dx^{r}} V(x)$ exist and be
  piecewise continuous. Then $\mu_{2m+1}$ and $\mu_{2m+2}$ both
  satisfy
\[
\sqrt{\mu} = (m+1)h + \sum_{k=1}^{r+1} \frac{A_{k}}{(m+1)^{k}h^{k}} + {\cal O}\l(\frac1{m^{r+2}h^{r+2}}\ri) + o\l(\frac1{m^{r+1}h^{r-2}}\ri)
\]
where the $A_{k}$ are independent of $m$ and involve $q(x)$ and its
derivatives up to order $r-1$. In particular,
\beq
A_{1} = \frac{1}{4\pi} \int_{-\pi}^{\pi} V(x)\,  dx\qmbox{,} A_{2} = 0
\qmbox{and}
A_{3} = \frac1{16\pi} \int_{-\pi}^{\pi} V^{2}(x) \, dx - A_{1}^{2}.
\Leq{eq-mucoeff}
\end{theorem}
\proof
We consider $V_{1}=V$ in \eqref{eq-t'our}.
Then $\mu_{1,n} = \mu_{n}$ and the case $r=1$ corresponds simply to
\eqref{eq-mu1approxin}.
To deal with $r\geq2$ we reconsider \eqref{eq-goodapprox}, which
is now
\beq
2(m+1)\pi = \int_{-\pi}^{\pi} \frac1{h} \sqrt{ \mu-V(x) } \, dx
  - \frac14 \int_{-\pi}^{\pi} \frac{ V'(x) }{ \mu - V(x) }
    \sin(2 \theta_{h}(x,\mu) ) \, dx
\Leq{eq-goodapprox1}
and $\mu$ is $\mu_{2m+1}$ or $\mu_{2m+2}$. By \eqref{eq-t'our},
with $V_{1} = V$, the second integral on the right in
\eqref{eq-goodapprox1} is
\begin{align}\hspace*{-.1cm}
\int_{\pi}^{\pi}\frac{h V'(x)}{(\mu-V(x))^{\frac32}}&\l(\theta'_{h}(x,\mu)
  + \frac14 \frac{ V_{1}'(x) }{ \mu - V_{1}(x) }
    \sin(2 \theta_{h}(x,\mu) ) \ri)
    \sin(2 \theta_{h}(x,\mu) )  \, dx \nonumber \\
= {} & \frac h2 \int_{-\pi}^{\pi}
  \l( \frac d{dx} \frac{V'(x)}{(\mu-V(x))^{\frac32}} \ri)
    \cos (2 \theta_{h}(x,\mu) ) \, dx \label{eq-al1} \\
  {} & + \frac h8 \int_{-\pi}^{\pi}
      \frac{V'^{2}(x)}{(\mu-V(x))^{\frac52}} \, dx
    - \frac h8 \int_{-\pi}^{\pi} \frac{V'^{2}(x)}{(\mu-V(x))^{\frac52}}
      \cos (4 \theta_{h}(x,\mu) ) \, dx \nonumber
\end{align}
after integrating by parts. The first term on the right here is $o\l(h
\mu^{-\frac32}\ri)$ by Lemma \ref{lemma-asymptsin2t}, the last is
$o\l(h \mu^{-\frac52}\ri)$ for the same reason and the central one is
${\cal O}\l(h \mu^{-\frac52}\ri)$. This, together with the binomial
expansion of $\sqrt{ \mu-V(x) }$ in the first term on the right of
\eqref{eq-goodapprox1} gives
\beq\hspace*{-1.1cm}
2(m+1)\pi = \frac{\mu^{\frac12}}{h} 2\pi
  - \frac{\int_{-\pi}^{\pi}V(x) \, dx}{2 h \mu^{\frac12}}
  - \frac{\int_{-\pi}^{\pi}V^{2}(x) \, dx}{8 h \mu^{\frac32}}
  + {\cal O}\l( \frac1{h\mu^{\frac 52}}\ri)
  + o\l(\frac h {\mu^{\frac32}}\ri)
\Leq{eq-gapp1}

To solve \eqref{eq-gapp1} for $\mu^{\frac12}$ in terms of $m$, we
write it as
\beq
\mu^{\frac12} = M + \mu^{-\frac12} A_{1}
  + \mu^{-\frac32}(A_{3}-A_{1}^{2})
  + {\cal O}\l( \frac1{m^{5} h^{5}}\ri)
  + o\l(\frac {1}{m^{3}h}\ri)
\Leq{eq-mordsol}
where $M=h(m+1)$. Then, taking the reciprocals we obtain
\begin{align}
\mu^{-\frac12} &= M^{-1} \l( 1 - \mu^{-\frac12} A_{1} M^{-1}
  + {\cal O}(h^{-4}m^{-4})\ri) = M^{-1} - M^{-3} A_{1}
  + {\cal O}(h^{-5}m^{-5}). \label{eq-al2}
\end{align}
And thus,
\beq
\mu^{-\frac32} = M^{-3}+{\cal O}(h^{-5}m^{-5}).
\Leq{eq-mu-3}

Substituting \eqref{eq-al2} and \eqref{eq-mu-3} into
\eqref{eq-mordsol} give the result for $r=2$.

To deal with $r=3$, we introduce $\theta'(x,\mu)$ into the integrals
in \eqref{eq-al1} involving $\cos (2 \theta_{h}(x,\mu) )$ and $\cos (4
\theta_{h}(x,\mu) )$, exactly as we did for \eqref{eq-goodapprox1}.
Then, if $\frac {d^{3}}{dx^{3}}V(x)$ exists and is piecewise
continuous, we can integrate by parts as before.  The binomial
expansions of $\frac1{h} \sqrt{ \mu-V(x) }$ and $( \mu-V(x)
)^{-\frac32}$ extend \eqref{eq-gapp1} to $o\l(h^{2}\mu^{-\frac52}\ri)
+ {\cal O}\l(h^{-1}\mu^{-\frac72}\ri)$ giving the result for $r=3$.
The process can be continued as long as $q(x)$ is sufficiently
differentiable for the integration by parts to be carried out, and the
theorem is proved.
\qed

\begin{remark}
  We can intend Theorem \ref{thm-hmu2} as the result of analytic
  perturbation theory of
\[
h^2(p(x)y'(x))' + y(x) = 0
\]
(derived from \eqref{eq-hsleq}) in terms of the parameter
$V_1(x)/\mu$. As a consequence we get $A_{2k} = 0$ for all $k\in\bN$.
\hfill$\Diamond$\end{remark}

\addcontentsline{toc}{section}{References}

\bibliographystyle{plain}
\bibliography{biblio}
\end{document}